\def\grad{{\bf \nabla}} \def\hbar{\mathchar'26\mkern-9muh}
 \def\wf{wave function~}

 \def\psiT{\Psi_{\rm T}}
\def\Eloc{{\cal E}}
\gdef\journal#1, #2, #3, 1#4#5#6{               % Journal reference.
    {\sl #1~}{\bf #2}, #3 (1#4#5#6)}            % off: name, vol, page,
    \def\commpureapplmath{\journal Comm. Pure Appl. Math., }
\def\commpureapplmath{\journal Comm. Pure Appl. Math., }

 \def\jcp{\journal J. Chem.
Phys., }         \def\prl{\journal Phys. Rev. Lett., }    \documentstyle[aps,floats]{revtex}
\begin{document}
\title{Many-body trial wave functions for atomic systems and ground
states of small noble gas clusters} \author{Andrei Mushinski and M.P.
Nightingale} \address{
  Department of Physics,\\ University of Rhode Island,\\ Kingston, RI
  02881.  } \maketitle

\begin{abstract} Clusters of sizes ranging from two to five are studied
by variational quantum Monte Carlo techniques.  The clusters consist of
Ar, Ne and hypothetical lighter (``$1 \over 2$-Ne") atoms.  A general
form of trial function is developed for which the variational bias is
considerably smaller than the statistical error of currently available
diffusion Monte Carlo estimates.  The trial functions are designed by a
careful analysis of long- and short-range behavior as a function of
inter-atomic distance;  at intermediate distances, on the order of the
average nearest neighbor distance, the trial functions are constructed
to have considerable variational freedom.  A systematic study of the
relative importance of $n$-body contributions to the quality of the
optimized trial wave function is made with $2\le n \le 5$.  Algebraic
invariants are employed to deal efficiently with the many-body
interactions.  \end{abstract}

\section{Introduction} \label{Introduction_section}

The experimental and theoretical study of clusters of a variety of
sizes and constituents has been pursued vigorously during the last
years.  Our interest in this field is predominantly motivated by the
expectation that it will be possible to perform a detailed comparisons
of experiments and theoretical predictions obtained by computational
physics methods.

In a series of papers various authors have investigated ground and
excited state properties of clusters in a wide range of sizes
\cite{Pan.Piep.Wir.86,Rama.Whaley.90,Rama.Whaley.91,%
rick:doll91,Barnett.Whaley.93}.  For these studies trial wave functions
were used with explicit two- and three-body contributions.  In this
paper we develop trial wave functions for the ground state of small,
bosonic, noble gas clusters. We focus primarily on atoms for which
quantum mechanical effects just start to play a role, such as Ar and
Ne.  Our current goal is to propose and test the quality and numerical
tractability of variational forms for the ground state wave functions
involving higher-order many-body contributions. The form of these trial
functions is derived from that proposed by Umrigar {\it et al.}
\cite{cyrus.trialfunctions} for electronic systems, modified to be
suitable for systems consisting of identical bosons.  Our goal for
future work is to use similar trial functions for excited states, where
quantum effects are expected to be more pronounced.  However, before
dealing with this considerably more complex problem, we have chosen to
decrease the mass of the atoms as a more practical way of enhancing the
quantum mechanical nature of our systems, and therefore, in addition to
Ar and Ne, we also studied ``$1\over 2$-Ne" clusters, which consist of
hypothetical atoms with the same inter-atomic potential as Ne, but only
half its mass.  In this way, for the time being at least, we also avoid
the problem that He clusters are expected to have either extremely
weakly bound states or no bound states at all for the systems as small
as we study.

We use the proposed trial wave functions to obtain variational Monte
Carlo estimates of the ground state energy.  The quality of the
optimized trial wave functions is such that we obtain variational
estimates of the ground state energy that in most cases have smaller
errors than the diffusion Monte Carlo estimates of
Ref.~\onlinecite{rick:doll91}, even though the latter, in contrast to
our variational estimates, have no systematic errors.

The layout of this paper is as follows.  The formulation of the problem
is in Section~\ref{Formulation_section}.
Section~\ref{Trial_Wave_Functions_section} contains a discussion of the
trial functions.  The basic variational form of the trial wave function
is derived within the two-body approximation in Section
\ref{Two-body_approximation_section}. Long-distance and short-distance
asymptotic properties are discussed in
Sections~\ref{Long-distance_section} and \ref{Short-distance_section}.
The intermediate-distance range requires most of the variational
freedom of the trial function.  This is provided by the power series
expansion discussed in Section~\ref{Intermediate-distance_section}.
Trial functions of high quality require many-body contributions.
Section \ref{Many-body_approximation_section} deals with those, and
also with the basis of fundamental polynomial invariants employed as an
efficient numerical tool to implement trial functions with Bose
symmetry as required for the noble gas clusters we consider.  The
numerical results, presented in Section~\ref{sec.results}, illustrate
the effects of truncation of the intermediate-range power series and of
higher-order many-body contributions.  Finally, technical details are
contained in Appendices~\ref{appendix_variance} and
\ref{appendix_invariants}.  Appendix ~\ref{appendix_variance} describes
the Monte Carlo algorithm, trial function optimization, numerical
differentiation, and gives optimized parameters of the Ne$_3$ wave
function as an explicit numerical example.  The invariants mentioned
above are tabulated in Appendix~\ref{appendix_invariants}.

\section{Formulation of the problem} \label{Formulation_section}

The purpose of this paper is to develop and study trial wave functions
for ground states of clusters consisting of a small number of identical
bosons interacting via a pair potential.   The position of the atom $i$
is given by ${\bf r}_i$ and the distance between atoms $i$ an $j$ is
denoted by $r_{ij}=|{\bf r}_j -{\bf r}_i|$.  As the starting point of
our discussion we assume the following dimensionless Hamiltonian for a
cluster of $N$ atoms:  \begin{equation} \label{Ham} H= - \frac{1}{2m}
\sum_{i=1}^{N} \grad_{i}^{2} + \sum_{(i,j)} V(r_{ij}), \end{equation}
where $(i,j)$ runs over all different pairs of atoms; $V$ is the
dimensionless Lennard-Jones potential \begin{equation} \label{LJ_def}
V(r) = {1 \over r^{12}} - {2 \over r^6}.  \end{equation}
Eq.~(\ref{Ham}) and (\ref{LJ_def}) have been made dimensionless by
expressing distances and energies in units of the diameter $\sigma$ and
the well depth $\epsilon$ of the Lennard-Jones potential, and if $\mu$
is the atomic mass, $m^{-1} \equiv \alpha=\hbar^2 /\mu \sigma^2
\epsilon$ is proportional to the square of the de Boer parameter
\cite{deBoer}.

Denote by ${\bf R}=({\bf r}_1,\dots,{\bf r}_N)$ a $3N$-dimensional
vector in configuration space and let $\psiT({\bf R},{\bf p})$ be the
trial wave function in the coordinate representation, where ${\bf p} =
(p_1,p_2,\dots)$ are adjustable parameters. We will omit the arguments
of $\psiT$, where it does not lead to confusion.  Following Umrigar
{\it et al.} \cite{cyrus.trialfunctions}, we optimize the parameters
${\bf p}$ of the trial \wf by minimizing the variance of the local
energy \begin{equation} \label{E_loc_def} \Eloc({\bf R},{\bf p}) =
\frac{1}{\psiT} H \psiT.  \end{equation} If $\psiT$ is the exact ground
state of the Hamiltonian, the local energy $\Eloc$ equals the
corresponding eigenvalue of $\psiT$, independent of $\bf R$.  The
existence of this zero-variance principle for the ideal case of a true
ground state motivates minimization of the variance of the local
energy.  Of course, the variance vanishes for any eigenstate of the
Hamiltonian, but by choosing a strictly positive trial function we can
guarantee that in fact we are approximating the ground state rather
than an excited state.

In principle, the parameters $\bf p$ are determined by minimization of
the variance \begin{equation} \label{chi_integral} \chi^2 ({\bf p}) =
\frac{\int\psiT^* (H-\tilde E_0)^2 \psiT ~d{\bf R}} {\int
|\psiT|^2~d{\bf R}}, \end{equation} where \begin{equation}
\label{e_integral} \tilde E_0({\bf p}) = {\int \psiT^* H \psiT \,d{\bf
R} \over \int |\psiT|^2\,d{\bf R}}, \end{equation} the variational
estimate of the ground state energy. In practice, minimization of
$\chi^2$, as given in Eq.~(\ref{chi_integral}), is performed with the
usual Monte Carlo scheme \cite{cyrus.trialfunctions}.  (We refer the
reader to Appendix~\ref{appendix_variance} for details of the algorithm
and its efficient implementation.)

We define a quantitative measure $Q$ of the quality of a trial \wf
$\psiT$ as \begin{equation} \label{Quality}
  Q = -\log_{10}\frac{ \chi  }
		 { |E_0| },
\end{equation} where the limit $Q \to\infty$ corresponds to an {\em
exact} solution of the time-independent Schr\"odinger equation.  In
terms of the quality $Q$, one can estimate the intrinsic error of the
variational energy $\tilde E_0$. The difference between an exact ground
state energy $E_{0}$ and the variational estimate $\tilde E_{0}$ is
bounded by $\chi$:  \begin{equation} \tilde E_{0} - \chi \leq  E_0 \leq
\tilde E_{0} \label{eq.chibound} \end{equation} (This inequality and
the next one are reviewed in Ref.~\onlinecite{Goedecker:bounds}.) If
one knows the energy $E_1$ of the lowest excited state of the same
symmetry as the trial wave function, a lower bound can be obtained that
is tighter for sufficiently small $\chi^2$, {\it viz.},
\begin{equation} \tilde E_{0} - \frac{\chi^2}{|E_1 - E_0|} \leq  E_{0}
\leq \tilde E_{0}.  \label{eq.chi2bound} \end{equation} Use of
Eqs.~(\ref{eq.chibound}) and (\ref{eq.chi2bound}) $E_0$ implicitly
makes the assumption that the trial wave function approximates the
ground state, rather than an excited state, an assumption that is
justified on physical grounds.

The relative error in the energy expressed in terms of this last
inequality (\ref{eq.chi2bound}) can be measured by \begin{equation}
\label{Quality'} Q'=-\log_{10} {\chi^2 \over (E_1-E_0)|E_0|}.
\end{equation} In estimates of $Q$ in Eq.~(\ref{Quality}) we replace
$E_0$ by the variational estimate $\tilde E_0$; in the case of $Q'$, we
also replace $E_1-E_0$ in Eq.~(\ref{Quality'}) by the approximate gap
obtained from the rather crude, harmonic approximation.

\section{Trial Wave Functions} \label{Trial_Wave_Functions_section}
%++++++++++++++++++++++++++++++

\subsection{Symmetries} \label{Symmetries_section}

A trial wave function for the ground state of a bosonic noble gas
cluster should be invariant under translation, rotation and particle
permutation.  We satisfy the first two of these requirements by
choosing as our coordinates the inter-atomic distances $r_{ij},\ 1\le
i<j \le N$.  This set of coordinates has ${1 \over 2}N(N-1)$ elements
rather than the required minimal number of $\max(3N-6,1)$ and since the
former exceeds the latter for $N\ge 5$, there will be dependences among
the inter-particle distances in that case. However, these dependences
will automatically be satisfied because the $r_{ij}$ will only assume
values derived from configurations given by the $3N$ variables $\bf
R$.  The third condition, particle exchange symmetry, has to be imposed
explicitly by considering only those functions that are invariant under
permutation of the indices of all $r_{ij}$.

In addition to these symmetry restrictions, we impose the condition of
positivity on the trial function.  In principle, $n$-body correlations
with all $n\le N$, should be incorporated in the trial wave function,
but these many-body effects are expected to become progressively less
important as $n$ increases.  Positivity and many-body correlations
suggest that the trial function be written as \begin{equation}
\label{Psi_thru_fF} \log \psiT = \sum_{(i,j)} u^{(2)}(r_{ij}) +
	     \sum_{(i,j,k)} u^{(3)}(r_{ij},~r_{jk},~r_{ki})+\cdots +
	     \sum_{(i_{1},\dots ,i_{N})} u^{(N)}(r_{i_1 i_2},\dots),
\end{equation} where the $u^{(n)}$ are real-valued, $n$-body
functions.

\subsection{Two-body approximation}
\label{Two-body_approximation_section}

The symmetries discussed above are not the only restrictions on the
form of the wave function given in Eq.~(\ref{Psi_thru_fF}).  Further
restrictions are derived from the asymptotic behavior of the local
energy for some particular cases of the $r_{ij} \to 0$ and $r_{ij} \to
\infty$ limits.  For the $r_{ij} \to 0$ limit we consider the special
case where the pair distance $r_{ij}$ of one arbitrary pair vanishes,
while all other pair distances remain finite and non-zero.  Collisions
involving more than two particles impose additional asymptotic
conditions on the wave function, but these are presumably less
important and will be ignored.  Also in the $r_{ij} \to \infty$ limit
we restrict ourselves to a special case:  one particle goes off to
infinity, while all others stay fixed.

First we derive the expression for the local energy using the two-body
approximation to the wave function,  i.e., \begin{equation}
\label{Psi_thru_f} \log \psiT =  \sum_{(i,j)} u^{(2)}(r_{ij}).
\end{equation} The general expression for the local energy is
\begin{equation} \label{E_loc_general} \Eloc = -\frac{\alpha}{2}
\sum_{i=1}^{N} \Bigl[\grad_{i}^{2} \log \psiT+ (\grad_{i} \log
\psiT)^{2}\Bigr] + \sum_{(i,j)} V(r_{ij}).  \end{equation} For a trial
wave function with only two-body correlations this equation reduces to
\begin{equation} \label{E_loc_full} \Eloc = - \frac{\alpha}{2}\sum_{i}
\Bigl[\sum_{j \ne i} u'(r_{ij}) {\bf e}_{ij} \Bigr]^{2} - \alpha
\sum_{(i,j)} \Bigl[u''(r_{ij}) + \frac{2}{r_{ij}}u'(r_{ij}) \Bigr]+
\sum_{(i,j)} \Bigl[ \frac{1}{r_{ij}^{12}}- \frac{2}{r_{ij}^6} \Bigr],
\end{equation} where $u'$ and $u''$ denote first- and second-order
derivatives of $u^{(2)}$; ${\bf e}_{ij}$ denotes the unit vector $({\bf
r}_{j} - {\bf r}_{i})/r_{ij}$.

The current approximation to the wave function contains no explicit
three-body correlations, yet the local energy contains three-body
contributions as shown explicitly by rewriting Eq.~(\ref{E_loc_full})
in the following form \begin{equation} \label{E_loc_UE} \Eloc =
\Eloc^{(2)} + \Eloc^{(3)} \equiv \sum_{(i,j)} \Eloc_{ij} +
\sum_{(i,j,k)} \Eloc_{ijk}, \end{equation} where \begin{equation}
\label{E_pair} \Eloc_{ij} = - \alpha \Bigl[ u''(r_{ij}) +
\frac{2}{r_{ij}}u'(r_{ij}) + u'^{2}(r_{ij})\Bigr] + \Bigl(
\frac{1}{r_{ij}^{12}}- \frac{2}{r_{ij}^6} \Bigr) \end{equation} is
local energy of a pair, and where, writing ${\bf u}_{ab} \equiv
u'(r_{ab}) {\bf e}_{ab}$, one has \begin{equation} \label{U_triplet}
\Eloc_{ikm} = \alpha (
 {\bf u}_{ki} \cdot {\bf u}_{im} + {\bf u}_{ik} \cdot {\bf u}_{km} +
 {\bf u}_{im} \cdot {\bf u}_{mk} ), \end{equation} which is the
`connected' local energy of a triplet, i.e., the contribution to the
local energy not accounted for by pair contributions.  $\Eloc^{(3)}$ in
Eq.~(\ref{E_loc_UE}) can be rewritten in terms of two-particle
interactions mediated by all other particles:  \begin{equation}
\label{W_def} \Eloc^{(3)} = \alpha \sum_{(ij)} \sum_{p(\ne i,j)} {\bf
u}_{ip} \cdot {\bf u}_{pj}.  \end{equation}

\subsubsection{Long-distance behavior} \label{Long-distance_section}

To derive the asymptotic properties at infinity we consider
configurations in which one atom goes off to infinity while the others
remain fixed.  Eq.~(\ref{E_loc_full}) shows that the conditions,
firstly, that the local energy remain finite, and, secondly, that the
atoms in the cluster be in a bound state, imply that $u'(r)$ approaches
a real, negative constant for $r \to \infty$.

\subsubsection{Short-distance behavior} \label{Short-distance_section}

As is well-known, good trial functions for electron systems satisfy the
`cusp' condition \cite{cuspcondition}, which guarantees that for
two-body collisions the contribution to the local energy due to the
divergent electron-electron (or electron-nucleus) Coulomb potential
energy is cancelled by a divergence of opposite sign in the kinetic
energy.  Analogously, as a first step in dealing with the divergence in
$\Eloc$ at short distances for the case of the Lennard-Jones potential,
we suppose that one pair distance, say $r_{12}$, is small compared to
unity and all other pair distances. Then the dominant contributions to
$\Eloc$ in Eq.~(\ref{E_loc_full}) come from the divergences at
$r_{12}=0$, and if one keeps only terms that contain $r_{12}$, the
singular part of $\Eloc$ can be approximated as a function of only the
one argument $r \equiv r_{12}$:  \begin{equation}
\label{E_loc_singular} \Eloc \approx - \alpha u''(r) - 2\alpha
(\frac{1}{r} + \xi ) u'(r) - \alpha u'^{2}(r) + \frac{1}{r^{12}} -
\frac{2}{r^{6}}, \end{equation} where $\xi$ plays the role of a `random
field' associated with the positions of particles $3,\dots ,N$:
\begin{equation} \xi = -\frac{1}{2} \sum_{i>2} \Bigl[ u'(r_{i1}){\bf
e}_{21}\cdot{\bf e}_{1i} + u'(r_{2i}){\bf e}_{12}\cdot{\bf e}_{2i}
\Bigr], \end{equation} a field that vanishes only for a diatomic
cluster.

We choose \begin{equation} u'(r) = b_{-6}/r^6 + b_{-1}/r + b_0 + b_4
r^4 + b_5 r^5 + {\rm O}(r^6) \label{eq.ub} \end{equation} with
coefficients to be determined to cancel the short-distance divergences
in the local energy. For the local energy this yields \begin{eqnarray}
\Eloc(r) &=&{{1 - \alpha b_{-6}^2 }\over {r^{12}}} -
   {{2 \alpha b_{-6} (b_{-1}-2)} \over {r^7}} - {{2 \alpha b_{-6} (b_0
   + \xi) +2 }\over {r^6} } -                    \nonumber \\
   &&{{\alpha ( 2 b_{-6} b_4 + b_{-1} + {b_{-1}^2} ) }\over {r^2} } -
   {{2 \alpha ( b_0 + b_{-6} b_5 + b_{-1} b_0 + b_{-1} \xi  ) }\over
   {r} }-     \nonumber \\ &&{\alpha b_0 ( b_0 + 2 \xi  )  }- 2 \alpha
   b_4 ( 3 + b_{-1} )  r^3 + {\rm O}(r^4) \label{LocEDiv}.
\end{eqnarray}

The following choice of the  $b_i$ for a given value of the random
field $\xi$ eliminates the power law divergences:  \begin{eqnarray}
b_{-6}&=& {1\over {\sqrt{\alpha}}},\ b_{-1}=2,\ b_0=-{1\over
{\sqrt{\alpha}}} - \xi,  \nonumber \\ b_4&=&-{{3 {\sqrt{\alpha}}}},\
b_5={3} + {{{\sqrt{\alpha}}\,\xi}}.  \label{BestFit} \end{eqnarray}
With the $b_i$ as given in Eqs.~(\ref{BestFit}), the $r^{-12},r^{-7}$
and $r^{-2}$ divergences in the local energy in Eq.~(\ref{LocEDiv}) are
eliminated simultaneously for all configurations of the $N$ atoms.
However, since the expressions for $b_0$ and $b_5$ depend on the random
field $\xi$, the best one can do with the remaining $r^{-6}$ and
$r^{-1}$ divergences is to have their amplitudes vanish in an average
sense. For the local energy this particular choice, i.e., replacing
$\xi$ by its average value, yields \begin{eqnarray} \Eloc(r)&=&-{{2
\sqrt\alpha \eta} \over {r^6}} -
	{{4 \alpha \eta} \over {r}} + \nonumber \\ && -{[(1-\sqrt\alpha
\eta)^2 - \alpha \xi^2] + 30 \alpha ^{3/2} r^3 + {\rm O}(r^4)},
\label{eq.best.Eloc} \end{eqnarray} with $\eta \equiv \xi -\langle \xi
\rangle$.  Many of the papers mentioned in the introduction contain
trial functions with singularities as in Eq.~(\ref{eq.ub}) and employ
these both for the Lennard-Jones and the Aziz potentials.  The main
difference is that in the derivation given above the coefficients
assume fixed values, while the nature of the terms is determined to
cancel divergences explicitly associated with the Lennard-Jones
potential. In our approach other potentials would require different
terms.

\subsubsection{Intermediate-distance behavior}
\label{Intermediate-distance_section}

The behavior of the trial function at distances of most physical
significance, distances of order unity, is conveniently expressed in
terms of a polynomial.  In order to avoid the problem that high-order
polynomial terms dominate the behavior of the pair correlation function
$u^{(2)}$ at infinity, we introduce a new (shifted) distance variable
$\hat r$ which approaches a constant for $r \to \infty$:
\begin{equation} \label{eq.truncvar}
    \hat r(r) = w\{1-e^{(r_0-r)/ w}\}.  \end{equation} Here $w$ and
$r_0$ in principle are variational parameters.  In practice, the
quality of the wave function depends only weakly on their values. They
are set to values of order unity and kept fixed during the optimization
of the trial function, which allows us to greatly improve the
efficiency of the optimization algorithm, as explained in detail in
Appendix~\ref{appendix_variance}.  The parameter $w$ is chosen to
reflect the length scale on which correlations exist in the cluster.
It should be noted that for values of $w\gg 1$ the optimization based
on a sample of states generated by Monte Carlo ({\it cf.}
Appendix~\ref{appendix_variance}) becomes unstable, since the trial
function develops too much variational freedom in regions of
configuration space associated with cluster conformations of very low
probability. The shift $r_0$ is included for the purpose of numerical
accuracy, so that high powers of $\hat r$ assume small values.

In terms of this new variable $\hat r$ we can combine all required
behaviors discussed above, by choosing the following variational
expression for $u^{(2)}$ for the trial function :  \begin{equation}
\label{eq.u2} u^{(2)}(r) = -\frac{1}{5 r^{5}\sqrt{\alpha}} -{\gamma
r\over \sqrt{\alpha}} + 2 \ln r +\sum_{p=1}^P c_{p} \hat r^p.
\end{equation} This form, with $\gamma$ and the $c_i$ as variational
parameters, provides the required variational freedom for the trial
function at intermediate distances, while it also displays the
asymptotic behavior derived above for small and large distances. Note
that Eq.~(\ref{eq.u2}) corresponds to keeping the parameters $b_i$ with
$i\ge 0$ as free parameters, rather than fixing their values by the
expressions given in Eq.~(\ref{BestFit}).

\subsection{Many-body approximation}
\label{Many-body_approximation_section}

The quality of a trial function obtained by exponentiation of the sum
over pairs of $u^{(2)}(r_{ij})$, as given in Eq.~(\ref{eq.u2}), cannot
be increased arbitrarily by increasing the degree $P$ of the polynomial
in $\hat r$.  At some point, the effect of the presence of high-order
two-body terms in the polynomial becomes smaller than the effect of the
absence of three-body contributions.  In principle, one has to allow
for $n$-body interactions with $n\le N$, and for this purpose we add
polynomials in all variables $\hat r_{ij}\,(i<j)$, defined in terms of
the original inter-atomic distances as in Eq.~(\ref{eq.truncvar}).  The
general form of these $n$-body polynomials for $n>2$ is
\begin{equation} \label{eq.un} u^{(n)}=\sum_{p=n-1}^P \mathop{{\sum}'}_
{\scriptstyle 0\le p_{12},p_{13},\dots,p_{N-1,N} \le p \atop
\scriptstyle p_{12}+p_{13}+\cdots +p_{N-1,N}=p}
c_{p_{12},p_{13},\dots,p_{N-1,N}} \hat r_{12}^{p_{12}} \hat
r_{13}^{p_{13}} \cdots \hat r_{N-1,N}^{p_{N-1,N}}, \end{equation} where
the prime on the summation indicates that the exponents $p_{ij}$ should
be chosen consistent with the condition that $u^{(n)}$ represent an
$n$-body interaction; that is, the set of indices $i$ for which a $j$
exists such that $p_{ij} \ne 0$ contains precisely $n$ elements.  Since
we are dealing with Bosons, the wave function should be symmetric under
particle permutation, i.e., $c_{p_{12},p_{13},\dots,p_{N-1,N}}=
c_{p'_{12},p'_{13},\dots,p'_{N-1,N}}$ if $\hat r_{12}^{p_{12}} \hat
r_{13}^{p_{13}} \cdots \hat r_{N-1,N}^{p_{N-1,N}}$ and $\hat
r_{12}^{p'_{12}} \hat r_{13}^{p'_{13}} \cdots \hat
r_{N-1,N}^{p'_{N-1,N}}$ can be obtained from each other by a particle
permutation.

Dealing explicitly with the particle permutation symmetry for general
$n$-body interactions in the context of polynomials of the form of
Eq.~(\ref{eq.un}) is rather cumbersome from the point of view of
programming, and also computationally expensive.  As an alternative, we
constructed, for each cluster size $N$, a basis of $M$ fundamental
polynomial invariants $I_{N1},\dots,I_{NM} $\cite{ref.Invariants}.  In
terms of the basis formed by the $I_{Nk}$ one can conveniently express
all polynomials in the variables $\hat r_{12},\hat r_{13},\dots$ that
are symmetric under all permutations of indices.  Each fundamental
invariant $I_{Nk}$ can be chosen to be a homogeneous, permutationally
symmetric polynomial of degree $d_k$ in the variables $\hat r_{ij}$.
Expressed in terms of these fundamental invariants, the logarithm of
the trial function reads \begin{equation} u({\bf R}) =
-\sum_{(ij)}\left( \frac{1}{5 \sqrt{\alpha}r_{ij}^{5}}+ {\gamma r_{ij}
\over \sqrt{\alpha}} - 2\ln r_{ij} \right) + \sum_{\scriptstyle
q_1,q_2,\dots,q_M  \atop \scriptstyle q_1 d_1+q_2 d_2+\cdots +q_M d_M
\le P} C_{q_1,\dots,q_M}I_{N1}^{q_1} \cdots I_{NM}^{q_M}.
\label{eq.u.invar} \end{equation} As a simple example we discuss the
case of a three particle cluster.  For $N=3$ a possible classic choice
for a basis of fundamental invariants is \begin{eqnarray} I_{31} &=&
\hat r_{12}+\hat r_{23}+\hat r_{13},\nonumber\\ I_{32} &=& \hat
r_{12}^2+\hat r_{23}^2+\hat r_{13}^2,\\ I_{33} &=& \hat r_{12} \hat
r_{23} \hat r_{13}\nonumber.  \end{eqnarray} Fundamental invariants for
the cases of three, four, and five particles are given in
Appendix~\ref{appendix_invariants}.

We end this section with a few comments regarding the efficiency of
employing a basis of fundamental invariants.  The number of terms of a
polynomial in $v$ variables of order $P$ is given by the binomial
coefficient $\left( v+p \atop p \right)$.  This implies that for a five
atom cluster the full many-body polynomial of degree five in the ten
variables $\hat r_{ij}\ (1\le i<j\le 5)$ has 3003 terms, if all
possible many-body interactions and a constant are included.  If one
evaluates such a polynomial by recursively applying Horner's rule to
all variables this requires 3003 multiplications, which we shall take
as the measure of the computational effort. Of course, this approach
would also require a scheme of equating polynomial coefficients to
impose the restriction to polynomials of the required symmetry.

Written as a polynomial in the fundamental invariants listed in
Appendix~\ref{appendix_invariants} there are only 64 terms.
%Found
%with DioSolve: 1+3+7+17+35=63 for orders 1 through 5 (Jan. 20,1994)
Since the evaluation of the invariants given in
Appendix~\ref{appendix_invariants} takes about 700 multiplications, use
of invariants speeds up the calculation by a factor almost equal to
four, a number that might be improved since we made no systematic
attempt to optimize the numerical efficiency of the bases of
fundamental invariants.

Simple use of Horner's rule applied recursively to the variable $\hat
r_{12},\hat r_{13},\dots$, makes no use at all of the Boson symmetry of
the polynomial.  As an alternative to employing a basis of fundamental
invariants, the particle permutation symmetry can be exploited by
constructing the terms in the polynomial diagrammatically.  This has
the advantage of providing naturally a separation of the polynomial in
contributions separated into $n$-body terms sorted according to
different values of $n$.  For $N=P=5$ example discussed above the use
of collecting and factorizing terms diagrammatically reduces the effort
by about a third compared to the brute force approach using Horner's
rule.

These considerations apply to the case of one single evaluation of the
trial wave function, but in practice this is not always the operation
of interest.  For example, during the optimization phase one only has
to compute the change in a trial function that results from the change
in the variational parameters (see Appendix~\ref{appendix_variance}).
For a variational or diffusion Monte Carlo calculation what may matter
is the change of the wave function in response to a change of the
coordinates of just one atom.  In such cases the increase in speed
associated with the use of invariants may be even greater.

In its simplest implementation use of invariants mixes up all many-body
interactions and destroys the separation into $n$-body interactions for
different values of $n$ and the hierarchy as implied in
Eq.~(\ref{Psi_thru_fF}).  There is a solution to this problem, but we
have not yet explored it in detail nor is it relevant within the
limited scope of this paper.

\section{Numerical results} \label{sec.results}

Our first numerical results address how much accuracy is gained in the
quality of the trial wave function as $n$-body terms with progressively
larger values of $n$ are included.  Figs.~\ref{fig.Ar3} through
\ref{fig.Ar5} display results for argon clusters of sizes three, four,
and five.  $Q$, our most conservative estimate of the quality of the
wave functions as defined in Eq.~(\ref{Quality}), is plotted versus the
power $P$ of the polynomials defined in Eqs.~(\ref{eq.u2}) and
(\ref{eq.un}), for different values of $n$.  The bottom curve, starting
at $P=1$, is for the case in which only two-body interactions are
present ($n=2$).  As shown, the quality levels off at a fixed value of
$Q$ with increasing $P$, an indication that in that regime the absence
of three-body terms in the logarithm of the trial wave function is the
dominant source of the variance of the local energy.  In the next curve
segment three-body terms are added ($n=3$). Here $P$ is redefined to
denote the order of three-body polynomial, which starts at $P=2$; in
the second curve segment the order of the polynomial that describes
two-body effects is kept constant at the highest $P$-value used in the
previous segment of the curve.  This process is repeated for $n$-body
terms with increasing $n$ until finally the complete $N$-body
polynomial for the $N$-atom cluster is included in the trial function.
At this point, the quality starts to go up roughly linearly with the
order of the polynomial.  Note, however, that the order $P$ of the
four-body polynomials is not sufficiently high in any of the figures
for the quality to have leveled off, as ultimately it must.  Analogous
plots for $\mbox{Ne}_4$ and $\mbox{Ne}_5$ are shown in
Figs.~\ref{fig.Ne4} and \ref{fig.Ne5}, and for ${1\over
2}\mbox{-Ne}_3$, ${1\over 2}\mbox{-Ne}_4$, and ${1\over 2}\mbox{-Ne}_5$
are shown in Figs.~\ref{fig.hNe3}, \ref{fig.hNe4} and \ref{fig.hNe5}.
%1
 \begin{figure} \centerline{\input argon3} \vskip 0.5cm
\caption[f1]{Quality $Q$, a measure of the accuracy of the optimized
trial function as defined in Eq.~(\ref{Quality}), as a function of the
power $P$ of two- and three-body polynomials, labeled by $n$, for
$\mbox{Ar}_3$.} \label{fig.Ar3} \end{figure}
%2
 \begin{figure}
\centerline{\input argon4} \vskip 0.5cm \caption[f1]{Quality $Q$, a
measure of the accuracy of the optimized trial function as defined in
Eq.~(\ref{Quality}), as a function of the power $P$ of two- , three-
and four-body polynomials, labeled by $n$, for $\mbox{Ar}_4$.}
\label{fig.Ar4} \end{figure}
%3
 \begin{figure} \centerline{\input
argon5} \vskip 0.5cm \caption[f1]{Quality $Q$, a measure of the
accuracy of the optimized trial function as defined in
Eq.~(\ref{Quality}), as a function of the power $P$ of two-, three-,
four- and five-body polynomials, labeled by $n$, for $\mbox{Ar}_5$.}
\label{fig.Ar5} \end{figure}
%4
 \begin{figure} \centerline{\input
neon4} \vskip 0.5cm \caption[f1]{Quality $Q$, a measure of the accuracy
of the optimized trial function as defined in Eq.~(\ref{Quality}), as a
function of power $P$ of two-, three- and four-body polynomials,
labeled by $n$, for $\mbox{Ne}_4$.} \label{fig.Ne4} \end{figure}
%5
\begin{figure} \centerline{\input neon5} \vskip 0.5cm
\caption[f1]{Quality $Q$, a measure of the accuracy of the optimized
trial function as defined in Eq.~(\ref{Quality}), as a function of
power $P$ of two-, three-, four- and five-body polynomials, labeled by
$n$, for $\mbox{Ne}_5$.} \label{fig.Ne5} \end{figure}
%6
\begin{figure}
\centerline{\input half3} \vskip 0.5cm \caption[f1]{Quality $Q$, a
measure of the accuracy of the optimized trial function as defined in
Eq.~(\ref{Quality}), as a function of power $P$ of two- and three-body
polynomials, labeled by $n$, for ${1 \over 2}\mbox{-Ne}_3$.}
\label{fig.hNe3} \end{figure}
%7
\begin{figure} \centerline{\input
half4} \vskip 0.5cm \caption[f1]{Quality $Q$, a measure of the accuracy
of the optimized trial function as defined in Eq.~(\ref{Quality}), as a
function of power $P$ of two-, three- and four-body polynomials,
labeled by $n$, for ${1 \over 2}\mbox{-Ne}_4$.} \label{fig.hNe4}
\end{figure}
%8
\begin{figure} \centerline{\input half5} \vskip 0.5cm
\caption[f1] {Quality $Q$, a measure of the accuracy of the optimized
trial function as defined in Eq.~(\ref{Quality}), as a function of
power $P$ of two-, three-, four- and five-body polynomials, labeled by
$n$, for ${1 \over 2}\mbox{-Ne}_5$.} \label{fig.hNe5} \end{figure}

Figs. \ref{fig.argon}, \ref{fig.neon}, and \ref{fig.half} display plots
of the quality $Q$ {\it vs.} the power $P$ for each type and size of
cluster for optimized wave functions including the full many-body
polynomials.  An interesting feature of these plots is that the $Q$
{\it vs.} $P$ curves for three- and four-atom ($N=3,4$) clusters almost
coincide but that they are distinct from the curves for $N=2$ and
$N=5$.  The $N=2$ clusters are unique in that the short-distance
divergences in the local energy have been fully removed, (note that
$\eta=0$ in Eq.~(\ref{eq.best.Eloc}) only for $N=2$) while also the
large-distance asymptotic properties of the trial function is superior
in this case.  Apart from this, the effect probably is geometric in
nature: clusters of sizes $N \le 4$ are fully symmetric in the
classical configuration of minimum energy and can be characterized by a
single inter-atomic distance.  For $N=5$ the cluster is frustrated: the
classical ground state is a  trigonal bipyramid and has three different
inter-atomic distances.  To test the above arguments we computed the
quality of five particle clusters in four dimensions, where the
classical configuration of minimum energy is fully symmetric and there
is no longer any frustration.  Indeed, we found that the curve for the
quality as a function of the polynomial power quantitatively agrees
with the curves for three and four particle clusters in three
dimensions.

Note that for very accurate optimized trial functions, the computed
variance of the local energy no longer decreases as more terms are
added to the complete polynomial.  A clear example of this is
$\mbox{Ar}_2$, in Fig.~\ref{fig.argon}. We attribute this to noisy
behavior of the local kinetic energy caused by round-off errors in the
numerical differentiation, which produce an effect of the order of
magnitude observed.  The data plotted in Figs.~\ref{fig.Ar3} through
Figs.~\ref{fig.hNe5} were obtained from the relatively small samples
that were also used to perform the parameter optimization.  Only small
differences were detected in those cases where we checked the variances
obtained from these small samples against those obtained from extensive
Monte Carlo runs. The data shown in Figs.~\ref{fig.argon} through
\ref{fig.half} were obtained from long runs (see below for details).

%9
\begin{figure} \centerline{\input argon} \vskip 0.5cm \caption[f1]
{Quality $Q$, a measure of the accuracy of the optimized trial function
as defined in Eq.~(\ref{Quality}), as a function of power $P$ of the
complete polynomial for argon clusters of sizes two through five.}
\label{fig.argon} \end{figure}
%10
 \begin{figure} \centerline{\input neon} \vskip 0.5cm
\caption[f1]{Quality $Q$,  a measure of the accuracy of the optimized
trial function as defined in Eq.~(\ref{Quality}), as a function of
power $P$ of the complete polynomial for neon clusters of sizes two
through five.} \label{fig.neon} \end{figure}
%11
 \begin{figure}
\centerline{\input half} \vskip 0.5cm \caption[f1]{Quality $Q$,  a
measure of the accuracy of the optimized trial function as defined in
Eq.~(\ref{Quality}), as a function of power $P$ of the complete
polynomial for ${1 \over 2}$-neon clusters of sizes two through five.}
\label{fig.half} \end{figure}

Our most accurate numerical estimates for the energies of various small
clusters are summarized in Table~\ref{tab.energies}.  The results were
obtained by standard variational Monte Carlo methods (see
Appendix~\ref{appendix_variance}) and appear under the heading $E_{\rm
vmc}$ in Table~\ref{tab.energies}.  The estimated averages were
obtained from runs of about $10^7$ Monte Carlo steps per atom.  The
energy auto-correlation time of the sampling algorithm was of the order
of ten steps in these units.  For comparison we included, under the
heading $E_{\rm dmc}$, the diffusion Monte Carlo results of
Ref.~\onlinecite{rick:doll91}, and the estimates of the harmonic
approximation, under $E_{\rm har}$.  Also included in
Table~\ref{tab.energies} are two estimates of the systematic
Rayleigh-Ritz variational errors in the results, and an estimate of the
statistical, Monte Carlo errors.  For the systematic errors we employ
the inequalities given in Eqs.~(\ref{eq.chibound}) and
(\ref{eq.chi2bound}).  As mentioned above, an estimate of the energy
gap in the spectrum is required for the error estimate based on the
second inequality, which provides a bound proportional to the variance
$\chi^2$ of the local energy rather than its standard deviation
$\chi$.  As an order of magnitude estimate of the gap, we employed the
harmonic approximation, and for completeness the required eigenvalues
of the Hessian matrix of the potential energy at the classical minimum
are listed in Table~\ref{tab.harmonic}.  We also mention in passing
that the classical energies are $E^{\rm class}_0
=-1,-3,-6,-9.103\,852\,415\,707\,556$ for $N=2,\dots,5$; all
inter-atomic distances are equal to unity except for $N=5$, where the
trigonal bipyramid has height $0.813\,335\,784\,076\,977$ and as base
an equilateral triangle with sides of length
$1.001\,453\,524\,076\,903$.  The quantity $Q''$ listed in the last of
column Table~\ref{tab.energies} is a measure of the statistical error
relative to the energy and is defined as in Eq.~(\ref{Quality}) with
$\chi$ replaced by the standard error of the Monte Carlo sample.  The
numerical values of the dimensionless inverse masses of the atoms we
considered are listed in Table~\ref{tab.masses}.

We note that, with the exception of the larger clusters, the
variational bias as measured by $Q'$ is smaller than the statistical
error measured by $Q''$, so that only for the larger clusters more
accurate results can be obtained by diffusion Monte Carlo for a
comparable amount of computational effort.  \begin{table}
\caption[tabresults]{ \small \parindent .5cm \narrower Variational
Monte Carlo estimates of the ground state energies $E_{\rm vmc}$ of
clusters containing up to five atoms compared with diffusion Monte
Carlo estimates $E_{\rm dmc}$ taken from Ref.~\onlinecite{rick:doll91}
and the harmonic approximation $E_{\rm har}$.  Standard errors in the
last digit are given in parentheses.  Estimates of the relative errors,
as discussed in the text, are given by the quality estimators $Q$ and
$Q'$ for the systematic variational errors and by $Q''$ for the
statistical errors. Also listed are estimates of the energy gap $\Delta
E_{\rm har}$ between the ground state and the first excited state of
the same symmetry.  No statistical error is quoted for $Q'$: the
dominant error in this case is the unknown error in $\Delta E_{\rm
har}$. In the column under the heading $P$, subsequent digits define
the orders of the $n$-body polynomials used in wave functions.
\par\vspace{5mm}} \begin{tabular} {lcllllllll}
\multicolumn{1}{c}{}&
\multicolumn{1}{c}{$N$} &
\multicolumn{1}{c}{$P$} &
\multicolumn{1}{c}{$E_{\rm vmc}$} &
\multicolumn{1}{c}{$E_{\rm dmc}$} &
\multicolumn{1}{c}{$E_{\rm har}$} &
\multicolumn{1}{c}{$\Delta E_{\rm har}$} &
\multicolumn{1}{c}{$Q$} &
\multicolumn{1}{c}{$Q'$} &
\multicolumn{1}{c}{$Q''$}\\
\tableline

%==========================================================================================================
%#+Mol	P	E(dE)			E_dmc		E_harm		Gap	Q(dQ)	 Q'	Q''
%==========================================================================================================
Ar& 2	&8	&$-$0.8489234312(1)	&$-$0.8485(2)	&$-$0.842	&0.3166	&6.93(8) &13.4	&9.93(8)\\
  & 3	&88	&$-$2.553335364(1)	&$-$2.555(1)	&$-$2.532	&0.3877	&6.36(8) &11.9	&9.36(8)\\
  & 4	&555	&$-$5.1182368(2)	&$-$5.113(1)	&$-$5.08	&0.3166	&4.37(7) &7.53	&7.38(7)\\
  & 5	&5555	&$-$7.78598(1)		&$-$7.791(1)	&$-$7.73	&0.2629	&2.85(3) &4.23	&5.84(3)\\
%==========================================================================================================
Ne& 2	&8	&$-$0.566783279(1)	&$-$0.5667(3)	&$-$0.49	&1.0105	&5.73(5) &11.7	&8.75(5)\\
  & 3	&88	&$-$1.7195589(3)	&$-$1.721(1)	&$-$1.5		&1.2376	&3.78(6) &7.40	&6.78(6)\\
  & 4	&555	&$-$3.464174(8)		&$-$3.461(1)	&$-$3.1		&1.0105	&2.61(6) &4.67	&5.61(6)\\
  & 5	&5555	&$-$5.29948(8)		&$-$5.299(1)	&$-$4.7		&0.8391	&1.80(4) &2.79	&4.80(4)\\
%==========================================================================================================
$1\over 2$-Ne& 2	&8	&$-$0.427526736(31)	&---		&$-$0.3		&1.4290	&4.1(1)  &8.77	&7.1(1)\\
  & 3	&88	&$-$1.308443(2)		&---		&$-$0.9		&1.7502	&2.92(6) &5.95	&5.94(6)\\
  & 4	&555	&$-$2.64356(3)		&---		&$-$1.8		&1.4290	&2.01(5) &3.74	&5.02(5)\\
  & 5	&5555	&$-$4.0669(1)		&---		&$-$2.9		&1.1867	&1.55(6) &2.55	&4.56(6)\\
%==========================================================================================================
\end{tabular} \label{tab.energies}
\end{table}
\begin{table}
\caption[tabharm]{ \small \parindent .5cm
\narrower Harmonic approximation: eigenvalues $\lambda_i$ of the
Hessian matrix evaluated at the classical minimum energy
configuration.  The energy levels are given by $E_{n_1,
n_2,\dots}=E_0^{\rm class}+ \sum_i (n_i+1/2) \sqrt(\lambda_i/m)$ with
$n_i=0,1,2,\dots$.  The numbers in parentheses are multiplicities of
the eigenvalues; the asterisks denote the lowest excited level that
corresponds to a state with the same symmetry as the ground state,
which defines the eigenvalue used to compute the gap $\Delta E_{\rm
har}$ in Table~\ref{tab.energies}.  \par\vspace{5mm}} \begin{tabular} {rrrr}
\multicolumn{1}{c}{$N=2$} &
\multicolumn{1}{c}{$N=3$} &
\multicolumn{1}{c}{$N=4$} &
\multicolumn{1}{c}{$N=5$} \\
\tableline
144 (*)	& 108 (2)	&72  (2)	& 43.8795887067966 (2)\\
	& 216 (*)	&144 (3)	& 99.3019911513688 (*)\\
	&		&288 (*)	&138.1042747649311 (2)\\
	&		&		&148.5071864137867 (2)\\
	&		&		&249.5948409655342 (1)\\
	&		&		&307.1635684436352 (1)\\
\end{tabular}
 \label{tab.harmonic} \end{table}
\begin{table}
\caption[tabmass]{ \small \parindent .5cm \narrower Inverse
dimensionless atomic mass $\alpha=m^{-1}$.  \par\vspace{5mm}} \begin{tabular} {cc}
	Ar		&	0.0006962	\\
	Ne		&	0.007092	\\
	${1\over 2}$-Ne	&	0.014184	\\
\end{tabular}
 \label{tab.masses} \end{table}
\section{Discussion}

We have presented a study of trial wave functions for small clusters.
The trial functions were designed to satisfy the correct asymptotic
behavior at small and large distances within the pair-correlation
approximation.  We demonstrated the computational tractability of wave
functions that explicitly contain many-body contributions with the
maximum number of interacting bodies for given cluster sizes and we
obtained a quantitative measure of the importance of these
contributions.  As a measure of the quality we used $Q$ as defined in
Eq.~(\ref{Quality}).  The estimate obtained from $Q'$ in
Eq.~(\ref{Quality'}) is typically twice as high, but it relies on the
gap computed obtained by the harmonic approximation.  It is interesting
to note in this context that in analogous computations for few-electron
systems the bound $\chi^2/\Delta E$ on the difference in energy between
the variational and the ground state energies is often more than an
order magnitude greater than the true difference \cite{cyrus.private}.

To deal with the many-body interactions we used bases of fundamental
polynomial invariants, but we did not yet perform a systematic study of
how to generate the computationally most efficient set of fundamental
invariants.  This holds in particular for larger systems where it is
important to truncate the hierarchy of $n$-body contributions in a way
that does not compromise the basic simplicity of the approach.

\acknowledgements It is a great pleasure to acknowledge numerous
discussions with David Freeman and Cyrus Umrigar without whose
stimulating input this work would not have been done.  This work was
supported by by the Office of Naval Research and by NSF Grants Nos.
DMR-9214669 and CHE-9203498.

\appendix \section{Trial function Optimization}
\label{appendix_variance} To generate a sequence of configurations we
used a standard Metropolis algorithm:  starting from a given
configuration the next one is constructed as follows.  First, a new
configuration is proposed by moving an atom $i$ (selected at random)
from ${\bf r}_{i}^{\rm old}$ to ${\bf r}_{i}^{\rm pr} \equiv {\bf
r}_{i}^{\rm old} + {\bf d}$ where ${\bf d}$ is a random vector sampled
uniformly from a cube of with linear dimension $d_{0}$ centered at the
origin, chosen to give an acceptance of about 50\%.  The proposed
configuration is accepted as the next configuration with probability
\begin{equation}
 A_{i} = \min \Bigl[1,~\frac{|\psiT({\bf R}^{\rm pr})|^2}
			   {|\psiT({\bf R}^{\rm old})|^2}\Bigr],
\end{equation} otherwise the old configuration is kept as the next
one.

This algorithm was used to generate the samples required for the
optimization of the trial function, as discussed below, and also to
compute the variational Monte Carlo estimates listed in
Table~\ref{tab.energies}.  We note that this particular version of the
Metropolis algorithm is rather primitive and we expect that a version
of the algorithm proposed by Umrigar \cite{cyrus.critical} would be
considerably more efficient.  Since the emphasis of the current work is
on the optimization of the wave function, the efficiency of the Monte
Carlo sampling algorithm was of no great concern.

For the optimization of the trial wave function we minimize the
variance of a fixed sample of $s$ configurations (with $s$ on the order
of 100 configurations per parameter) ${\bf R}_i$ sampled from
$|\Psi_{\rm T}({\bf R},{\bf p}_0)|^2$, where ${\bf p}_0$ is an initial
estimate of the parameters \cite{cyrus.trialfunctions}.  The variance
is given by \begin{equation} \label{chi_sq_rew} \chi^2 ({\bf p}) =
\frac{ \sum_i^s [ \Eloc({\bf p}, {\bf R}_{i}) - \overline{\Eloc} ({\bf
p})]^2 W_i} {\sum_i^s  W_i} \end{equation} where re-weighting factors
are defined as \begin{equation} W_i = \frac{|\psiT({\bf p}, {\bf
R}_{i})|^2 } {|\psiT({\bf p}_0, {\bf R}_{i})|^2 } \end{equation} and
\begin{equation} \label{mean_rew} \overline{\Eloc} ({\bf p}) = \frac{
\sum_i^s  \Eloc({\bf p}, {\bf R}_{i}) W_i} {\sum_i^s  W_i}.
\end{equation} The optimization can be regarded as a least-squares
parameter fit, for which the Levenberg-Marquardt algorithm can be
used.  If the initial estimate ${\bf p}_0$ of the variational
parameters is poor, in the sense that the variance of the weights $W_i$
becomes large, the procedure is restarted, and a new set of
configurations is sampled using the distribution defined by the wave
function with the current, improved parameter estimates.

With the exception of the parameters $w$ and $r_0$ used in the
definition of the inter-atomic distance variables $\hat r_{ij}$,
defined in Eq.~(\ref{eq.truncvar}), all parameters appear linearly in
the logarithm of the wave function.  The latter can therefore be
written in the form \begin{equation} \label{linform} \log \psiT ({\bf
p},{\bf R}) = {\bf A} \cdot {\bf f}({\bf R}), \end{equation} where $\bf
A$ is a parameter vector and $\bf f$ is a conjugate vector of
functions.  These functions can be chosen to depend only on the
configuration variables $\bf R$, but not the variational parameters
(except $w$ and $r_0$).  For a given sample of configurations, this way
of writing the wave function allows one to perform the computationally
expensive evaluation of the $\bf f$ for each configuration in the
sample no more than once for each parameter optimization.  We note that
variation of $w$ and $r_0$ would preclude this separation of
configuration variables and parameters, but it was found that changes
in $w$ in the range from one to two could be compensated by changes in
the parameters appearing in $\bf A$ ranging from 10\% to 20\% without
significant increase in the minimized variance of the local energy.
Similarly, changes of $r_0$ roughly from 1 to 1.2 have insignificant
effects.

Finally, as a simple example included for those readers who would want
to reproduce some of our results we tabulate in Tab~\ref{tab.Ne3Inv}
the optimized parameters for the Ne$_3$ cluster.  The trial function
presented utilizes the complete two- and three-body polynomial expanded
to order eight.

\begin{table} \caption[Ne3Inv]{ \small \parindent .5cm \narrower
Coefficients $C_{ijk}$ of the polynomial in the invariants $I_{3,1},
I_{3,2}$ and $I_{3,3}$ for a ${\rm Ne}_3$ cluster.  The trial wave
function is of the form of Eq.~(\ref{eq.u.invar}) with $N=3$,
$\alpha=0.007\,092\,000$, $r_0=1.1$, $w=1.120\,918$, and
$\gamma=0.907\,544\,158$.  \par\vspace{5mm}} \begin{tabular} {rl}
\multicolumn{1}{c}{$C_{ijk}$}&
\multicolumn{1}{c}{$(i,j,k)$}\\
\tableline
   2.244\,332	&(2,0,0)\\
  $-$1.512\,491	&(3,0,0)\\
  $-$0.980\,342	&(4,0,0)\\
  $-$0.981\,160	&(5,0,0)\\
  $-$1.730\,223	&(6,0,0)\\
  $-$3.759\,748	&(7,0,0)\\
  $-$1.729\,545	&(8,0,0)\\
  $-$2.742\,988	&(0,1,0)\\
   2.191\,407	&(1,1,0)\\
   1.950\,623	&(2,1,0)\\
   0.412\,871	&(3,1,0)\\
   2.180\,581	&(4,1,0)\\
  14.877\,507	&(5,1,0)\\
   2.827\,592	&(6,1,0)\\
  $-$1.361\,489	&(0,2,0)\\
   1.316\,542	&(1,2,0)\\
   3.565\,016	&(2,2,0)\\
 $-$23.619\,189	&(3,2,0)\\
   1.027\,512	&(4,2,0)\\
  $-$5.148\,204	&(0,3,0)\\
  14.682\,555	&(1,3,0)\\
   3.380\,246	&(2,3,0)\\
  $-$9.469\,800	&(0,4,0)\\
  32.051\,098	&(0,0,1)\\
  19.134\,118	&(1,0,1)\\
  12.411\,076	&(2,0,1)\\
  49.151\,573	&(3,0,1)\\
  11.797\,876	&(4,0,1)\\
  41.574\,293	&(5,0,1)\\
  25.096\,391	&(0,1,1)\\
 $-$45.475\,383	&(1,1,1)\\
 $-$27.791\,103	&(2,1,1)\\
  11.415\,828	&(3,1,1)\\
  58.250\,660	&(0,2,1)\\
$-$106.076\,304	&(1,2,1)\\
$-$318.164\,801	&(0,0,2)\\
 404.995\,936	&(1,0,2)\\
$-$246.486\,099	&(2,0,2)\\
$-$520.003\,142	&(0,1,2)\\
\end{tabular}
\label{tab.Ne3Inv} \end{table}

To calculate the kinetic energy we used Eq.~(\ref{E_loc_general}).
Since some of the numbers presented reflect the finite accuracy of the
numerical derivatives we present some further detail here. For
numerical the second-order derivatives appearing in the kinetic energy
we used three-point central difference scheme with finite difference
$\epsilon_2$ given by \begin{equation} \epsilon_2=\max
(|x|,0.1)10^{-d/4}, \end{equation} where $d=15$ is our machine floating
point precision \cite{dbl_float} and $x$ is the point at which the
derivative is computed.  For the gradient we used a two-point scheme
with \begin{equation} \epsilon_1=1.9 \max (|x|,0.1)10^{-d/3}.
\end{equation} This allows us to get relative numerical accuracy for
the second-order derivatives of about $10^{-7}$. Of course, the factor
1.9 can be safely omitted, it is included to help the reader to
reproduce our numerical result as close as possible.  Also, we note
that our scheme uses five points to approximate both derivatives.  A
five-point difference scheme might be used and yield more accurate
results.  We did not investigate this approach nor the analytic
computation of derivatives.
%1
\section{Invariants}
\label{appendix_invariants}

In Tables \ref{tab.invar4} and \ref{tab.invar5}  we list the
fundamental invariants for clusters of four and five particles.
Fig.~\ref{fig.diagrams} is a graphical representation of the
connectivity of the invariants. To construct a single monomial term of
an invariant from a given diagram one labels the vertices from
$1,2,\dots ,N$.  Each pair of vertices $i$ and $j$ connected by $l$
edges represents a factor $\hat r_{ij}^l$.  To get the full invariant,
sum over all permutations of site labels with the restriction that
different permutations yielding the same monomial are counted only
once. The system for $N=3$ is known to be complete.  We have not
attempted to prove that the systems given here form complete bases for
$N=4$ and $N=5$, but we have checked that the sets are complete up to
and including polynomials order eight and five respectively.  Also it
is likely that systems can be constructed that are computationally
superior and even have fewer fundamental invariants.

It should be noted the set for $N=4$ has syzygies (polynomials in the
invariants that vanish identically). E.g., there are homogeneous
polynomials of order eight that can be expressed in a one-parameter
infinity of different ways as polynomials in the invariants.  In
practice this means for the trial functions that the coefficients of
the polynomials in the invariants are not completely independent.
Since the number of of redundant parameters is small, this causes no
problem during the parameter optimization and we did not explore the
syzygies systematically.  \clearpage \begin{table} \caption[Inv4]{
\small \parindent .5cm \narrower Fundamental invariants for 4 particle
clusters.  \par\vspace{5mm}} \begin{tabular} {ll} $I_{4,1}$&$x_{12}+x_{13}+x_{14}+x_{23}+x_{24}+x_{34}$\\
$I_{4,2}$&$x_{14}x_{23}+x_{13}x_{24}+x_{12}x_{34}$\\
$I_{4,3}$&$x_{12}^2+x_{13}^2+x_{14}^2+x_{23}^2+x_{24}^2+x_{34}^2$\\
$I_{4,4}$&$x_{12}x_{13}x_{14}+x_{12}x_{23}x_{24}+x_{13}x_{23}x_{34}+x_{14}x_{24}x_{34}$\\
$I_{4,5}$&$x_{12}x_{13}x_{23}+x_{12}x_{14}x_{24}+x_{13}x_{14}x_{34}+x_{23}x_{24}x_{34}$\\
$I_{4,6}$&$x_{12}^3+x_{13}^3+x_{14}^3+x_{23}^3+x_{24}^3+x_{34}^3$\\
$I_{4,7}$&$x_{14}^2x_{23}^2+x_{13}^2x_{24}^2+x_{12}^2x_{34}^2$\\
$I_{4,8}$&$x_{12}^4+x_{13}^4+x_{14}^4+x_{23}^4+x_{24}^4+x_{34}^4$\\
$I_{4,9}$&$x_{12}^5+x_{13}^5+x_{14}^5+x_{23}^5+x_{24}^5+x_{34}^5$\\

\end{tabular} \label{tab.invar4} \end{table}
%2
\begin{table}
\caption[Inv4]{ \small \parindent .5cm \narrower Fundamental invariants
for 5 particle clusters.  The second column gives the number of terms
in the invariant to its right.  \par\vspace{5mm}} \tiny{
\begin{tabular} {lll} $I_ {5,1}  $&10&$x_{12}+x_{13}+x_{14}+x_{15}+x_{23}+x_{24}+x_{25}+x_{34}+x_{35}+x_{45}$\\
$I_ {5,2}  $&10&$x_{12}^2+x_{13}^2+x_{14}^2+x_{15}^2+\cdots$\\%x_{23}^2+x_{24}^2+x_{25}^2+x_{34}^2+x_{35}^2+x_{45}^2$\\
$I_ {5,3}  $&15&$x_{14}x_{23}+x_{15}x_{23}+x_{13}x_{24}+x_{15}x_{24}+\cdots$\\%x_{13}x_{25}+x_{14}x_{25}+x_{12}x_{34}+x_{15}x_{34}+x_{25}x_{34}+x_{12}x_{35}+x_{14}x_{35}+x_{24}x_{35}+x_{12}x_{45}+x_{13}x_{45}+x_{23}x_{45}$\\
$I_ {5,4}  $&10&$x_{12}^3+x_{13}^3+x_{14}^3+x_{15}^3+x_{23}^3+x_{24}^3+x_{25}^3+x_{34}^3+x_{35}^3+x_{45}^3$\\
$I_ {5,5}  $&10&$x_{12}x_{13}x_{23}+x_{12}x_{14}x_{24}+x_{12}x_{15}x_{25}+x_{13}x_{14}x_{34}+\cdots$\\%x_{23}x_{24}x_{34}+x_{13}x_{15}x_{35}+x_{23}x_{25}x_{35}+x_{14}x_{15}x_{45}+x_{24}x_{25}x_{45}+x_{34}x_{35}x_{45}$\\
$I_ {5,6}  $&20&$x_{12}x_{13}x_{14}+x_{12}x_{13}x_{15}+x_{12}x_{14}x_{15}+x_{13}x_{14}x_{15}+\cdots$\\%x_{12}x_{23}x_{24}+x_{12}x_{23}x_{25}+x_{12}x_{24}x_{25}+x_{23}x_{24}x_{25}+x_{13}x_{23}x_{34}+x_{14}x_{24}x_{34}+x_{13}x_{23}x_{35}+x_{15}x_{25}x_{35}+x_{13}x_{34}x_{35}+x_{23}x_{34}x_{35}+x_{14}x_{24}x_{45}+x_{15}x_{25}x_{45}+x_{14}x_{34}x_{45}+x_{24}x_{34}x_{45}+x_{15}x_{35}x_{45}+x_{25}x_{35}x_{45}$\\
$I_ {5,7}  $&30&$x_{14}^2x_{23}+x_{15}^2x_{23}+x_{14}x_{23}^2+x_{15}x_{23}^2+\cdots$\\%x_{13}^2x_{24}+x_{15}^2x_{24}+x_{13}x_{24}^2+x_{15}x_{24}^2+x_{13}^2x_{25}+x_{14}^2x_{25}+x_{13}x_{25}^2+x_{14}x_{25}^2+x_{12}^2x_{34}+x_{15}^2x_{34}+x_{25}^2x_{34}+x_{12}x_{34}^2+x_{15}x_{34}^2+x_{25}x_{34}^2+x_{12}^2x_{35}+x_{14}^2x_{35}+x_{24}^2x_{35}+x_{12}x_{35}^2+x_{14}x_{35}^2+x_{24}x_{35}^2+x_{12}^2x_{45}+x_{13}^2x_{45}+x_{23}^2x_{45}+x_{12}x_{45}^2+x_{13}x_{45}^2+x_{23}x_{45}^2$\\
$I_ {5,8}  $& 5&$x_{12}x_{13}x_{14}x_{15}+x_{12}x_{23}x_{24}x_{25}+x_{13}x_{23}x_{34}x_{35}+x_{14}x_{24}x_{34}x_{45}+x_{15}x_{25}x_{35}x_{45}$\\
$I_ {5,9}  $&10&$x_{12}^4+x_{13}^4+x_{14}^4+x_{15}^4+x_{23}^4+x_{24}^4+x_{25}^4+x_{34}^4+x_{35}^4+x_{45}^4$\\
$I_{5,10}$&10&$x_{15}x_{23}x_{24}x_{34}+x_{13}x_{14}x_{25}x_{34}+x_{12}x_{15}x_{25}x_{34}+x_{12}x_{14}x_{24}x_{35}+\cdots$\\%x_{13}x_{15}x_{24}x_{35}+x_{14}x_{23}x_{25}x_{35}+x_{12}x_{13}x_{23}x_{45}+x_{14}x_{15}x_{23}x_{45}+x_{13}x_{24}x_{25}x_{45}+x_{12}x_{34}x_{35}x_{45}$\\
$I_{5,11}$&15&$x_{14}^2x_{23}^2+x_{15}^2x_{23}^2+x_{13}^2x_{24}^2+x_{15}^2x_{24}^2+\cdots$\\%x_{13}^2x_{25}^2+x_{14}^2x_{25}^2+x_{12}^2x_{34}^2+x_{15}^2x_{34}^2+x_{25}^2x_{34}^2+x_{12}^2x_{35}^2+x_{14}^2x_{35}^2+x_{24}^2x_{35}^2+x_{12}^2x_{45}^2+x_{13}^2x_{45}^2+x_{23}^2x_{45}^2$\\
$I_{5,12}$&15&$x_{13}x_{14}x_{23}x_{24}+x_{13}x_{15}x_{23}x_{25}+x_{14}x_{15}x_{24}x_{25}+x_{12}x_{14}x_{23}x_{34}+\cdots$\\%x_{12}x_{13}x_{24}x_{34}+x_{12}x_{15}x_{23}x_{35}+x_{12}x_{13}x_{25}x_{35}+x_{14}x_{15}x_{34}x_{35}+x_{24}x_{25}x_{34}x_{35}+x_{12}x_{15}x_{24}x_{45}+x_{12}x_{14}x_{25}x_{45}+x_{13}x_{15}x_{34}x_{45}+x_{23}x_{25}x_{34}x_{45}+x_{13}x_{14}x_{35}x_{45}+x_{23}x_{24}x_{35}x_{45}$\\
$I_{5,13}$&30&$x_{12}^2x_{13}x_{23}+x_{12}x_{13}^2x_{23}+x_{12}x_{13}x_{23}^2+x_{12}^2x_{14}x_{24}+\cdots$\\%x_{12}x_{14}^2x_{24}+x_{12}x_{14}x_{24}^2+x_{12}^2x_{15}x_{25}+x_{12}x_{15}^2x_{25}+x_{12}x_{15}x_{25}^2+x_{13}^2x_{14}x_{34}+x_{13}x_{14}^2x_{34}+x_{23}^2x_{24}x_{34}+x_{23}x_{24}^2x_{34}+x_{13}x_{14}x_{34}^2+x_{23}x_{24}x_{34}^2+x_{13}^2x_{15}x_{35}+x_{13}x_{15}^2x_{35}+x_{23}^2x_{25}x_{35}+x_{23}x_{25}^2x_{35}+x_{13}x_{15}x_{35}^2+x_{23}x_{25}x_{35}^2+x_{14}^2x_{15}x_{45}+x_{14}x_{15}^2x_{45}+x_{24}^2x_{25}x_{45}+x_{24}x_{25}^2x_{45}+x_{34}^2x_{35}x_{45}+x_{34}x_{35}^2x_{45}+x_{14}x_{15}x_{45}^2+x_{24}x_{25}x_{45}^2+x_{34}x_{35}x_{45}^2$\\
$I_{5,14}$&30&$x_{14}^3x_{23}+x_{15}^3x_{23}+x_{14}x_{23}^3+x_{15}x_{23}^3+\cdots$\\%x_{13}^3x_{24}+x_{15}^3x_{24}+x_{13}x_{24}^3+x_{15}x_{24}^3+x_{13}^3x_{25}+x_{14}^3x_{25}+x_{13}x_{25}^3+x_{14}x_{25}^3+x_{12}^3x_{34}+x_{15}^3x_{34}+x_{25}^3x_{34}+x_{12}x_{34}^3+x_{15}x_{34}^3+x_{25}x_{34}^3+x_{12}^3x_{35}+x_{14}^3x_{35}+x_{24}^3x_{35}+x_{12}x_{35}^3+x_{14}x_{35}^3+x_{24}x_{35}^3+x_{12}^3x_{45}+x_{13}^3x_{45}+x_{23}^3x_{45}+x_{12}x_{45}^3+x_{13}x_{45}^3+x_{23}x_{45}^3$\\
$I_{5,15}$&10&$x_{12}^5+x_{13}^5+x_{14}^5+x_{15}^5+x_{23}^5+x_{24}^5+x_{25}^5+x_{34}^5+x_{35}^5+x_{45}^5$\\
$I_{5,16}$&10&$x_{15}^2x_{23}x_{24}x_{34}+x_{13}x_{14}x_{25}^2x_{34}+x_{12}x_{15}x_{25}x_{34}^2+x_{13}x_{15}x_{24}^2x_{35}+\cdots$\\%x_{14}^2x_{23}x_{25}x_{35}+x_{12}x_{14}x_{24}x_{35}^2+x_{14}x_{15}x_{23}^2x_{45}+x_{13}^2x_{24}x_{25}x_{45}+x_{12}^2x_{34}x_{35}x_{45}+x_{12}x_{13}x_{23}x_{45}^2$\\
$I_{5,17}$&12&$x_{14}x_{15}x_{23}x_{25}x_{34}+x_{13}x_{15}x_{24}x_{25}x_{34}+x_{14}x_{15}x_{23}x_{24}x_{35}+x_{13}x_{14}x_{24}x_{25}x_{35}+\cdots$\\%x_{12}x_{15}x_{24}x_{34}x_{35}+x_{12}x_{14}x_{25}x_{34}x_{35}+x_{13}x_{15}x_{23}x_{24}x_{45}+x_{13}x_{14}x_{23}x_{25}x_{45}+x_{12}x_{15}x_{23}x_{34}x_{45}+x_{12}x_{13}x_{25}x_{34}x_{45}+x_{12}x_{14}x_{23}x_{35}x_{45}+x_{12}x_{13}x_{24}x_{35}x_{45}$\\
$I_{5,18}$&20&$x_{12}^2x_{13}x_{14}x_{15}+x_{12}x_{13}^2x_{14}x_{15}+x_{12}x_{13}x_{14}^2x_{15}+x_{12}x_{13}x_{14}x_{15}^2+\cdots$\\%x_{12}^2x_{23}x_{24}x_{25}+x_{12}x_{23}^2x_{24}x_{25}+x_{12}x_{23}x_{24}^2x_{25}+x_{12}x_{23}x_{24}x_{25}^2+x_{13}^2x_{23}x_{34}x_{35}+x_{13}x_{23}^2x_{34}x_{35}+x_{13}x_{23}x_{34}^2x_{35}+x_{13}x_{23}x_{34}x_{35}^2+x_{14}^2x_{24}x_{34}x_{45}+x_{14}x_{24}^2x_{34}x_{45}+x_{14}x_{24}x_{34}^2x_{45}+x_{15}^2x_{25}x_{35}x_{45}+x_{15}x_{25}^2x_{35}x_{45}+x_{15}x_{25}x_{35}^2x_{45}+x_{14}x_{24}x_{34}x_{45}^2+x_{15}x_{25}x_{35}x_{45}^2$\\
$I_{5,19}$&30&$x_{12}^3x_{13}x_{23}+x_{12}x_{13}^3x_{23}+x_{12}x_{13}x_{23}^3+x_{12}^3x_{14}x_{24}+\cdots$\\%x_{12}x_{14}^3x_{24}+x_{12}x_{14}x_{24}^3+x_{12}^3x_{15}x_{25}+x_{12}x_{15}^3x_{25}+x_{12}x_{15}x_{25}^3+x_{13}^3x_{14}x_{34}+x_{13}x_{14}^3x_{34}+x_{23}^3x_{24}x_{34}+x_{23}x_{24}^3x_{34}+x_{13}x_{14}x_{34}^3+x_{23}x_{24}x_{34}^3+x_{13}^3x_{15}x_{35}+x_{13}x_{15}^3x_{35}+x_{23}^3x_{25}x_{35}+x_{23}x_{25}^3x_{35}+x_{13}x_{15}x_{35}^3+x_{23}x_{25}x_{35}^3+x_{14}^3x_{15}x_{45}+x_{14}x_{15}^3x_{45}+x_{24}^3x_{25}x_{45}+x_{24}x_{25}^3x_{45}+x_{34}^3x_{35}x_{45}+x_{34}x_{35}^3x_{45}+x_{14}x_{15}x_{45}^3+x_{24}x_{25}x_{45}^3+x_{34}x_{35}x_{45}^3$\\
$I_{5,20}$&30&$x_{12}^2x_{13}^2x_{23}+x_{12}^2x_{13}x_{23}^2+x_{12}x_{13}^2x_{23}^2+x_{12}^2x_{14}^2x_{24}+\cdots$\\%x_{12}^2x_{14}x_{24}^2+x_{12}x_{14}^2x_{24}^2+x_{12}^2x_{15}^2x_{25}+x_{12}^2x_{15}x_{25}^2+x_{12}x_{15}^2x_{25}^2+x_{13}^2x_{14}^2x_{34}+x_{23}^2x_{24}^2x_{34}+x_{13}^2x_{14}x_{34}^2+x_{13}x_{14}^2x_{34}^2+x_{23}^2x_{24}x_{34}^2+x_{23}x_{24}^2x_{34}^2+x_{13}^2x_{15}^2x_{35}+x_{23}^2x_{25}^2x_{35}+x_{13}^2x_{15}x_{35}^2+x_{13}x_{15}^2x_{35}^2+x_{23}^2x_{25}x_{35}^2+x_{23}x_{25}^2x_{35}^2+x_{14}^2x_{15}^2x_{45}+x_{24}^2x_{25}^2x_{45}+x_{34}^2x_{35}^2x_{45}+x_{14}^2x_{15}x_{45}^2+x_{14}x_{15}^2x_{45}^2+x_{24}^2x_{25}x_{45}^2+x_{24}x_{25}^2x_{45}^2+x_{34}^2x_{35}x_{45}^2+x_{34}x_{35}^2x_{45}^2$\\
$I_{5,21}$&30&$x_{14}^4x_{23}+x_{15}^4x_{23}+x_{14}x_{23}^4+x_{15}x_{23}^4+\cdots$\\%x_{13}^4x_{24}+x_{15}^4x_{24}+x_{13}x_{24}^4+x_{15}x_{24}^4+x_{13}^4x_{25}+x_{14}^4x_{25}+x_{13}x_{25}^4+x_{14}x_{25}^4+x_{12}^4x_{34}+x_{15}^4x_{34}+x_{25}^4x_{34}+x_{12}x_{34}^4+x_{15}x_{34}^4+x_{25}x_{34}^4+x_{12}^4x_{35}+x_{14}^4x_{35}+x_{24}^4x_{35}+x_{12}x_{35}^4+x_{14}x_{35}^4+x_{24}x_{35}^4+x_{12}^4x_{45}+x_{13}^4x_{45}+x_{23}^4x_{45}+x_{12}x_{45}^4+x_{13}x_{45}^4+x_{23}x_{45}^4$\\
$I_{5,22}$&30&$x_{14}^2x_{15}^2x_{23}+x_{13}^2x_{15}^2x_{24}+x_{15}x_{23}^2x_{24}^2+x_{13}^2x_{14}^2x_{25}+\cdots$\\%x_{14}x_{23}^2x_{25}^2+x_{13}x_{24}^2x_{25}^2+x_{12}^2x_{15}^2x_{34}+x_{12}^2x_{25}^2x_{34}+x_{15}^2x_{25}^2x_{34}+x_{15}x_{23}^2x_{34}^2+x_{15}x_{24}^2x_{34}^2+x_{13}^2x_{25}x_{34}^2+x_{14}^2x_{25}x_{34}^2+x_{12}^2x_{14}^2x_{35}+x_{12}^2x_{24}^2x_{35}+x_{14}^2x_{24}^2x_{35}+x_{14}x_{23}^2x_{35}^2+x_{13}^2x_{24}x_{35}^2+x_{15}^2x_{24}x_{35}^2+x_{14}x_{25}^2x_{35}^2+x_{12}x_{34}^2x_{35}^2+x_{12}^2x_{13}^2x_{45}+x_{12}^2x_{23}^2x_{45}+x_{13}^2x_{23}^2x_{45}+x_{14}^2x_{23}x_{45}^2+x_{15}^2x_{23}x_{45}^2+x_{13}x_{24}^2x_{45}^2+x_{13}x_{25}^2x_{45}^2+x_{12}x_{34}^2x_{45}^2+x_{12}x_{35}^2x_{45}^2$\\
$I_{5,23}$&30&$x_{14}^3x_{23}^2+x_{15}^3x_{23}^2+x_{14}^2x_{23}^3+x_{15}^2x_{23}^3+\cdots$\\%x_{13}^3x_{24}^2+x_{15}^3x_{24}^2+x_{13}^2x_{24}^3+x_{15}^2x_{24}^3+x_{13}^3x_{25}^2+x_{14}^3x_{25}^2+x_{13}^2x_{25}^3+x_{14}^2x_{25}^3+x_{12}^3x_{34}^2+x_{15}^3x_{34}^2+x_{25}^3x_{34}^2+x_{12}^2x_{34}^3+x_{15}^2x_{34}^3+x_{25}^2x_{34}^3+x_{12}^3x_{35}^2+x_{14}^3x_{35}^2+x_{24}^3x_{35}^2+x_{12}^2x_{35}^3+x_{14}^2x_{35}^3+x_{24}^2x_{35}^3+x_{12}^3x_{45}^2+x_{13}^3x_{45}^2+x_{23}^3x_{45}^2+x_{12}^2x_{45}^3+x_{13}^2x_{45}^3+x_{23}^2x_{45}^3$\\
$I_{5,24}$&30&$x_{14}x_{15}x_{23}^3+x_{15}^3x_{23}x_{24}+x_{13}x_{15}x_{24}^3+x_{14}^3x_{23}x_{25}+\cdots$\\%x_{13}^3x_{24}x_{25}+x_{13}x_{14}x_{25}^3+x_{15}^3x_{23}x_{34}+x_{15}^3x_{24}x_{34}+x_{13}x_{25}^3x_{34}+x_{14}x_{25}^3x_{34}+x_{12}x_{15}x_{34}^3+x_{12}x_{25}x_{34}^3+x_{15}x_{25}x_{34}^3+x_{14}^3x_{23}x_{35}+x_{13}x_{24}^3x_{35}+x_{15}x_{24}^3x_{35}+x_{14}^3x_{25}x_{35}+x_{12}^3x_{34}x_{35}+x_{12}x_{14}x_{35}^3+x_{12}x_{24}x_{35}^3+x_{14}x_{24}x_{35}^3+x_{14}x_{23}^3x_{45}+x_{15}x_{23}^3x_{45}+x_{13}^3x_{24}x_{45}+x_{13}^3x_{25}x_{45}+x_{12}^3x_{34}x_{45}+x_{12}^3x_{35}x_{45}+x_{12}x_{13}x_{45}^3+x_{12}x_{23}x_{45}^3+x_{13}x_{23}x_{45}^3$\\

\end{tabular}}
\label{tab.invar5} \end{table} \clearpage
%\documentstyle[preprint,aps]{revtex}
%\begin{document}
\begin{figure}
\caption[diag]{\small \parindent .5cm \narrower
Graphical representation of fundamental invariants for clusters of 
sizes $N=3$, $N=4$, and $N=5$.
\par\vspace{5mm}}
\label{fig.diagrams}
	{\Large
	% GNUPLOT: LaTeX picture
\setlength{\unitlength}{0.240900pt}
\ifx\plotpoint\undefined\newsavebox{\plotpoint}\fi
\sbox{\plotpoint}{\rule[-0.175pt]{0.350pt}{0.350pt}}%
\begin{picture}(250,300)(0,100)
\tenrm
\sbox{\plotpoint}{\rule[-0.175pt]{0.350pt}{0.350pt}}%
\put(370,95){\makebox(0,0){$I_{3,1}$}}
\put(370,346){\circle*{40}}
\put(297,221){\circle*{40}}
\put(443,221){\circle*{40}}
\put(297,221){\usebox{\plotpoint}}
\put(297,221){\rule[-0.175pt]{35.171pt}{0.350pt}}
\sbox{\plotpoint}{\rule[-0.500pt]{1.000pt}{1.000pt}}%
\end{picture}
	% GNUPLOT: LaTeX picture
\setlength{\unitlength}{0.240900pt}
\ifx\plotpoint\undefined\newsavebox{\plotpoint}\fi
\sbox{\plotpoint}{\rule[-0.175pt]{0.350pt}{0.350pt}}%
\begin{picture}(250,300)(0,100)
\tenrm
\sbox{\plotpoint}{\rule[-0.175pt]{0.350pt}{0.350pt}}%
\put(370,95){\makebox(0,0){$I_{3,2}$}}
\put(370,346){\circle*{40}}
\put(297,221){\circle*{40}}
\put(443,221){\circle*{40}}
\put(311,229){\usebox{\plotpoint}}
\put(311,229){\rule[-0.175pt]{28.426pt}{0.350pt}}
\put(289,217){\usebox{\plotpoint}}
\put(289,217){\rule[-0.175pt]{39.026pt}{0.350pt}}
\sbox{\plotpoint}{\rule[-0.250pt]{0.500pt}{0.500pt}}%
\end{picture}
	% GNUPLOT: LaTeX picture
\setlength{\unitlength}{0.240900pt}
\ifx\plotpoint\undefined\newsavebox{\plotpoint}\fi
\sbox{\plotpoint}{\rule[-0.175pt]{0.350pt}{0.350pt}}%
\begin{picture}(250,300)(0,100)
\tenrm
\sbox{\plotpoint}{\rule[-0.175pt]{0.350pt}{0.350pt}}%
\put(370,95){\makebox(0,0){$I_{3,3}$}}
\put(370,346){\circle*{40}}
\put(297,221){\circle*{40}}
\put(443,221){\circle*{40}}
\put(370,346){\usebox{\plotpoint}}
\put(370,344){\rule[-0.175pt]{0.350pt}{0.413pt}}
\put(369,342){\rule[-0.175pt]{0.350pt}{0.413pt}}
\put(368,340){\rule[-0.175pt]{0.350pt}{0.413pt}}
\put(367,339){\rule[-0.175pt]{0.350pt}{0.413pt}}
\put(366,337){\rule[-0.175pt]{0.350pt}{0.413pt}}
\put(365,335){\rule[-0.175pt]{0.350pt}{0.413pt}}
\put(364,334){\rule[-0.175pt]{0.350pt}{0.413pt}}
\put(363,332){\rule[-0.175pt]{0.350pt}{0.413pt}}
\put(362,330){\rule[-0.175pt]{0.350pt}{0.413pt}}
\put(361,328){\rule[-0.175pt]{0.350pt}{0.413pt}}
\put(360,327){\rule[-0.175pt]{0.350pt}{0.413pt}}
\put(359,325){\rule[-0.175pt]{0.350pt}{0.413pt}}
\put(358,323){\rule[-0.175pt]{0.350pt}{0.413pt}}
\put(357,322){\rule[-0.175pt]{0.350pt}{0.413pt}}
\put(356,320){\rule[-0.175pt]{0.350pt}{0.413pt}}
\put(355,318){\rule[-0.175pt]{0.350pt}{0.413pt}}
\put(354,316){\rule[-0.175pt]{0.350pt}{0.413pt}}
\put(353,315){\rule[-0.175pt]{0.350pt}{0.413pt}}
\put(352,313){\rule[-0.175pt]{0.350pt}{0.413pt}}
\put(351,311){\rule[-0.175pt]{0.350pt}{0.413pt}}
\put(350,310){\rule[-0.175pt]{0.350pt}{0.413pt}}
\put(349,308){\rule[-0.175pt]{0.350pt}{0.413pt}}
\put(348,306){\rule[-0.175pt]{0.350pt}{0.413pt}}
\put(347,304){\rule[-0.175pt]{0.350pt}{0.413pt}}
\put(346,303){\rule[-0.175pt]{0.350pt}{0.413pt}}
\put(345,301){\rule[-0.175pt]{0.350pt}{0.413pt}}
\put(344,299){\rule[-0.175pt]{0.350pt}{0.413pt}}
\put(343,298){\rule[-0.175pt]{0.350pt}{0.413pt}}
\put(342,296){\rule[-0.175pt]{0.350pt}{0.413pt}}
\put(341,294){\rule[-0.175pt]{0.350pt}{0.413pt}}
\put(340,292){\rule[-0.175pt]{0.350pt}{0.413pt}}
\put(339,291){\rule[-0.175pt]{0.350pt}{0.413pt}}
\put(338,289){\rule[-0.175pt]{0.350pt}{0.413pt}}
\put(337,287){\rule[-0.175pt]{0.350pt}{0.413pt}}
\put(336,286){\rule[-0.175pt]{0.350pt}{0.413pt}}
\put(335,284){\rule[-0.175pt]{0.350pt}{0.413pt}}
\put(334,282){\rule[-0.175pt]{0.350pt}{0.413pt}}
\put(333,280){\rule[-0.175pt]{0.350pt}{0.413pt}}
\put(332,279){\rule[-0.175pt]{0.350pt}{0.413pt}}
\put(331,277){\rule[-0.175pt]{0.350pt}{0.413pt}}
\put(330,275){\rule[-0.175pt]{0.350pt}{0.413pt}}
\put(329,274){\rule[-0.175pt]{0.350pt}{0.413pt}}
\put(328,272){\rule[-0.175pt]{0.350pt}{0.413pt}}
\put(327,270){\rule[-0.175pt]{0.350pt}{0.413pt}}
\put(326,268){\rule[-0.175pt]{0.350pt}{0.413pt}}
\put(325,267){\rule[-0.175pt]{0.350pt}{0.413pt}}
\put(324,265){\rule[-0.175pt]{0.350pt}{0.413pt}}
\put(323,263){\rule[-0.175pt]{0.350pt}{0.413pt}}
\put(322,262){\rule[-0.175pt]{0.350pt}{0.413pt}}
\put(321,260){\rule[-0.175pt]{0.350pt}{0.413pt}}
\put(320,258){\rule[-0.175pt]{0.350pt}{0.413pt}}
\put(319,256){\rule[-0.175pt]{0.350pt}{0.413pt}}
\put(318,255){\rule[-0.175pt]{0.350pt}{0.412pt}}
\put(317,253){\rule[-0.175pt]{0.350pt}{0.412pt}}
\put(316,251){\rule[-0.175pt]{0.350pt}{0.412pt}}
\put(315,250){\rule[-0.175pt]{0.350pt}{0.412pt}}
\put(314,248){\rule[-0.175pt]{0.350pt}{0.412pt}}
\put(313,246){\rule[-0.175pt]{0.350pt}{0.412pt}}
\put(312,244){\rule[-0.175pt]{0.350pt}{0.412pt}}
\put(311,243){\rule[-0.175pt]{0.350pt}{0.412pt}}
\put(310,241){\rule[-0.175pt]{0.350pt}{0.412pt}}
\put(309,239){\rule[-0.175pt]{0.350pt}{0.412pt}}
\put(308,238){\rule[-0.175pt]{0.350pt}{0.412pt}}
\put(307,236){\rule[-0.175pt]{0.350pt}{0.412pt}}
\put(306,234){\rule[-0.175pt]{0.350pt}{0.412pt}}
\put(305,232){\rule[-0.175pt]{0.350pt}{0.412pt}}
\put(304,231){\rule[-0.175pt]{0.350pt}{0.412pt}}
\put(303,229){\rule[-0.175pt]{0.350pt}{0.412pt}}
\put(302,227){\rule[-0.175pt]{0.350pt}{0.412pt}}
\put(301,226){\rule[-0.175pt]{0.350pt}{0.412pt}}
\put(300,224){\rule[-0.175pt]{0.350pt}{0.412pt}}
\put(299,222){\rule[-0.175pt]{0.350pt}{0.412pt}}
\put(298,221){\rule[-0.175pt]{0.350pt}{0.412pt}}
\put(370,346){\usebox{\plotpoint}}
\put(370,344){\rule[-0.175pt]{0.350pt}{0.413pt}}
\put(371,342){\rule[-0.175pt]{0.350pt}{0.413pt}}
\put(372,340){\rule[-0.175pt]{0.350pt}{0.413pt}}
\put(373,339){\rule[-0.175pt]{0.350pt}{0.413pt}}
\put(374,337){\rule[-0.175pt]{0.350pt}{0.413pt}}
\put(375,335){\rule[-0.175pt]{0.350pt}{0.413pt}}
\put(376,334){\rule[-0.175pt]{0.350pt}{0.413pt}}
\put(377,332){\rule[-0.175pt]{0.350pt}{0.413pt}}
\put(378,330){\rule[-0.175pt]{0.350pt}{0.413pt}}
\put(379,328){\rule[-0.175pt]{0.350pt}{0.413pt}}
\put(380,327){\rule[-0.175pt]{0.350pt}{0.413pt}}
\put(381,325){\rule[-0.175pt]{0.350pt}{0.413pt}}
\put(382,323){\rule[-0.175pt]{0.350pt}{0.413pt}}
\put(383,322){\rule[-0.175pt]{0.350pt}{0.413pt}}
\put(384,320){\rule[-0.175pt]{0.350pt}{0.413pt}}
\put(385,318){\rule[-0.175pt]{0.350pt}{0.413pt}}
\put(386,316){\rule[-0.175pt]{0.350pt}{0.413pt}}
\put(387,315){\rule[-0.175pt]{0.350pt}{0.413pt}}
\put(388,313){\rule[-0.175pt]{0.350pt}{0.413pt}}
\put(389,311){\rule[-0.175pt]{0.350pt}{0.413pt}}
\put(390,310){\rule[-0.175pt]{0.350pt}{0.413pt}}
\put(391,308){\rule[-0.175pt]{0.350pt}{0.413pt}}
\put(392,306){\rule[-0.175pt]{0.350pt}{0.413pt}}
\put(393,304){\rule[-0.175pt]{0.350pt}{0.413pt}}
\put(394,303){\rule[-0.175pt]{0.350pt}{0.413pt}}
\put(395,301){\rule[-0.175pt]{0.350pt}{0.413pt}}
\put(396,299){\rule[-0.175pt]{0.350pt}{0.413pt}}
\put(397,298){\rule[-0.175pt]{0.350pt}{0.413pt}}
\put(398,296){\rule[-0.175pt]{0.350pt}{0.413pt}}
\put(399,294){\rule[-0.175pt]{0.350pt}{0.413pt}}
\put(400,292){\rule[-0.175pt]{0.350pt}{0.413pt}}
\put(401,291){\rule[-0.175pt]{0.350pt}{0.413pt}}
\put(402,289){\rule[-0.175pt]{0.350pt}{0.413pt}}
\put(403,287){\rule[-0.175pt]{0.350pt}{0.413pt}}
\put(404,286){\rule[-0.175pt]{0.350pt}{0.413pt}}
\put(405,284){\rule[-0.175pt]{0.350pt}{0.413pt}}
\put(406,282){\rule[-0.175pt]{0.350pt}{0.413pt}}
\put(407,280){\rule[-0.175pt]{0.350pt}{0.413pt}}
\put(408,279){\rule[-0.175pt]{0.350pt}{0.413pt}}
\put(409,277){\rule[-0.175pt]{0.350pt}{0.413pt}}
\put(410,275){\rule[-0.175pt]{0.350pt}{0.413pt}}
\put(411,274){\rule[-0.175pt]{0.350pt}{0.413pt}}
\put(412,272){\rule[-0.175pt]{0.350pt}{0.413pt}}
\put(413,270){\rule[-0.175pt]{0.350pt}{0.413pt}}
\put(414,268){\rule[-0.175pt]{0.350pt}{0.413pt}}
\put(415,267){\rule[-0.175pt]{0.350pt}{0.413pt}}
\put(416,265){\rule[-0.175pt]{0.350pt}{0.413pt}}
\put(417,263){\rule[-0.175pt]{0.350pt}{0.413pt}}
\put(418,262){\rule[-0.175pt]{0.350pt}{0.413pt}}
\put(419,260){\rule[-0.175pt]{0.350pt}{0.413pt}}
\put(420,258){\rule[-0.175pt]{0.350pt}{0.413pt}}
\put(421,256){\rule[-0.175pt]{0.350pt}{0.413pt}}
\put(422,255){\rule[-0.175pt]{0.350pt}{0.412pt}}
\put(423,253){\rule[-0.175pt]{0.350pt}{0.412pt}}
\put(424,251){\rule[-0.175pt]{0.350pt}{0.412pt}}
\put(425,250){\rule[-0.175pt]{0.350pt}{0.412pt}}
\put(426,248){\rule[-0.175pt]{0.350pt}{0.412pt}}
\put(427,246){\rule[-0.175pt]{0.350pt}{0.412pt}}
\put(428,244){\rule[-0.175pt]{0.350pt}{0.412pt}}
\put(429,243){\rule[-0.175pt]{0.350pt}{0.412pt}}
\put(430,241){\rule[-0.175pt]{0.350pt}{0.412pt}}
\put(431,239){\rule[-0.175pt]{0.350pt}{0.412pt}}
\put(432,238){\rule[-0.175pt]{0.350pt}{0.412pt}}
\put(433,236){\rule[-0.175pt]{0.350pt}{0.412pt}}
\put(434,234){\rule[-0.175pt]{0.350pt}{0.412pt}}
\put(435,232){\rule[-0.175pt]{0.350pt}{0.412pt}}
\put(436,231){\rule[-0.175pt]{0.350pt}{0.412pt}}
\put(437,229){\rule[-0.175pt]{0.350pt}{0.412pt}}
\put(438,227){\rule[-0.175pt]{0.350pt}{0.412pt}}
\put(439,226){\rule[-0.175pt]{0.350pt}{0.412pt}}
\put(440,224){\rule[-0.175pt]{0.350pt}{0.412pt}}
\put(441,222){\rule[-0.175pt]{0.350pt}{0.412pt}}
\put(442,221){\rule[-0.175pt]{0.350pt}{0.412pt}}
\put(297,221){\usebox{\plotpoint}}
\put(297,221){\rule[-0.175pt]{35.171pt}{0.350pt}}
\sbox{\plotpoint}{\rule[-0.250pt]{0.500pt}{0.500pt}}%
\end{picture}
	% GNUPLOT: LaTeX picture
\setlength{\unitlength}{0.240900pt}
\ifx\plotpoint\undefined\newsavebox{\plotpoint}\fi
\sbox{\plotpoint}{\rule[-0.175pt]{0.350pt}{0.350pt}}%
\begin{picture}(250,300)(0,100)
\tenrm
\sbox{\plotpoint}{\rule[-0.175pt]{0.350pt}{0.350pt}}%
\put(370,95){\makebox(0,0){$I_{4,1}$}}
\put(430,322){\circle*{40}}
\put(310,322){\circle*{40}}
\put(310,203){\circle*{40}}
\put(430,203){\circle*{40}}
\put(310,203){\usebox{\plotpoint}}
\put(310,203){\rule[-0.175pt]{28.908pt}{0.350pt}}
\sbox{\plotpoint}{\rule[-0.500pt]{1.000pt}{1.000pt}}%
\end{picture}
	% GNUPLOT: LaTeX picture
\setlength{\unitlength}{0.240900pt}
\ifx\plotpoint\undefined\newsavebox{\plotpoint}\fi
\sbox{\plotpoint}{\rule[-0.175pt]{0.350pt}{0.350pt}}%
\begin{picture}(250,300)(0,100)
\tenrm
\sbox{\plotpoint}{\rule[-0.175pt]{0.350pt}{0.350pt}}%
\put(370,95){\makebox(0,0){$I_{4,2}$}}
\put(430,322){\circle*{40}}
\put(310,322){\circle*{40}}
\put(310,203){\circle*{40}}
\put(430,203){\circle*{40}}
\put(430,322){\usebox{\plotpoint}}
\put(310,322){\rule[-0.175pt]{28.908pt}{0.350pt}}
\put(310,203){\usebox{\plotpoint}}
\put(310,203){\rule[-0.175pt]{28.908pt}{0.350pt}}
\sbox{\plotpoint}{\rule[-0.250pt]{0.500pt}{0.500pt}}%
\end{picture}
	% GNUPLOT: LaTeX picture
\setlength{\unitlength}{0.240900pt}
\ifx\plotpoint\undefined\newsavebox{\plotpoint}\fi
\sbox{\plotpoint}{\rule[-0.175pt]{0.350pt}{0.350pt}}%
\begin{picture}(250,300)(0,100)
\tenrm
\sbox{\plotpoint}{\rule[-0.175pt]{0.350pt}{0.350pt}}%
\put(370,95){\makebox(0,0){$I_{4,3}$}}
\put(430,322){\circle*{40}}
\put(310,322){\circle*{40}}
\put(310,203){\circle*{40}}
\put(430,203){\circle*{40}}
\put(303,197){\usebox{\plotpoint}}
\put(303,197){\rule[-0.175pt]{32.281pt}{0.350pt}}
\put(317,210){\usebox{\plotpoint}}
\put(317,210){\rule[-0.175pt]{25.535pt}{0.350pt}}
\sbox{\plotpoint}{\rule[-0.250pt]{0.500pt}{0.500pt}}%
\end{picture}
	}
\end{figure}
\begin{figure}
	{\Large
	% GNUPLOT: LaTeX picture
\setlength{\unitlength}{0.240900pt}
\ifx\plotpoint\undefined\newsavebox{\plotpoint}\fi
\sbox{\plotpoint}{\rule[-0.175pt]{0.350pt}{0.350pt}}%
\begin{picture}(250,300)(0,100)
\tenrm
\sbox{\plotpoint}{\rule[-0.175pt]{0.350pt}{0.350pt}}%
\put(370,95){\makebox(0,0){$I_{4,4}$}}
\put(430,322){\circle*{40}}
\put(310,322){\circle*{40}}
\put(310,203){\circle*{40}}
\put(430,203){\circle*{40}}
\put(430,322){\usebox{\plotpoint}}
\put(310,322){\rule[-0.175pt]{28.908pt}{0.350pt}}
\put(430,322){\usebox{\plotpoint}}
\put(430,203){\rule[-0.175pt]{0.350pt}{28.667pt}}
\put(430,322){\usebox{\plotpoint}}
\put(428,322){\usebox{\plotpoint}}
\put(427,321){\usebox{\plotpoint}}
\put(426,320){\usebox{\plotpoint}}
\put(425,319){\usebox{\plotpoint}}
\put(424,318){\usebox{\plotpoint}}
\put(423,317){\usebox{\plotpoint}}
\put(422,316){\usebox{\plotpoint}}
\put(421,315){\usebox{\plotpoint}}
\put(420,314){\usebox{\plotpoint}}
\put(419,313){\usebox{\plotpoint}}
\put(418,312){\usebox{\plotpoint}}
\put(417,311){\usebox{\plotpoint}}
\put(416,310){\usebox{\plotpoint}}
\put(415,309){\usebox{\plotpoint}}
\put(414,308){\usebox{\plotpoint}}
\put(413,307){\usebox{\plotpoint}}
\put(412,306){\usebox{\plotpoint}}
\put(411,305){\usebox{\plotpoint}}
\put(410,304){\usebox{\plotpoint}}
\put(409,303){\usebox{\plotpoint}}
\put(408,302){\usebox{\plotpoint}}
\put(407,301){\usebox{\plotpoint}}
\put(406,300){\usebox{\plotpoint}}
\put(405,299){\usebox{\plotpoint}}
\put(404,298){\usebox{\plotpoint}}
\put(403,297){\usebox{\plotpoint}}
\put(402,296){\usebox{\plotpoint}}
\put(401,295){\usebox{\plotpoint}}
\put(400,294){\usebox{\plotpoint}}
\put(399,293){\usebox{\plotpoint}}
\put(398,292){\usebox{\plotpoint}}
\put(397,291){\usebox{\plotpoint}}
\put(396,290){\usebox{\plotpoint}}
\put(395,289){\usebox{\plotpoint}}
\put(394,288){\usebox{\plotpoint}}
\put(393,287){\usebox{\plotpoint}}
\put(392,286){\usebox{\plotpoint}}
\put(391,285){\usebox{\plotpoint}}
\put(390,284){\usebox{\plotpoint}}
\put(389,283){\usebox{\plotpoint}}
\put(388,282){\usebox{\plotpoint}}
\put(387,281){\usebox{\plotpoint}}
\put(386,280){\usebox{\plotpoint}}
\put(385,279){\usebox{\plotpoint}}
\put(384,278){\usebox{\plotpoint}}
\put(383,277){\usebox{\plotpoint}}
\put(382,276){\usebox{\plotpoint}}
\put(381,275){\usebox{\plotpoint}}
\put(380,274){\usebox{\plotpoint}}
\put(379,273){\usebox{\plotpoint}}
\put(378,272){\usebox{\plotpoint}}
\put(377,271){\usebox{\plotpoint}}
\put(376,270){\usebox{\plotpoint}}
\put(375,269){\usebox{\plotpoint}}
\put(374,268){\usebox{\plotpoint}}
\put(373,267){\usebox{\plotpoint}}
\put(372,266){\usebox{\plotpoint}}
\put(371,265){\usebox{\plotpoint}}
\put(370,264){\usebox{\plotpoint}}
\put(369,263){\usebox{\plotpoint}}
\put(368,262){\usebox{\plotpoint}}
\put(367,261){\usebox{\plotpoint}}
\put(366,260){\usebox{\plotpoint}}
\put(365,259){\usebox{\plotpoint}}
\put(364,258){\usebox{\plotpoint}}
\put(363,257){\usebox{\plotpoint}}
\put(362,256){\usebox{\plotpoint}}
\put(361,255){\usebox{\plotpoint}}
\put(360,254){\usebox{\plotpoint}}
\put(359,253){\usebox{\plotpoint}}
\put(358,252){\usebox{\plotpoint}}
\put(357,251){\usebox{\plotpoint}}
\put(356,250){\usebox{\plotpoint}}
\put(355,249){\usebox{\plotpoint}}
\put(354,248){\usebox{\plotpoint}}
\put(353,247){\usebox{\plotpoint}}
\put(352,246){\usebox{\plotpoint}}
\put(351,245){\usebox{\plotpoint}}
\put(350,244){\usebox{\plotpoint}}
\put(349,243){\usebox{\plotpoint}}
\put(348,242){\usebox{\plotpoint}}
\put(347,241){\usebox{\plotpoint}}
\put(346,240){\usebox{\plotpoint}}
\put(345,239){\usebox{\plotpoint}}
\put(344,238){\usebox{\plotpoint}}
\put(343,237){\usebox{\plotpoint}}
\put(342,236){\usebox{\plotpoint}}
\put(341,235){\usebox{\plotpoint}}
\put(340,234){\usebox{\plotpoint}}
\put(339,233){\usebox{\plotpoint}}
\put(338,232){\usebox{\plotpoint}}
\put(337,231){\usebox{\plotpoint}}
\put(336,230){\usebox{\plotpoint}}
\put(335,229){\usebox{\plotpoint}}
\put(334,228){\usebox{\plotpoint}}
\put(333,227){\usebox{\plotpoint}}
\put(332,226){\usebox{\plotpoint}}
\put(331,225){\usebox{\plotpoint}}
\put(330,224){\usebox{\plotpoint}}
\put(329,223){\usebox{\plotpoint}}
\put(328,222){\usebox{\plotpoint}}
\put(327,221){\usebox{\plotpoint}}
\put(326,220){\usebox{\plotpoint}}
\put(325,219){\usebox{\plotpoint}}
\put(324,218){\usebox{\plotpoint}}
\put(323,217){\usebox{\plotpoint}}
\put(322,216){\usebox{\plotpoint}}
\put(321,215){\usebox{\plotpoint}}
\put(320,214){\usebox{\plotpoint}}
\put(319,213){\usebox{\plotpoint}}
\put(318,212){\usebox{\plotpoint}}
\put(317,211){\usebox{\plotpoint}}
\put(316,210){\usebox{\plotpoint}}
\put(315,209){\usebox{\plotpoint}}
\put(314,208){\usebox{\plotpoint}}
\put(313,207){\usebox{\plotpoint}}
\put(312,206){\usebox{\plotpoint}}
\put(311,205){\usebox{\plotpoint}}
\put(310,204){\usebox{\plotpoint}}
\put(310,203){\usebox{\plotpoint}}
\sbox{\plotpoint}{\rule[-0.250pt]{0.500pt}{0.500pt}}%
\end{picture}
	% GNUPLOT: LaTeX picture
\setlength{\unitlength}{0.240900pt}
\ifx\plotpoint\undefined\newsavebox{\plotpoint}\fi
\sbox{\plotpoint}{\rule[-0.175pt]{0.350pt}{0.350pt}}%
\begin{picture}(250,300)(0,100)
\tenrm
\sbox{\plotpoint}{\rule[-0.175pt]{0.350pt}{0.350pt}}%
\put(370,95){\makebox(0,0){$I_{4,5}$}}
\put(430,322){\circle*{40}}
\put(310,322){\circle*{40}}
\put(310,203){\circle*{40}}
\put(430,203){\circle*{40}}
\put(430,322){\usebox{\plotpoint}}
\put(310,322){\rule[-0.175pt]{28.908pt}{0.350pt}}
\put(310,322){\usebox{\plotpoint}}
\put(310,203){\rule[-0.175pt]{0.350pt}{28.667pt}}
\put(430,322){\usebox{\plotpoint}}
\put(428,322){\usebox{\plotpoint}}
\put(427,321){\usebox{\plotpoint}}
\put(426,320){\usebox{\plotpoint}}
\put(425,319){\usebox{\plotpoint}}
\put(424,318){\usebox{\plotpoint}}
\put(423,317){\usebox{\plotpoint}}
\put(422,316){\usebox{\plotpoint}}
\put(421,315){\usebox{\plotpoint}}
\put(420,314){\usebox{\plotpoint}}
\put(419,313){\usebox{\plotpoint}}
\put(418,312){\usebox{\plotpoint}}
\put(417,311){\usebox{\plotpoint}}
\put(416,310){\usebox{\plotpoint}}
\put(415,309){\usebox{\plotpoint}}
\put(414,308){\usebox{\plotpoint}}
\put(413,307){\usebox{\plotpoint}}
\put(412,306){\usebox{\plotpoint}}
\put(411,305){\usebox{\plotpoint}}
\put(410,304){\usebox{\plotpoint}}
\put(409,303){\usebox{\plotpoint}}
\put(408,302){\usebox{\plotpoint}}
\put(407,301){\usebox{\plotpoint}}
\put(406,300){\usebox{\plotpoint}}
\put(405,299){\usebox{\plotpoint}}
\put(404,298){\usebox{\plotpoint}}
\put(403,297){\usebox{\plotpoint}}
\put(402,296){\usebox{\plotpoint}}
\put(401,295){\usebox{\plotpoint}}
\put(400,294){\usebox{\plotpoint}}
\put(399,293){\usebox{\plotpoint}}
\put(398,292){\usebox{\plotpoint}}
\put(397,291){\usebox{\plotpoint}}
\put(396,290){\usebox{\plotpoint}}
\put(395,289){\usebox{\plotpoint}}
\put(394,288){\usebox{\plotpoint}}
\put(393,287){\usebox{\plotpoint}}
\put(392,286){\usebox{\plotpoint}}
\put(391,285){\usebox{\plotpoint}}
\put(390,284){\usebox{\plotpoint}}
\put(389,283){\usebox{\plotpoint}}
\put(388,282){\usebox{\plotpoint}}
\put(387,281){\usebox{\plotpoint}}
\put(386,280){\usebox{\plotpoint}}
\put(385,279){\usebox{\plotpoint}}
\put(384,278){\usebox{\plotpoint}}
\put(383,277){\usebox{\plotpoint}}
\put(382,276){\usebox{\plotpoint}}
\put(381,275){\usebox{\plotpoint}}
\put(380,274){\usebox{\plotpoint}}
\put(379,273){\usebox{\plotpoint}}
\put(378,272){\usebox{\plotpoint}}
\put(377,271){\usebox{\plotpoint}}
\put(376,270){\usebox{\plotpoint}}
\put(375,269){\usebox{\plotpoint}}
\put(374,268){\usebox{\plotpoint}}
\put(373,267){\usebox{\plotpoint}}
\put(372,266){\usebox{\plotpoint}}
\put(371,265){\usebox{\plotpoint}}
\put(370,264){\usebox{\plotpoint}}
\put(369,263){\usebox{\plotpoint}}
\put(368,262){\usebox{\plotpoint}}
\put(367,261){\usebox{\plotpoint}}
\put(366,260){\usebox{\plotpoint}}
\put(365,259){\usebox{\plotpoint}}
\put(364,258){\usebox{\plotpoint}}
\put(363,257){\usebox{\plotpoint}}
\put(362,256){\usebox{\plotpoint}}
\put(361,255){\usebox{\plotpoint}}
\put(360,254){\usebox{\plotpoint}}
\put(359,253){\usebox{\plotpoint}}
\put(358,252){\usebox{\plotpoint}}
\put(357,251){\usebox{\plotpoint}}
\put(356,250){\usebox{\plotpoint}}
\put(355,249){\usebox{\plotpoint}}
\put(354,248){\usebox{\plotpoint}}
\put(353,247){\usebox{\plotpoint}}
\put(352,246){\usebox{\plotpoint}}
\put(351,245){\usebox{\plotpoint}}
\put(350,244){\usebox{\plotpoint}}
\put(349,243){\usebox{\plotpoint}}
\put(348,242){\usebox{\plotpoint}}
\put(347,241){\usebox{\plotpoint}}
\put(346,240){\usebox{\plotpoint}}
\put(345,239){\usebox{\plotpoint}}
\put(344,238){\usebox{\plotpoint}}
\put(343,237){\usebox{\plotpoint}}
\put(342,236){\usebox{\plotpoint}}
\put(341,235){\usebox{\plotpoint}}
\put(340,234){\usebox{\plotpoint}}
\put(339,233){\usebox{\plotpoint}}
\put(338,232){\usebox{\plotpoint}}
\put(337,231){\usebox{\plotpoint}}
\put(336,230){\usebox{\plotpoint}}
\put(335,229){\usebox{\plotpoint}}
\put(334,228){\usebox{\plotpoint}}
\put(333,227){\usebox{\plotpoint}}
\put(332,226){\usebox{\plotpoint}}
\put(331,225){\usebox{\plotpoint}}
\put(330,224){\usebox{\plotpoint}}
\put(329,223){\usebox{\plotpoint}}
\put(328,222){\usebox{\plotpoint}}
\put(327,221){\usebox{\plotpoint}}
\put(326,220){\usebox{\plotpoint}}
\put(325,219){\usebox{\plotpoint}}
\put(324,218){\usebox{\plotpoint}}
\put(323,217){\usebox{\plotpoint}}
\put(322,216){\usebox{\plotpoint}}
\put(321,215){\usebox{\plotpoint}}
\put(320,214){\usebox{\plotpoint}}
\put(319,213){\usebox{\plotpoint}}
\put(318,212){\usebox{\plotpoint}}
\put(317,211){\usebox{\plotpoint}}
\put(316,210){\usebox{\plotpoint}}
\put(315,209){\usebox{\plotpoint}}
\put(314,208){\usebox{\plotpoint}}
\put(313,207){\usebox{\plotpoint}}
\put(312,206){\usebox{\plotpoint}}
\put(311,205){\usebox{\plotpoint}}
\put(310,204){\usebox{\plotpoint}}
\put(310,203){\usebox{\plotpoint}}
\sbox{\plotpoint}{\rule[-0.250pt]{0.500pt}{0.500pt}}%
\end{picture}
	% GNUPLOT: LaTeX picture
\setlength{\unitlength}{0.240900pt}
\ifx\plotpoint\undefined\newsavebox{\plotpoint}\fi
\sbox{\plotpoint}{\rule[-0.175pt]{0.350pt}{0.350pt}}%
\begin{picture}(250,300)(0,100)
\tenrm
\sbox{\plotpoint}{\rule[-0.175pt]{0.350pt}{0.350pt}}%
\put(370,95){\makebox(0,0){$I_{4,6}$}}
\put(430,322){\circle*{40}}
\put(310,322){\circle*{40}}
\put(310,203){\circle*{40}}
\put(430,203){\circle*{40}}
\put(310,203){\usebox{\plotpoint}}
\put(310,203){\rule[-0.175pt]{28.908pt}{0.350pt}}
\put(303,197){\usebox{\plotpoint}}
\put(303,197){\rule[-0.175pt]{32.281pt}{0.350pt}}
\put(317,210){\usebox{\plotpoint}}
\put(317,210){\rule[-0.175pt]{25.535pt}{0.350pt}}
\sbox{\plotpoint}{\rule[-0.250pt]{0.500pt}{0.500pt}}%
\end{picture}
	% GNUPLOT: LaTeX picture
\setlength{\unitlength}{0.240900pt}
\ifx\plotpoint\undefined\newsavebox{\plotpoint}\fi
\sbox{\plotpoint}{\rule[-0.175pt]{0.350pt}{0.350pt}}%
\begin{picture}(250,300)(0,100)
\tenrm
\sbox{\plotpoint}{\rule[-0.175pt]{0.350pt}{0.350pt}}%
\put(370,95){\makebox(0,0){$I_{4,7}$}}
\put(430,322){\circle*{40}}
\put(310,322){\circle*{40}}
\put(310,203){\circle*{40}}
\put(430,203){\circle*{40}}
\put(437,328){\usebox{\plotpoint}}
\put(303,328){\rule[-0.175pt]{32.281pt}{0.350pt}}
\put(423,315){\usebox{\plotpoint}}
\put(317,315){\rule[-0.175pt]{25.535pt}{0.350pt}}
\put(303,197){\usebox{\plotpoint}}
\put(303,197){\rule[-0.175pt]{32.281pt}{0.350pt}}
\put(317,210){\usebox{\plotpoint}}
\put(317,210){\rule[-0.175pt]{25.535pt}{0.350pt}}
\sbox{\plotpoint}{\rule[-0.250pt]{0.500pt}{0.500pt}}%
\end{picture}
	% GNUPLOT: LaTeX picture
\setlength{\unitlength}{0.240900pt}
\ifx\plotpoint\undefined\newsavebox{\plotpoint}\fi
\sbox{\plotpoint}{\rule[-0.175pt]{0.350pt}{0.350pt}}%
\begin{picture}(250,300)(0,100)
\tenrm
\sbox{\plotpoint}{\rule[-0.175pt]{0.350pt}{0.350pt}}%
\put(370,95){\makebox(0,0){$I_{4,8}$}}
\put(430,322){\circle*{40}}
\put(310,322){\circle*{40}}
\put(310,203){\circle*{40}}
\put(430,203){\circle*{40}}
\put(296,190){\usebox{\plotpoint}}
\put(296,190){\rule[-0.175pt]{35.653pt}{0.350pt}}
\put(324,217){\usebox{\plotpoint}}
\put(324,217){\rule[-0.175pt]{22.163pt}{0.350pt}}
\put(303,197){\usebox{\plotpoint}}
\put(303,197){\rule[-0.175pt]{32.281pt}{0.350pt}}
\put(317,210){\usebox{\plotpoint}}
\put(317,210){\rule[-0.175pt]{25.535pt}{0.350pt}}
\sbox{\plotpoint}{\rule[-0.250pt]{0.500pt}{0.500pt}}%
\end{picture}
	% GNUPLOT: LaTeX picture
\setlength{\unitlength}{0.240900pt}
\ifx\plotpoint\undefined\newsavebox{\plotpoint}\fi
\sbox{\plotpoint}{\rule[-0.175pt]{0.350pt}{0.350pt}}%
\begin{picture}(250,300)(0,100)
\tenrm
\sbox{\plotpoint}{\rule[-0.175pt]{0.350pt}{0.350pt}}%
\put(370,95){\makebox(0,0){$I_{4,9}$}}
\put(430,322){\circle*{40}}
\put(310,322){\circle*{40}}
\put(310,203){\circle*{40}}
\put(430,203){\circle*{40}}
\put(296,190){\usebox{\plotpoint}}
\put(296,190){\rule[-0.175pt]{35.653pt}{0.350pt}}
\put(310,203){\usebox{\plotpoint}}
\put(310,203){\rule[-0.175pt]{28.908pt}{0.350pt}}
\put(324,217){\usebox{\plotpoint}}
\put(324,217){\rule[-0.175pt]{22.163pt}{0.350pt}}
\put(303,197){\usebox{\plotpoint}}
\put(303,197){\rule[-0.175pt]{32.281pt}{0.350pt}}
\put(317,210){\usebox{\plotpoint}}
\put(317,210){\rule[-0.175pt]{25.535pt}{0.350pt}}
\end{picture}
	}
\end{figure}
\begin{figure}
	{\Large
	% GNUPLOT: LaTeX picture
\setlength{\unitlength}{0.240900pt}
\ifx\plotpoint\undefined\newsavebox{\plotpoint}\fi
\sbox{\plotpoint}{\rule[-0.175pt]{0.350pt}{0.350pt}}%
\begin{picture}(250,300)(0,100)
\tenrm
\sbox{\plotpoint}{\rule[-0.175pt]{0.350pt}{0.350pt}}%
\put(370,95){\makebox(0,0){$I_{5,1}$}}
\put(370,346){\circle*{40}}
\put(289,288){\circle*{40}}
\put(320,195){\circle*{40}}
\put(420,195){\circle*{40}}
\put(451,288){\circle*{40}}
\put(320,195){\usebox{\plotpoint}}
\put(320,195){\rule[-0.175pt]{24.090pt}{0.350pt}}
\sbox{\plotpoint}{\rule[-0.500pt]{1.000pt}{1.000pt}}%
\end{picture}
	% GNUPLOT: LaTeX picture
\setlength{\unitlength}{0.240900pt}
\ifx\plotpoint\undefined\newsavebox{\plotpoint}\fi
\sbox{\plotpoint}{\rule[-0.175pt]{0.350pt}{0.350pt}}%
\begin{picture}(250,300)(0,100)
\tenrm
\sbox{\plotpoint}{\rule[-0.175pt]{0.350pt}{0.350pt}}%
\put(370,95){\makebox(0,0){$I_{5,2}$}}
\put(370,346){\circle*{40}}
\put(289,288){\circle*{40}}
\put(320,195){\circle*{40}}
\put(420,195){\circle*{40}}
\put(451,288){\circle*{40}}
\put(315,188){\usebox{\plotpoint}}
\put(315,188){\rule[-0.175pt]{26.499pt}{0.350pt}}
\put(325,202){\usebox{\plotpoint}}
\put(325,202){\rule[-0.175pt]{21.681pt}{0.350pt}}
\sbox{\plotpoint}{\rule[-0.250pt]{0.500pt}{0.500pt}}%
\end{picture}
	% GNUPLOT: LaTeX picture
\setlength{\unitlength}{0.240900pt}
\ifx\plotpoint\undefined\newsavebox{\plotpoint}\fi
\sbox{\plotpoint}{\rule[-0.175pt]{0.350pt}{0.350pt}}%
\begin{picture}(250,300)(0,100)
\tenrm
\sbox{\plotpoint}{\rule[-0.175pt]{0.350pt}{0.350pt}}%
\put(370,95){\makebox(0,0){$I_{5,3}$}}
\put(370,346){\circle*{40}}
\put(289,288){\circle*{40}}
\put(320,195){\circle*{40}}
\put(420,195){\circle*{40}}
\put(451,288){\circle*{40}}
\put(289,288){\usebox{\plotpoint}}
\put(289,285){\rule[-0.175pt]{0.350pt}{0.723pt}}
\put(290,282){\rule[-0.175pt]{0.350pt}{0.723pt}}
\put(291,279){\rule[-0.175pt]{0.350pt}{0.723pt}}
\put(292,276){\rule[-0.175pt]{0.350pt}{0.723pt}}
\put(293,273){\rule[-0.175pt]{0.350pt}{0.723pt}}
\put(294,270){\rule[-0.175pt]{0.350pt}{0.723pt}}
\put(295,267){\rule[-0.175pt]{0.350pt}{0.723pt}}
\put(296,264){\rule[-0.175pt]{0.350pt}{0.723pt}}
\put(297,261){\rule[-0.175pt]{0.350pt}{0.723pt}}
\put(298,258){\rule[-0.175pt]{0.350pt}{0.723pt}}
\put(299,255){\rule[-0.175pt]{0.350pt}{0.723pt}}
\put(300,252){\rule[-0.175pt]{0.350pt}{0.723pt}}
\put(301,249){\rule[-0.175pt]{0.350pt}{0.723pt}}
\put(302,246){\rule[-0.175pt]{0.350pt}{0.723pt}}
\put(303,243){\rule[-0.175pt]{0.350pt}{0.723pt}}
\put(304,240){\rule[-0.175pt]{0.350pt}{0.723pt}}
\put(305,237){\rule[-0.175pt]{0.350pt}{0.723pt}}
\put(306,234){\rule[-0.175pt]{0.350pt}{0.723pt}}
\put(307,231){\rule[-0.175pt]{0.350pt}{0.723pt}}
\put(308,228){\rule[-0.175pt]{0.350pt}{0.723pt}}
\put(309,225){\rule[-0.175pt]{0.350pt}{0.723pt}}
\put(310,222){\rule[-0.175pt]{0.350pt}{0.723pt}}
\put(311,219){\rule[-0.175pt]{0.350pt}{0.723pt}}
\put(312,216){\rule[-0.175pt]{0.350pt}{0.723pt}}
\put(313,213){\rule[-0.175pt]{0.350pt}{0.723pt}}
\put(314,210){\rule[-0.175pt]{0.350pt}{0.723pt}}
\put(315,207){\rule[-0.175pt]{0.350pt}{0.723pt}}
\put(316,204){\rule[-0.175pt]{0.350pt}{0.723pt}}
\put(317,201){\rule[-0.175pt]{0.350pt}{0.723pt}}
\put(318,198){\rule[-0.175pt]{0.350pt}{0.723pt}}
\put(319,195){\rule[-0.175pt]{0.350pt}{0.723pt}}
\put(420,195){\usebox{\plotpoint}}
\put(420,195){\rule[-0.175pt]{0.350pt}{0.723pt}}
\put(421,198){\rule[-0.175pt]{0.350pt}{0.723pt}}
\put(422,201){\rule[-0.175pt]{0.350pt}{0.723pt}}
\put(423,204){\rule[-0.175pt]{0.350pt}{0.723pt}}
\put(424,207){\rule[-0.175pt]{0.350pt}{0.723pt}}
\put(425,210){\rule[-0.175pt]{0.350pt}{0.723pt}}
\put(426,213){\rule[-0.175pt]{0.350pt}{0.723pt}}
\put(427,216){\rule[-0.175pt]{0.350pt}{0.723pt}}
\put(428,219){\rule[-0.175pt]{0.350pt}{0.723pt}}
\put(429,222){\rule[-0.175pt]{0.350pt}{0.723pt}}
\put(430,225){\rule[-0.175pt]{0.350pt}{0.723pt}}
\put(431,228){\rule[-0.175pt]{0.350pt}{0.723pt}}
\put(432,231){\rule[-0.175pt]{0.350pt}{0.723pt}}
\put(433,234){\rule[-0.175pt]{0.350pt}{0.723pt}}
\put(434,237){\rule[-0.175pt]{0.350pt}{0.723pt}}
\put(435,240){\rule[-0.175pt]{0.350pt}{0.723pt}}
\put(436,243){\rule[-0.175pt]{0.350pt}{0.723pt}}
\put(437,246){\rule[-0.175pt]{0.350pt}{0.723pt}}
\put(438,249){\rule[-0.175pt]{0.350pt}{0.723pt}}
\put(439,252){\rule[-0.175pt]{0.350pt}{0.723pt}}
\put(440,255){\rule[-0.175pt]{0.350pt}{0.723pt}}
\put(441,258){\rule[-0.175pt]{0.350pt}{0.723pt}}
\put(442,261){\rule[-0.175pt]{0.350pt}{0.723pt}}
\put(443,264){\rule[-0.175pt]{0.350pt}{0.723pt}}
\put(444,267){\rule[-0.175pt]{0.350pt}{0.723pt}}
\put(445,270){\rule[-0.175pt]{0.350pt}{0.723pt}}
\put(446,273){\rule[-0.175pt]{0.350pt}{0.723pt}}
\put(447,276){\rule[-0.175pt]{0.350pt}{0.723pt}}
\put(448,279){\rule[-0.175pt]{0.350pt}{0.723pt}}
\put(449,282){\rule[-0.175pt]{0.350pt}{0.723pt}}
\put(450,285){\rule[-0.175pt]{0.350pt}{0.723pt}}
\sbox{\plotpoint}{\rule[-0.250pt]{0.500pt}{0.500pt}}%
\end{picture}
	% GNUPLOT: LaTeX picture
\setlength{\unitlength}{0.240900pt}
\ifx\plotpoint\undefined\newsavebox{\plotpoint}\fi
\sbox{\plotpoint}{\rule[-0.175pt]{0.350pt}{0.350pt}}%
\begin{picture}(250,300)(0,100)
\tenrm
\sbox{\plotpoint}{\rule[-0.175pt]{0.350pt}{0.350pt}}%
\put(370,95){\makebox(0,0){$I_{5,4}$}}
\put(370,346){\circle*{40}}
\put(289,288){\circle*{40}}
\put(320,195){\circle*{40}}
\put(420,195){\circle*{40}}
\put(451,288){\circle*{40}}
\put(315,188){\usebox{\plotpoint}}
\put(315,188){\rule[-0.175pt]{26.499pt}{0.350pt}}
\put(320,195){\usebox{\plotpoint}}
\put(320,195){\rule[-0.175pt]{24.090pt}{0.350pt}}
\put(325,202){\usebox{\plotpoint}}
\put(325,202){\rule[-0.175pt]{21.681pt}{0.350pt}}
\sbox{\plotpoint}{\rule[-0.250pt]{0.500pt}{0.500pt}}%
\end{picture}
	% GNUPLOT: LaTeX picture
\setlength{\unitlength}{0.240900pt}
\ifx\plotpoint\undefined\newsavebox{\plotpoint}\fi
\sbox{\plotpoint}{\rule[-0.175pt]{0.350pt}{0.350pt}}%
\begin{picture}(250,300)(0,100)
\tenrm
\sbox{\plotpoint}{\rule[-0.175pt]{0.350pt}{0.350pt}}%
\put(370,95){\makebox(0,0){$I_{5,5}$}}
\put(370,346){\circle*{40}}
\put(289,288){\circle*{40}}
\put(320,195){\circle*{40}}
\put(420,195){\circle*{40}}
\put(451,288){\circle*{40}}
\put(370,346){\usebox{\plotpoint}}
\put(368,346){\usebox{\plotpoint}}
\put(367,345){\usebox{\plotpoint}}
\put(365,344){\usebox{\plotpoint}}
\put(364,343){\usebox{\plotpoint}}
\put(363,342){\usebox{\plotpoint}}
\put(361,341){\usebox{\plotpoint}}
\put(360,340){\usebox{\plotpoint}}
\put(358,339){\usebox{\plotpoint}}
\put(357,338){\usebox{\plotpoint}}
\put(356,337){\usebox{\plotpoint}}
\put(354,336){\usebox{\plotpoint}}
\put(353,335){\usebox{\plotpoint}}
\put(351,334){\usebox{\plotpoint}}
\put(350,333){\usebox{\plotpoint}}
\put(349,332){\usebox{\plotpoint}}
\put(347,331){\usebox{\plotpoint}}
\put(346,330){\usebox{\plotpoint}}
\put(344,329){\usebox{\plotpoint}}
\put(343,328){\usebox{\plotpoint}}
\put(342,327){\usebox{\plotpoint}}
\put(340,326){\usebox{\plotpoint}}
\put(339,325){\usebox{\plotpoint}}
\put(337,324){\usebox{\plotpoint}}
\put(336,323){\usebox{\plotpoint}}
\put(335,322){\usebox{\plotpoint}}
\put(333,321){\usebox{\plotpoint}}
\put(332,320){\usebox{\plotpoint}}
\put(330,319){\usebox{\plotpoint}}
\put(329,318){\usebox{\plotpoint}}
\put(328,317){\usebox{\plotpoint}}
\put(326,316){\usebox{\plotpoint}}
\put(325,315){\usebox{\plotpoint}}
\put(323,314){\usebox{\plotpoint}}
\put(322,313){\usebox{\plotpoint}}
\put(321,312){\usebox{\plotpoint}}
\put(319,311){\usebox{\plotpoint}}
\put(318,310){\usebox{\plotpoint}}
\put(316,309){\usebox{\plotpoint}}
\put(315,308){\usebox{\plotpoint}}
\put(314,307){\usebox{\plotpoint}}
\put(312,306){\usebox{\plotpoint}}
\put(311,305){\usebox{\plotpoint}}
\put(309,304){\usebox{\plotpoint}}
\put(308,303){\usebox{\plotpoint}}
\put(307,302){\usebox{\plotpoint}}
\put(305,301){\usebox{\plotpoint}}
\put(304,300){\usebox{\plotpoint}}
\put(302,299){\usebox{\plotpoint}}
\put(301,298){\usebox{\plotpoint}}
\put(300,297){\usebox{\plotpoint}}
\put(298,296){\usebox{\plotpoint}}
\put(297,295){\usebox{\plotpoint}}
\put(295,294){\usebox{\plotpoint}}
\put(294,293){\usebox{\plotpoint}}
\put(293,292){\usebox{\plotpoint}}
\put(291,291){\usebox{\plotpoint}}
\put(290,290){\usebox{\plotpoint}}
\put(289,289){\usebox{\plotpoint}}
\put(289,288){\usebox{\plotpoint}}
\put(370,346){\usebox{\plotpoint}}
\put(370,346){\usebox{\plotpoint}}
\put(371,345){\usebox{\plotpoint}}
\put(372,344){\usebox{\plotpoint}}
\put(374,343){\usebox{\plotpoint}}
\put(375,342){\usebox{\plotpoint}}
\put(376,341){\usebox{\plotpoint}}
\put(378,340){\usebox{\plotpoint}}
\put(379,339){\usebox{\plotpoint}}
\put(381,338){\usebox{\plotpoint}}
\put(382,337){\usebox{\plotpoint}}
\put(383,336){\usebox{\plotpoint}}
\put(385,335){\usebox{\plotpoint}}
\put(386,334){\usebox{\plotpoint}}
\put(388,333){\usebox{\plotpoint}}
\put(389,332){\usebox{\plotpoint}}
\put(390,331){\usebox{\plotpoint}}
\put(392,330){\usebox{\plotpoint}}
\put(393,329){\usebox{\plotpoint}}
\put(395,328){\usebox{\plotpoint}}
\put(396,327){\usebox{\plotpoint}}
\put(397,326){\usebox{\plotpoint}}
\put(399,325){\usebox{\plotpoint}}
\put(400,324){\usebox{\plotpoint}}
\put(402,323){\usebox{\plotpoint}}
\put(403,322){\usebox{\plotpoint}}
\put(404,321){\usebox{\plotpoint}}
\put(406,320){\usebox{\plotpoint}}
\put(407,319){\usebox{\plotpoint}}
\put(409,318){\usebox{\plotpoint}}
\put(410,317){\usebox{\plotpoint}}
\put(411,316){\usebox{\plotpoint}}
\put(413,315){\usebox{\plotpoint}}
\put(414,314){\usebox{\plotpoint}}
\put(416,313){\usebox{\plotpoint}}
\put(417,312){\usebox{\plotpoint}}
\put(418,311){\usebox{\plotpoint}}
\put(420,310){\usebox{\plotpoint}}
\put(421,309){\usebox{\plotpoint}}
\put(423,308){\usebox{\plotpoint}}
\put(424,307){\usebox{\plotpoint}}
\put(425,306){\usebox{\plotpoint}}
\put(427,305){\usebox{\plotpoint}}
\put(428,304){\usebox{\plotpoint}}
\put(430,303){\usebox{\plotpoint}}
\put(431,302){\usebox{\plotpoint}}
\put(432,301){\usebox{\plotpoint}}
\put(434,300){\usebox{\plotpoint}}
\put(435,299){\usebox{\plotpoint}}
\put(437,298){\usebox{\plotpoint}}
\put(438,297){\usebox{\plotpoint}}
\put(439,296){\usebox{\plotpoint}}
\put(441,295){\usebox{\plotpoint}}
\put(442,294){\usebox{\plotpoint}}
\put(444,293){\usebox{\plotpoint}}
\put(445,292){\usebox{\plotpoint}}
\put(446,291){\usebox{\plotpoint}}
\put(448,290){\usebox{\plotpoint}}
\put(449,289){\usebox{\plotpoint}}
\put(450,288){\usebox{\plotpoint}}
\put(289,288){\usebox{\plotpoint}}
\put(289,288){\rule[-0.175pt]{39.026pt}{0.350pt}}
\sbox{\plotpoint}{\rule[-0.250pt]{0.500pt}{0.500pt}}%
\end{picture}
	% GNUPLOT: LaTeX picture
\setlength{\unitlength}{0.240900pt}
\ifx\plotpoint\undefined\newsavebox{\plotpoint}\fi
\sbox{\plotpoint}{\rule[-0.175pt]{0.350pt}{0.350pt}}%
\begin{picture}(250,300)(0,100)
\tenrm
\sbox{\plotpoint}{\rule[-0.175pt]{0.350pt}{0.350pt}}%
\put(370,95){\makebox(0,0){$I_{5,6}$}}
\put(370,346){\circle*{40}}
\put(289,288){\circle*{40}}
\put(320,195){\circle*{40}}
\put(420,195){\circle*{40}}
\put(451,288){\circle*{40}}
\put(370,346){\usebox{\plotpoint}}
\put(368,346){\usebox{\plotpoint}}
\put(367,345){\usebox{\plotpoint}}
\put(365,344){\usebox{\plotpoint}}
\put(364,343){\usebox{\plotpoint}}
\put(363,342){\usebox{\plotpoint}}
\put(361,341){\usebox{\plotpoint}}
\put(360,340){\usebox{\plotpoint}}
\put(358,339){\usebox{\plotpoint}}
\put(357,338){\usebox{\plotpoint}}
\put(356,337){\usebox{\plotpoint}}
\put(354,336){\usebox{\plotpoint}}
\put(353,335){\usebox{\plotpoint}}
\put(351,334){\usebox{\plotpoint}}
\put(350,333){\usebox{\plotpoint}}
\put(349,332){\usebox{\plotpoint}}
\put(347,331){\usebox{\plotpoint}}
\put(346,330){\usebox{\plotpoint}}
\put(344,329){\usebox{\plotpoint}}
\put(343,328){\usebox{\plotpoint}}
\put(342,327){\usebox{\plotpoint}}
\put(340,326){\usebox{\plotpoint}}
\put(339,325){\usebox{\plotpoint}}
\put(337,324){\usebox{\plotpoint}}
\put(336,323){\usebox{\plotpoint}}
\put(335,322){\usebox{\plotpoint}}
\put(333,321){\usebox{\plotpoint}}
\put(332,320){\usebox{\plotpoint}}
\put(330,319){\usebox{\plotpoint}}
\put(329,318){\usebox{\plotpoint}}
\put(328,317){\usebox{\plotpoint}}
\put(326,316){\usebox{\plotpoint}}
\put(325,315){\usebox{\plotpoint}}
\put(323,314){\usebox{\plotpoint}}
\put(322,313){\usebox{\plotpoint}}
\put(321,312){\usebox{\plotpoint}}
\put(319,311){\usebox{\plotpoint}}
\put(318,310){\usebox{\plotpoint}}
\put(316,309){\usebox{\plotpoint}}
\put(315,308){\usebox{\plotpoint}}
\put(314,307){\usebox{\plotpoint}}
\put(312,306){\usebox{\plotpoint}}
\put(311,305){\usebox{\plotpoint}}
\put(309,304){\usebox{\plotpoint}}
\put(308,303){\usebox{\plotpoint}}
\put(307,302){\usebox{\plotpoint}}
\put(305,301){\usebox{\plotpoint}}
\put(304,300){\usebox{\plotpoint}}
\put(302,299){\usebox{\plotpoint}}
\put(301,298){\usebox{\plotpoint}}
\put(300,297){\usebox{\plotpoint}}
\put(298,296){\usebox{\plotpoint}}
\put(297,295){\usebox{\plotpoint}}
\put(295,294){\usebox{\plotpoint}}
\put(294,293){\usebox{\plotpoint}}
\put(293,292){\usebox{\plotpoint}}
\put(291,291){\usebox{\plotpoint}}
\put(290,290){\usebox{\plotpoint}}
\put(289,289){\usebox{\plotpoint}}
\put(289,288){\usebox{\plotpoint}}
\put(370,346){\usebox{\plotpoint}}
\put(370,342){\rule[-0.175pt]{0.350pt}{0.728pt}}
\put(369,339){\rule[-0.175pt]{0.350pt}{0.728pt}}
\put(368,336){\rule[-0.175pt]{0.350pt}{0.728pt}}
\put(367,333){\rule[-0.175pt]{0.350pt}{0.728pt}}
\put(366,330){\rule[-0.175pt]{0.350pt}{0.728pt}}
\put(365,327){\rule[-0.175pt]{0.350pt}{0.728pt}}
\put(364,324){\rule[-0.175pt]{0.350pt}{0.728pt}}
\put(363,321){\rule[-0.175pt]{0.350pt}{0.728pt}}
\put(362,318){\rule[-0.175pt]{0.350pt}{0.728pt}}
\put(361,315){\rule[-0.175pt]{0.350pt}{0.728pt}}
\put(360,312){\rule[-0.175pt]{0.350pt}{0.728pt}}
\put(359,309){\rule[-0.175pt]{0.350pt}{0.728pt}}
\put(358,306){\rule[-0.175pt]{0.350pt}{0.728pt}}
\put(357,303){\rule[-0.175pt]{0.350pt}{0.728pt}}
\put(356,300){\rule[-0.175pt]{0.350pt}{0.728pt}}
\put(355,297){\rule[-0.175pt]{0.350pt}{0.728pt}}
\put(354,294){\rule[-0.175pt]{0.350pt}{0.728pt}}
\put(353,291){\rule[-0.175pt]{0.350pt}{0.728pt}}
\put(352,288){\rule[-0.175pt]{0.350pt}{0.728pt}}
\put(351,285){\rule[-0.175pt]{0.350pt}{0.728pt}}
\put(350,282){\rule[-0.175pt]{0.350pt}{0.728pt}}
\put(349,279){\rule[-0.175pt]{0.350pt}{0.728pt}}
\put(348,276){\rule[-0.175pt]{0.350pt}{0.728pt}}
\put(347,273){\rule[-0.175pt]{0.350pt}{0.728pt}}
\put(346,270){\rule[-0.175pt]{0.350pt}{0.728pt}}
\put(345,267){\rule[-0.175pt]{0.350pt}{0.728pt}}
\put(344,264){\rule[-0.175pt]{0.350pt}{0.728pt}}
\put(343,261){\rule[-0.175pt]{0.350pt}{0.728pt}}
\put(342,258){\rule[-0.175pt]{0.350pt}{0.728pt}}
\put(341,255){\rule[-0.175pt]{0.350pt}{0.728pt}}
\put(340,252){\rule[-0.175pt]{0.350pt}{0.728pt}}
\put(339,249){\rule[-0.175pt]{0.350pt}{0.728pt}}
\put(338,246){\rule[-0.175pt]{0.350pt}{0.728pt}}
\put(337,243){\rule[-0.175pt]{0.350pt}{0.728pt}}
\put(336,240){\rule[-0.175pt]{0.350pt}{0.728pt}}
\put(335,237){\rule[-0.175pt]{0.350pt}{0.728pt}}
\put(334,234){\rule[-0.175pt]{0.350pt}{0.728pt}}
\put(333,231){\rule[-0.175pt]{0.350pt}{0.728pt}}
\put(332,228){\rule[-0.175pt]{0.350pt}{0.728pt}}
\put(331,225){\rule[-0.175pt]{0.350pt}{0.728pt}}
\put(330,222){\rule[-0.175pt]{0.350pt}{0.728pt}}
\put(329,219){\rule[-0.175pt]{0.350pt}{0.728pt}}
\put(328,216){\rule[-0.175pt]{0.350pt}{0.728pt}}
\put(327,213){\rule[-0.175pt]{0.350pt}{0.728pt}}
\put(326,210){\rule[-0.175pt]{0.350pt}{0.728pt}}
\put(325,207){\rule[-0.175pt]{0.350pt}{0.728pt}}
\put(324,204){\rule[-0.175pt]{0.350pt}{0.728pt}}
\put(323,201){\rule[-0.175pt]{0.350pt}{0.728pt}}
\put(322,198){\rule[-0.175pt]{0.350pt}{0.728pt}}
\put(321,195){\rule[-0.175pt]{0.350pt}{0.728pt}}
\put(320,195){\usebox{\plotpoint}}
\put(370,346){\usebox{\plotpoint}}
\put(370,342){\rule[-0.175pt]{0.350pt}{0.728pt}}
\put(371,339){\rule[-0.175pt]{0.350pt}{0.728pt}}
\put(372,336){\rule[-0.175pt]{0.350pt}{0.728pt}}
\put(373,333){\rule[-0.175pt]{0.350pt}{0.728pt}}
\put(374,330){\rule[-0.175pt]{0.350pt}{0.728pt}}
\put(375,327){\rule[-0.175pt]{0.350pt}{0.728pt}}
\put(376,324){\rule[-0.175pt]{0.350pt}{0.728pt}}
\put(377,321){\rule[-0.175pt]{0.350pt}{0.728pt}}
\put(378,318){\rule[-0.175pt]{0.350pt}{0.728pt}}
\put(379,315){\rule[-0.175pt]{0.350pt}{0.728pt}}
\put(380,312){\rule[-0.175pt]{0.350pt}{0.728pt}}
\put(381,309){\rule[-0.175pt]{0.350pt}{0.728pt}}
\put(382,306){\rule[-0.175pt]{0.350pt}{0.728pt}}
\put(383,303){\rule[-0.175pt]{0.350pt}{0.728pt}}
\put(384,300){\rule[-0.175pt]{0.350pt}{0.728pt}}
\put(385,297){\rule[-0.175pt]{0.350pt}{0.728pt}}
\put(386,294){\rule[-0.175pt]{0.350pt}{0.728pt}}
\put(387,291){\rule[-0.175pt]{0.350pt}{0.728pt}}
\put(388,288){\rule[-0.175pt]{0.350pt}{0.728pt}}
\put(389,285){\rule[-0.175pt]{0.350pt}{0.728pt}}
\put(390,282){\rule[-0.175pt]{0.350pt}{0.728pt}}
\put(391,279){\rule[-0.175pt]{0.350pt}{0.728pt}}
\put(392,276){\rule[-0.175pt]{0.350pt}{0.728pt}}
\put(393,273){\rule[-0.175pt]{0.350pt}{0.728pt}}
\put(394,270){\rule[-0.175pt]{0.350pt}{0.728pt}}
\put(395,267){\rule[-0.175pt]{0.350pt}{0.728pt}}
\put(396,264){\rule[-0.175pt]{0.350pt}{0.728pt}}
\put(397,261){\rule[-0.175pt]{0.350pt}{0.728pt}}
\put(398,258){\rule[-0.175pt]{0.350pt}{0.728pt}}
\put(399,255){\rule[-0.175pt]{0.350pt}{0.728pt}}
\put(400,252){\rule[-0.175pt]{0.350pt}{0.728pt}}
\put(401,249){\rule[-0.175pt]{0.350pt}{0.728pt}}
\put(402,246){\rule[-0.175pt]{0.350pt}{0.728pt}}
\put(403,243){\rule[-0.175pt]{0.350pt}{0.728pt}}
\put(404,240){\rule[-0.175pt]{0.350pt}{0.728pt}}
\put(405,237){\rule[-0.175pt]{0.350pt}{0.728pt}}
\put(406,234){\rule[-0.175pt]{0.350pt}{0.728pt}}
\put(407,231){\rule[-0.175pt]{0.350pt}{0.728pt}}
\put(408,228){\rule[-0.175pt]{0.350pt}{0.728pt}}
\put(409,225){\rule[-0.175pt]{0.350pt}{0.728pt}}
\put(410,222){\rule[-0.175pt]{0.350pt}{0.728pt}}
\put(411,219){\rule[-0.175pt]{0.350pt}{0.728pt}}
\put(412,216){\rule[-0.175pt]{0.350pt}{0.728pt}}
\put(413,213){\rule[-0.175pt]{0.350pt}{0.728pt}}
\put(414,210){\rule[-0.175pt]{0.350pt}{0.728pt}}
\put(415,207){\rule[-0.175pt]{0.350pt}{0.728pt}}
\put(416,204){\rule[-0.175pt]{0.350pt}{0.728pt}}
\put(417,201){\rule[-0.175pt]{0.350pt}{0.728pt}}
\put(418,198){\rule[-0.175pt]{0.350pt}{0.728pt}}
\put(419,195){\rule[-0.175pt]{0.350pt}{0.728pt}}
\put(420,195){\usebox{\plotpoint}}
\sbox{\plotpoint}{\rule[-0.250pt]{0.500pt}{0.500pt}}%
\end{picture}
	}
\end{figure}
\begin{figure}
	{\Large
	% GNUPLOT: LaTeX picture
\setlength{\unitlength}{0.240900pt}
\ifx\plotpoint\undefined\newsavebox{\plotpoint}\fi
\sbox{\plotpoint}{\rule[-0.175pt]{0.350pt}{0.350pt}}%
\begin{picture}(250,300)(0,100)
\tenrm
\sbox{\plotpoint}{\rule[-0.175pt]{0.350pt}{0.350pt}}%
\put(370,95){\makebox(0,0){$I_{5,7}$}}
\put(370,346){\circle*{40}}
\put(289,288){\circle*{40}}
\put(320,195){\circle*{40}}
\put(420,195){\circle*{40}}
\put(451,288){\circle*{40}}
\put(370,346){\usebox{\plotpoint}}
\put(368,346){\usebox{\plotpoint}}
\put(367,345){\usebox{\plotpoint}}
\put(365,344){\usebox{\plotpoint}}
\put(364,343){\usebox{\plotpoint}}
\put(363,342){\usebox{\plotpoint}}
\put(361,341){\usebox{\plotpoint}}
\put(360,340){\usebox{\plotpoint}}
\put(358,339){\usebox{\plotpoint}}
\put(357,338){\usebox{\plotpoint}}
\put(356,337){\usebox{\plotpoint}}
\put(354,336){\usebox{\plotpoint}}
\put(353,335){\usebox{\plotpoint}}
\put(351,334){\usebox{\plotpoint}}
\put(350,333){\usebox{\plotpoint}}
\put(349,332){\usebox{\plotpoint}}
\put(347,331){\usebox{\plotpoint}}
\put(346,330){\usebox{\plotpoint}}
\put(344,329){\usebox{\plotpoint}}
\put(343,328){\usebox{\plotpoint}}
\put(342,327){\usebox{\plotpoint}}
\put(340,326){\usebox{\plotpoint}}
\put(339,325){\usebox{\plotpoint}}
\put(337,324){\usebox{\plotpoint}}
\put(336,323){\usebox{\plotpoint}}
\put(335,322){\usebox{\plotpoint}}
\put(333,321){\usebox{\plotpoint}}
\put(332,320){\usebox{\plotpoint}}
\put(330,319){\usebox{\plotpoint}}
\put(329,318){\usebox{\plotpoint}}
\put(328,317){\usebox{\plotpoint}}
\put(326,316){\usebox{\plotpoint}}
\put(325,315){\usebox{\plotpoint}}
\put(323,314){\usebox{\plotpoint}}
\put(322,313){\usebox{\plotpoint}}
\put(321,312){\usebox{\plotpoint}}
\put(319,311){\usebox{\plotpoint}}
\put(318,310){\usebox{\plotpoint}}
\put(316,309){\usebox{\plotpoint}}
\put(315,308){\usebox{\plotpoint}}
\put(314,307){\usebox{\plotpoint}}
\put(312,306){\usebox{\plotpoint}}
\put(311,305){\usebox{\plotpoint}}
\put(309,304){\usebox{\plotpoint}}
\put(308,303){\usebox{\plotpoint}}
\put(307,302){\usebox{\plotpoint}}
\put(305,301){\usebox{\plotpoint}}
\put(304,300){\usebox{\plotpoint}}
\put(302,299){\usebox{\plotpoint}}
\put(301,298){\usebox{\plotpoint}}
\put(300,297){\usebox{\plotpoint}}
\put(298,296){\usebox{\plotpoint}}
\put(297,295){\usebox{\plotpoint}}
\put(295,294){\usebox{\plotpoint}}
\put(294,293){\usebox{\plotpoint}}
\put(293,292){\usebox{\plotpoint}}
\put(291,291){\usebox{\plotpoint}}
\put(290,290){\usebox{\plotpoint}}
\put(289,289){\usebox{\plotpoint}}
\put(289,288){\usebox{\plotpoint}}
\put(315,188){\usebox{\plotpoint}}
\put(315,188){\rule[-0.175pt]{26.499pt}{0.350pt}}
\put(325,202){\usebox{\plotpoint}}
\put(325,202){\rule[-0.175pt]{21.681pt}{0.350pt}}
\sbox{\plotpoint}{\rule[-0.250pt]{0.500pt}{0.500pt}}%
\end{picture}
	% GNUPLOT: LaTeX picture
\setlength{\unitlength}{0.240900pt}
\ifx\plotpoint\undefined\newsavebox{\plotpoint}\fi
\sbox{\plotpoint}{\rule[-0.175pt]{0.350pt}{0.350pt}}%
\begin{picture}(250,300)(0,100)
\tenrm
\sbox{\plotpoint}{\rule[-0.175pt]{0.350pt}{0.350pt}}%
\put(370,95){\makebox(0,0){$I_{5,8}$}}
\put(370,346){\circle*{40}}
\put(289,288){\circle*{40}}
\put(320,195){\circle*{40}}
\put(420,195){\circle*{40}}
\put(451,288){\circle*{40}}
\put(370,346){\usebox{\plotpoint}}
\put(368,346){\usebox{\plotpoint}}
\put(367,345){\usebox{\plotpoint}}
\put(365,344){\usebox{\plotpoint}}
\put(364,343){\usebox{\plotpoint}}
\put(363,342){\usebox{\plotpoint}}
\put(361,341){\usebox{\plotpoint}}
\put(360,340){\usebox{\plotpoint}}
\put(358,339){\usebox{\plotpoint}}
\put(357,338){\usebox{\plotpoint}}
\put(356,337){\usebox{\plotpoint}}
\put(354,336){\usebox{\plotpoint}}
\put(353,335){\usebox{\plotpoint}}
\put(351,334){\usebox{\plotpoint}}
\put(350,333){\usebox{\plotpoint}}
\put(349,332){\usebox{\plotpoint}}
\put(347,331){\usebox{\plotpoint}}
\put(346,330){\usebox{\plotpoint}}
\put(344,329){\usebox{\plotpoint}}
\put(343,328){\usebox{\plotpoint}}
\put(342,327){\usebox{\plotpoint}}
\put(340,326){\usebox{\plotpoint}}
\put(339,325){\usebox{\plotpoint}}
\put(337,324){\usebox{\plotpoint}}
\put(336,323){\usebox{\plotpoint}}
\put(335,322){\usebox{\plotpoint}}
\put(333,321){\usebox{\plotpoint}}
\put(332,320){\usebox{\plotpoint}}
\put(330,319){\usebox{\plotpoint}}
\put(329,318){\usebox{\plotpoint}}
\put(328,317){\usebox{\plotpoint}}
\put(326,316){\usebox{\plotpoint}}
\put(325,315){\usebox{\plotpoint}}
\put(323,314){\usebox{\plotpoint}}
\put(322,313){\usebox{\plotpoint}}
\put(321,312){\usebox{\plotpoint}}
\put(319,311){\usebox{\plotpoint}}
\put(318,310){\usebox{\plotpoint}}
\put(316,309){\usebox{\plotpoint}}
\put(315,308){\usebox{\plotpoint}}
\put(314,307){\usebox{\plotpoint}}
\put(312,306){\usebox{\plotpoint}}
\put(311,305){\usebox{\plotpoint}}
\put(309,304){\usebox{\plotpoint}}
\put(308,303){\usebox{\plotpoint}}
\put(307,302){\usebox{\plotpoint}}
\put(305,301){\usebox{\plotpoint}}
\put(304,300){\usebox{\plotpoint}}
\put(302,299){\usebox{\plotpoint}}
\put(301,298){\usebox{\plotpoint}}
\put(300,297){\usebox{\plotpoint}}
\put(298,296){\usebox{\plotpoint}}
\put(297,295){\usebox{\plotpoint}}
\put(295,294){\usebox{\plotpoint}}
\put(294,293){\usebox{\plotpoint}}
\put(293,292){\usebox{\plotpoint}}
\put(291,291){\usebox{\plotpoint}}
\put(290,290){\usebox{\plotpoint}}
\put(289,289){\usebox{\plotpoint}}
\put(289,288){\usebox{\plotpoint}}
\put(370,346){\usebox{\plotpoint}}
\put(370,342){\rule[-0.175pt]{0.350pt}{0.728pt}}
\put(369,339){\rule[-0.175pt]{0.350pt}{0.728pt}}
\put(368,336){\rule[-0.175pt]{0.350pt}{0.728pt}}
\put(367,333){\rule[-0.175pt]{0.350pt}{0.728pt}}
\put(366,330){\rule[-0.175pt]{0.350pt}{0.728pt}}
\put(365,327){\rule[-0.175pt]{0.350pt}{0.728pt}}
\put(364,324){\rule[-0.175pt]{0.350pt}{0.728pt}}
\put(363,321){\rule[-0.175pt]{0.350pt}{0.728pt}}
\put(362,318){\rule[-0.175pt]{0.350pt}{0.728pt}}
\put(361,315){\rule[-0.175pt]{0.350pt}{0.728pt}}
\put(360,312){\rule[-0.175pt]{0.350pt}{0.728pt}}
\put(359,309){\rule[-0.175pt]{0.350pt}{0.728pt}}
\put(358,306){\rule[-0.175pt]{0.350pt}{0.728pt}}
\put(357,303){\rule[-0.175pt]{0.350pt}{0.728pt}}
\put(356,300){\rule[-0.175pt]{0.350pt}{0.728pt}}
\put(355,297){\rule[-0.175pt]{0.350pt}{0.728pt}}
\put(354,294){\rule[-0.175pt]{0.350pt}{0.728pt}}
\put(353,291){\rule[-0.175pt]{0.350pt}{0.728pt}}
\put(352,288){\rule[-0.175pt]{0.350pt}{0.728pt}}
\put(351,285){\rule[-0.175pt]{0.350pt}{0.728pt}}
\put(350,282){\rule[-0.175pt]{0.350pt}{0.728pt}}
\put(349,279){\rule[-0.175pt]{0.350pt}{0.728pt}}
\put(348,276){\rule[-0.175pt]{0.350pt}{0.728pt}}
\put(347,273){\rule[-0.175pt]{0.350pt}{0.728pt}}
\put(346,270){\rule[-0.175pt]{0.350pt}{0.728pt}}
\put(345,267){\rule[-0.175pt]{0.350pt}{0.728pt}}
\put(344,264){\rule[-0.175pt]{0.350pt}{0.728pt}}
\put(343,261){\rule[-0.175pt]{0.350pt}{0.728pt}}
\put(342,258){\rule[-0.175pt]{0.350pt}{0.728pt}}
\put(341,255){\rule[-0.175pt]{0.350pt}{0.728pt}}
\put(340,252){\rule[-0.175pt]{0.350pt}{0.728pt}}
\put(339,249){\rule[-0.175pt]{0.350pt}{0.728pt}}
\put(338,246){\rule[-0.175pt]{0.350pt}{0.728pt}}
\put(337,243){\rule[-0.175pt]{0.350pt}{0.728pt}}
\put(336,240){\rule[-0.175pt]{0.350pt}{0.728pt}}
\put(335,237){\rule[-0.175pt]{0.350pt}{0.728pt}}
\put(334,234){\rule[-0.175pt]{0.350pt}{0.728pt}}
\put(333,231){\rule[-0.175pt]{0.350pt}{0.728pt}}
\put(332,228){\rule[-0.175pt]{0.350pt}{0.728pt}}
\put(331,225){\rule[-0.175pt]{0.350pt}{0.728pt}}
\put(330,222){\rule[-0.175pt]{0.350pt}{0.728pt}}
\put(329,219){\rule[-0.175pt]{0.350pt}{0.728pt}}
\put(328,216){\rule[-0.175pt]{0.350pt}{0.728pt}}
\put(327,213){\rule[-0.175pt]{0.350pt}{0.728pt}}
\put(326,210){\rule[-0.175pt]{0.350pt}{0.728pt}}
\put(325,207){\rule[-0.175pt]{0.350pt}{0.728pt}}
\put(324,204){\rule[-0.175pt]{0.350pt}{0.728pt}}
\put(323,201){\rule[-0.175pt]{0.350pt}{0.728pt}}
\put(322,198){\rule[-0.175pt]{0.350pt}{0.728pt}}
\put(321,195){\rule[-0.175pt]{0.350pt}{0.728pt}}
\put(320,195){\usebox{\plotpoint}}
\put(370,346){\usebox{\plotpoint}}
\put(370,342){\rule[-0.175pt]{0.350pt}{0.728pt}}
\put(371,339){\rule[-0.175pt]{0.350pt}{0.728pt}}
\put(372,336){\rule[-0.175pt]{0.350pt}{0.728pt}}
\put(373,333){\rule[-0.175pt]{0.350pt}{0.728pt}}
\put(374,330){\rule[-0.175pt]{0.350pt}{0.728pt}}
\put(375,327){\rule[-0.175pt]{0.350pt}{0.728pt}}
\put(376,324){\rule[-0.175pt]{0.350pt}{0.728pt}}
\put(377,321){\rule[-0.175pt]{0.350pt}{0.728pt}}
\put(378,318){\rule[-0.175pt]{0.350pt}{0.728pt}}
\put(379,315){\rule[-0.175pt]{0.350pt}{0.728pt}}
\put(380,312){\rule[-0.175pt]{0.350pt}{0.728pt}}
\put(381,309){\rule[-0.175pt]{0.350pt}{0.728pt}}
\put(382,306){\rule[-0.175pt]{0.350pt}{0.728pt}}
\put(383,303){\rule[-0.175pt]{0.350pt}{0.728pt}}
\put(384,300){\rule[-0.175pt]{0.350pt}{0.728pt}}
\put(385,297){\rule[-0.175pt]{0.350pt}{0.728pt}}
\put(386,294){\rule[-0.175pt]{0.350pt}{0.728pt}}
\put(387,291){\rule[-0.175pt]{0.350pt}{0.728pt}}
\put(388,288){\rule[-0.175pt]{0.350pt}{0.728pt}}
\put(389,285){\rule[-0.175pt]{0.350pt}{0.728pt}}
\put(390,282){\rule[-0.175pt]{0.350pt}{0.728pt}}
\put(391,279){\rule[-0.175pt]{0.350pt}{0.728pt}}
\put(392,276){\rule[-0.175pt]{0.350pt}{0.728pt}}
\put(393,273){\rule[-0.175pt]{0.350pt}{0.728pt}}
\put(394,270){\rule[-0.175pt]{0.350pt}{0.728pt}}
\put(395,267){\rule[-0.175pt]{0.350pt}{0.728pt}}
\put(396,264){\rule[-0.175pt]{0.350pt}{0.728pt}}
\put(397,261){\rule[-0.175pt]{0.350pt}{0.728pt}}
\put(398,258){\rule[-0.175pt]{0.350pt}{0.728pt}}
\put(399,255){\rule[-0.175pt]{0.350pt}{0.728pt}}
\put(400,252){\rule[-0.175pt]{0.350pt}{0.728pt}}
\put(401,249){\rule[-0.175pt]{0.350pt}{0.728pt}}
\put(402,246){\rule[-0.175pt]{0.350pt}{0.728pt}}
\put(403,243){\rule[-0.175pt]{0.350pt}{0.728pt}}
\put(404,240){\rule[-0.175pt]{0.350pt}{0.728pt}}
\put(405,237){\rule[-0.175pt]{0.350pt}{0.728pt}}
\put(406,234){\rule[-0.175pt]{0.350pt}{0.728pt}}
\put(407,231){\rule[-0.175pt]{0.350pt}{0.728pt}}
\put(408,228){\rule[-0.175pt]{0.350pt}{0.728pt}}
\put(409,225){\rule[-0.175pt]{0.350pt}{0.728pt}}
\put(410,222){\rule[-0.175pt]{0.350pt}{0.728pt}}
\put(411,219){\rule[-0.175pt]{0.350pt}{0.728pt}}
\put(412,216){\rule[-0.175pt]{0.350pt}{0.728pt}}
\put(413,213){\rule[-0.175pt]{0.350pt}{0.728pt}}
\put(414,210){\rule[-0.175pt]{0.350pt}{0.728pt}}
\put(415,207){\rule[-0.175pt]{0.350pt}{0.728pt}}
\put(416,204){\rule[-0.175pt]{0.350pt}{0.728pt}}
\put(417,201){\rule[-0.175pt]{0.350pt}{0.728pt}}
\put(418,198){\rule[-0.175pt]{0.350pt}{0.728pt}}
\put(419,195){\rule[-0.175pt]{0.350pt}{0.728pt}}
\put(420,195){\usebox{\plotpoint}}
\put(370,346){\usebox{\plotpoint}}
\put(370,346){\usebox{\plotpoint}}
\put(371,345){\usebox{\plotpoint}}
\put(372,344){\usebox{\plotpoint}}
\put(374,343){\usebox{\plotpoint}}
\put(375,342){\usebox{\plotpoint}}
\put(376,341){\usebox{\plotpoint}}
\put(378,340){\usebox{\plotpoint}}
\put(379,339){\usebox{\plotpoint}}
\put(381,338){\usebox{\plotpoint}}
\put(382,337){\usebox{\plotpoint}}
\put(383,336){\usebox{\plotpoint}}
\put(385,335){\usebox{\plotpoint}}
\put(386,334){\usebox{\plotpoint}}
\put(388,333){\usebox{\plotpoint}}
\put(389,332){\usebox{\plotpoint}}
\put(390,331){\usebox{\plotpoint}}
\put(392,330){\usebox{\plotpoint}}
\put(393,329){\usebox{\plotpoint}}
\put(395,328){\usebox{\plotpoint}}
\put(396,327){\usebox{\plotpoint}}
\put(397,326){\usebox{\plotpoint}}
\put(399,325){\usebox{\plotpoint}}
\put(400,324){\usebox{\plotpoint}}
\put(402,323){\usebox{\plotpoint}}
\put(403,322){\usebox{\plotpoint}}
\put(404,321){\usebox{\plotpoint}}
\put(406,320){\usebox{\plotpoint}}
\put(407,319){\usebox{\plotpoint}}
\put(409,318){\usebox{\plotpoint}}
\put(410,317){\usebox{\plotpoint}}
\put(411,316){\usebox{\plotpoint}}
\put(413,315){\usebox{\plotpoint}}
\put(414,314){\usebox{\plotpoint}}
\put(416,313){\usebox{\plotpoint}}
\put(417,312){\usebox{\plotpoint}}
\put(418,311){\usebox{\plotpoint}}
\put(420,310){\usebox{\plotpoint}}
\put(421,309){\usebox{\plotpoint}}
\put(423,308){\usebox{\plotpoint}}
\put(424,307){\usebox{\plotpoint}}
\put(425,306){\usebox{\plotpoint}}
\put(427,305){\usebox{\plotpoint}}
\put(428,304){\usebox{\plotpoint}}
\put(430,303){\usebox{\plotpoint}}
\put(431,302){\usebox{\plotpoint}}
\put(432,301){\usebox{\plotpoint}}
\put(434,300){\usebox{\plotpoint}}
\put(435,299){\usebox{\plotpoint}}
\put(437,298){\usebox{\plotpoint}}
\put(438,297){\usebox{\plotpoint}}
\put(439,296){\usebox{\plotpoint}}
\put(441,295){\usebox{\plotpoint}}
\put(442,294){\usebox{\plotpoint}}
\put(444,293){\usebox{\plotpoint}}
\put(445,292){\usebox{\plotpoint}}
\put(446,291){\usebox{\plotpoint}}
\put(448,290){\usebox{\plotpoint}}
\put(449,289){\usebox{\plotpoint}}
\put(450,288){\usebox{\plotpoint}}
\sbox{\plotpoint}{\rule[-0.250pt]{0.500pt}{0.500pt}}%
\end{picture}
	% GNUPLOT: LaTeX picture
\setlength{\unitlength}{0.240900pt}
\ifx\plotpoint\undefined\newsavebox{\plotpoint}\fi
\sbox{\plotpoint}{\rule[-0.175pt]{0.350pt}{0.350pt}}%
\begin{picture}(250,300)(0,100)
\tenrm
\sbox{\plotpoint}{\rule[-0.175pt]{0.350pt}{0.350pt}}%
\put(370,95){\makebox(0,0){$I_{5,9}$}}
\put(370,346){\circle*{40}}
\put(289,288){\circle*{40}}
\put(320,195){\circle*{40}}
\put(420,195){\circle*{40}}
\put(451,288){\circle*{40}}
\put(310,181){\usebox{\plotpoint}}
\put(310,181){\rule[-0.175pt]{28.908pt}{0.350pt}}
\put(330,208){\usebox{\plotpoint}}
\put(330,208){\rule[-0.175pt]{19.272pt}{0.350pt}}
\put(315,188){\usebox{\plotpoint}}
\put(315,188){\rule[-0.175pt]{26.499pt}{0.350pt}}
\put(325,202){\usebox{\plotpoint}}
\put(325,202){\rule[-0.175pt]{21.681pt}{0.350pt}}
\sbox{\plotpoint}{\rule[-0.250pt]{0.500pt}{0.500pt}}%
\end{picture}
	% GNUPLOT: LaTeX picture
\setlength{\unitlength}{0.240900pt}
\ifx\plotpoint\undefined\newsavebox{\plotpoint}\fi
\sbox{\plotpoint}{\rule[-0.175pt]{0.350pt}{0.350pt}}%
\begin{picture}(250,300)(0,100)
\tenrm
\sbox{\plotpoint}{\rule[-0.175pt]{0.350pt}{0.350pt}}%
\put(370,95){\makebox(0,0){$I_{5,10}$}}
\put(370,346){\circle*{40}}
\put(289,288){\circle*{40}}
\put(320,195){\circle*{40}}
\put(420,195){\circle*{40}}
\put(451,288){\circle*{40}}
\put(370,346){\usebox{\plotpoint}}
\put(368,346){\usebox{\plotpoint}}
\put(367,345){\usebox{\plotpoint}}
\put(365,344){\usebox{\plotpoint}}
\put(364,343){\usebox{\plotpoint}}
\put(363,342){\usebox{\plotpoint}}
\put(361,341){\usebox{\plotpoint}}
\put(360,340){\usebox{\plotpoint}}
\put(358,339){\usebox{\plotpoint}}
\put(357,338){\usebox{\plotpoint}}
\put(356,337){\usebox{\plotpoint}}
\put(354,336){\usebox{\plotpoint}}
\put(353,335){\usebox{\plotpoint}}
\put(351,334){\usebox{\plotpoint}}
\put(350,333){\usebox{\plotpoint}}
\put(349,332){\usebox{\plotpoint}}
\put(347,331){\usebox{\plotpoint}}
\put(346,330){\usebox{\plotpoint}}
\put(344,329){\usebox{\plotpoint}}
\put(343,328){\usebox{\plotpoint}}
\put(342,327){\usebox{\plotpoint}}
\put(340,326){\usebox{\plotpoint}}
\put(339,325){\usebox{\plotpoint}}
\put(337,324){\usebox{\plotpoint}}
\put(336,323){\usebox{\plotpoint}}
\put(335,322){\usebox{\plotpoint}}
\put(333,321){\usebox{\plotpoint}}
\put(332,320){\usebox{\plotpoint}}
\put(330,319){\usebox{\plotpoint}}
\put(329,318){\usebox{\plotpoint}}
\put(328,317){\usebox{\plotpoint}}
\put(326,316){\usebox{\plotpoint}}
\put(325,315){\usebox{\plotpoint}}
\put(323,314){\usebox{\plotpoint}}
\put(322,313){\usebox{\plotpoint}}
\put(321,312){\usebox{\plotpoint}}
\put(319,311){\usebox{\plotpoint}}
\put(318,310){\usebox{\plotpoint}}
\put(316,309){\usebox{\plotpoint}}
\put(315,308){\usebox{\plotpoint}}
\put(314,307){\usebox{\plotpoint}}
\put(312,306){\usebox{\plotpoint}}
\put(311,305){\usebox{\plotpoint}}
\put(309,304){\usebox{\plotpoint}}
\put(308,303){\usebox{\plotpoint}}
\put(307,302){\usebox{\plotpoint}}
\put(305,301){\usebox{\plotpoint}}
\put(304,300){\usebox{\plotpoint}}
\put(302,299){\usebox{\plotpoint}}
\put(301,298){\usebox{\plotpoint}}
\put(300,297){\usebox{\plotpoint}}
\put(298,296){\usebox{\plotpoint}}
\put(297,295){\usebox{\plotpoint}}
\put(295,294){\usebox{\plotpoint}}
\put(294,293){\usebox{\plotpoint}}
\put(293,292){\usebox{\plotpoint}}
\put(291,291){\usebox{\plotpoint}}
\put(290,290){\usebox{\plotpoint}}
\put(289,289){\usebox{\plotpoint}}
\put(289,288){\usebox{\plotpoint}}
\put(370,346){\usebox{\plotpoint}}
\put(370,346){\usebox{\plotpoint}}
\put(371,345){\usebox{\plotpoint}}
\put(372,344){\usebox{\plotpoint}}
\put(374,343){\usebox{\plotpoint}}
\put(375,342){\usebox{\plotpoint}}
\put(376,341){\usebox{\plotpoint}}
\put(378,340){\usebox{\plotpoint}}
\put(379,339){\usebox{\plotpoint}}
\put(381,338){\usebox{\plotpoint}}
\put(382,337){\usebox{\plotpoint}}
\put(383,336){\usebox{\plotpoint}}
\put(385,335){\usebox{\plotpoint}}
\put(386,334){\usebox{\plotpoint}}
\put(388,333){\usebox{\plotpoint}}
\put(389,332){\usebox{\plotpoint}}
\put(390,331){\usebox{\plotpoint}}
\put(392,330){\usebox{\plotpoint}}
\put(393,329){\usebox{\plotpoint}}
\put(395,328){\usebox{\plotpoint}}
\put(396,327){\usebox{\plotpoint}}
\put(397,326){\usebox{\plotpoint}}
\put(399,325){\usebox{\plotpoint}}
\put(400,324){\usebox{\plotpoint}}
\put(402,323){\usebox{\plotpoint}}
\put(403,322){\usebox{\plotpoint}}
\put(404,321){\usebox{\plotpoint}}
\put(406,320){\usebox{\plotpoint}}
\put(407,319){\usebox{\plotpoint}}
\put(409,318){\usebox{\plotpoint}}
\put(410,317){\usebox{\plotpoint}}
\put(411,316){\usebox{\plotpoint}}
\put(413,315){\usebox{\plotpoint}}
\put(414,314){\usebox{\plotpoint}}
\put(416,313){\usebox{\plotpoint}}
\put(417,312){\usebox{\plotpoint}}
\put(418,311){\usebox{\plotpoint}}
\put(420,310){\usebox{\plotpoint}}
\put(421,309){\usebox{\plotpoint}}
\put(423,308){\usebox{\plotpoint}}
\put(424,307){\usebox{\plotpoint}}
\put(425,306){\usebox{\plotpoint}}
\put(427,305){\usebox{\plotpoint}}
\put(428,304){\usebox{\plotpoint}}
\put(430,303){\usebox{\plotpoint}}
\put(431,302){\usebox{\plotpoint}}
\put(432,301){\usebox{\plotpoint}}
\put(434,300){\usebox{\plotpoint}}
\put(435,299){\usebox{\plotpoint}}
\put(437,298){\usebox{\plotpoint}}
\put(438,297){\usebox{\plotpoint}}
\put(439,296){\usebox{\plotpoint}}
\put(441,295){\usebox{\plotpoint}}
\put(442,294){\usebox{\plotpoint}}
\put(444,293){\usebox{\plotpoint}}
\put(445,292){\usebox{\plotpoint}}
\put(446,291){\usebox{\plotpoint}}
\put(448,290){\usebox{\plotpoint}}
\put(449,289){\usebox{\plotpoint}}
\put(450,288){\usebox{\plotpoint}}
\put(289,288){\usebox{\plotpoint}}
\put(289,288){\rule[-0.175pt]{39.026pt}{0.350pt}}
\put(320,195){\usebox{\plotpoint}}
\put(320,195){\rule[-0.175pt]{24.090pt}{0.350pt}}
\sbox{\plotpoint}{\rule[-0.250pt]{0.500pt}{0.500pt}}%
\end{picture}
	% GNUPLOT: LaTeX picture
\setlength{\unitlength}{0.240900pt}
\ifx\plotpoint\undefined\newsavebox{\plotpoint}\fi
\sbox{\plotpoint}{\rule[-0.175pt]{0.350pt}{0.350pt}}%
\begin{picture}(250,300)(0,100)
\tenrm
\sbox{\plotpoint}{\rule[-0.175pt]{0.350pt}{0.350pt}}%
\put(370,95){\makebox(0,0){$I_{5,11}$}}
\put(370,346){\circle*{40}}
\put(289,288){\circle*{40}}
\put(320,195){\circle*{40}}
\put(420,195){\circle*{40}}
\put(451,288){\circle*{40}}
\put(281,291){\usebox{\plotpoint}}
\put(281,287){\rule[-0.175pt]{0.350pt}{0.730pt}}
\put(282,284){\rule[-0.175pt]{0.350pt}{0.730pt}}
\put(283,281){\rule[-0.175pt]{0.350pt}{0.730pt}}
\put(284,278){\rule[-0.175pt]{0.350pt}{0.730pt}}
\put(285,275){\rule[-0.175pt]{0.350pt}{0.730pt}}
\put(286,272){\rule[-0.175pt]{0.350pt}{0.730pt}}
\put(287,269){\rule[-0.175pt]{0.350pt}{0.730pt}}
\put(288,266){\rule[-0.175pt]{0.350pt}{0.730pt}}
\put(289,263){\rule[-0.175pt]{0.350pt}{0.730pt}}
\put(290,260){\rule[-0.175pt]{0.350pt}{0.730pt}}
\put(291,257){\rule[-0.175pt]{0.350pt}{0.730pt}}
\put(292,254){\rule[-0.175pt]{0.350pt}{0.730pt}}
\put(293,251){\rule[-0.175pt]{0.350pt}{0.730pt}}
\put(294,248){\rule[-0.175pt]{0.350pt}{0.730pt}}
\put(295,245){\rule[-0.175pt]{0.350pt}{0.730pt}}
\put(296,242){\rule[-0.175pt]{0.350pt}{0.730pt}}
\put(297,239){\rule[-0.175pt]{0.350pt}{0.730pt}}
\put(298,236){\rule[-0.175pt]{0.350pt}{0.730pt}}
\put(299,233){\rule[-0.175pt]{0.350pt}{0.730pt}}
\put(300,230){\rule[-0.175pt]{0.350pt}{0.730pt}}
\put(301,227){\rule[-0.175pt]{0.350pt}{0.730pt}}
\put(302,224){\rule[-0.175pt]{0.350pt}{0.730pt}}
\put(303,221){\rule[-0.175pt]{0.350pt}{0.730pt}}
\put(304,218){\rule[-0.175pt]{0.350pt}{0.730pt}}
\put(305,215){\rule[-0.175pt]{0.350pt}{0.730pt}}
\put(306,212){\rule[-0.175pt]{0.350pt}{0.730pt}}
\put(307,209){\rule[-0.175pt]{0.350pt}{0.730pt}}
\put(308,206){\rule[-0.175pt]{0.350pt}{0.730pt}}
\put(309,203){\rule[-0.175pt]{0.350pt}{0.730pt}}
\put(310,200){\rule[-0.175pt]{0.350pt}{0.730pt}}
\put(311,197){\rule[-0.175pt]{0.350pt}{0.730pt}}
\put(312,194){\rule[-0.175pt]{0.350pt}{0.730pt}}
\put(313,191){\rule[-0.175pt]{0.350pt}{0.730pt}}
\put(314,188){\rule[-0.175pt]{0.350pt}{0.730pt}}
\put(297,286){\usebox{\plotpoint}}
\put(297,283){\rule[-0.175pt]{0.350pt}{0.723pt}}
\put(298,280){\rule[-0.175pt]{0.350pt}{0.723pt}}
\put(299,277){\rule[-0.175pt]{0.350pt}{0.723pt}}
\put(300,274){\rule[-0.175pt]{0.350pt}{0.723pt}}
\put(301,271){\rule[-0.175pt]{0.350pt}{0.723pt}}
\put(302,268){\rule[-0.175pt]{0.350pt}{0.723pt}}
\put(303,265){\rule[-0.175pt]{0.350pt}{0.723pt}}
\put(304,262){\rule[-0.175pt]{0.350pt}{0.723pt}}
\put(305,259){\rule[-0.175pt]{0.350pt}{0.723pt}}
\put(306,256){\rule[-0.175pt]{0.350pt}{0.723pt}}
\put(307,253){\rule[-0.175pt]{0.350pt}{0.723pt}}
\put(308,250){\rule[-0.175pt]{0.350pt}{0.723pt}}
\put(309,247){\rule[-0.175pt]{0.350pt}{0.723pt}}
\put(310,244){\rule[-0.175pt]{0.350pt}{0.723pt}}
\put(311,241){\rule[-0.175pt]{0.350pt}{0.723pt}}
\put(312,238){\rule[-0.175pt]{0.350pt}{0.723pt}}
\put(313,235){\rule[-0.175pt]{0.350pt}{0.723pt}}
\put(314,232){\rule[-0.175pt]{0.350pt}{0.723pt}}
\put(315,229){\rule[-0.175pt]{0.350pt}{0.723pt}}
\put(316,226){\rule[-0.175pt]{0.350pt}{0.723pt}}
\put(317,223){\rule[-0.175pt]{0.350pt}{0.723pt}}
\put(318,220){\rule[-0.175pt]{0.350pt}{0.723pt}}
\put(319,217){\rule[-0.175pt]{0.350pt}{0.723pt}}
\put(320,214){\rule[-0.175pt]{0.350pt}{0.723pt}}
\put(321,211){\rule[-0.175pt]{0.350pt}{0.723pt}}
\put(322,208){\rule[-0.175pt]{0.350pt}{0.723pt}}
\put(323,205){\rule[-0.175pt]{0.350pt}{0.723pt}}
\put(324,202){\rule[-0.175pt]{0.350pt}{0.723pt}}
\put(425,188){\usebox{\plotpoint}}
\put(425,188){\rule[-0.175pt]{0.350pt}{0.730pt}}
\put(426,191){\rule[-0.175pt]{0.350pt}{0.730pt}}
\put(427,194){\rule[-0.175pt]{0.350pt}{0.730pt}}
\put(428,197){\rule[-0.175pt]{0.350pt}{0.730pt}}
\put(429,200){\rule[-0.175pt]{0.350pt}{0.730pt}}
\put(430,203){\rule[-0.175pt]{0.350pt}{0.730pt}}
\put(431,206){\rule[-0.175pt]{0.350pt}{0.730pt}}
\put(432,209){\rule[-0.175pt]{0.350pt}{0.730pt}}
\put(433,212){\rule[-0.175pt]{0.350pt}{0.730pt}}
\put(434,215){\rule[-0.175pt]{0.350pt}{0.730pt}}
\put(435,218){\rule[-0.175pt]{0.350pt}{0.730pt}}
\put(436,221){\rule[-0.175pt]{0.350pt}{0.730pt}}
\put(437,224){\rule[-0.175pt]{0.350pt}{0.730pt}}
\put(438,227){\rule[-0.175pt]{0.350pt}{0.730pt}}
\put(439,230){\rule[-0.175pt]{0.350pt}{0.730pt}}
\put(440,233){\rule[-0.175pt]{0.350pt}{0.730pt}}
\put(441,236){\rule[-0.175pt]{0.350pt}{0.730pt}}
\put(442,239){\rule[-0.175pt]{0.350pt}{0.730pt}}
\put(443,242){\rule[-0.175pt]{0.350pt}{0.730pt}}
\put(444,245){\rule[-0.175pt]{0.350pt}{0.730pt}}
\put(445,248){\rule[-0.175pt]{0.350pt}{0.730pt}}
\put(446,251){\rule[-0.175pt]{0.350pt}{0.730pt}}
\put(447,254){\rule[-0.175pt]{0.350pt}{0.730pt}}
\put(448,257){\rule[-0.175pt]{0.350pt}{0.730pt}}
\put(449,260){\rule[-0.175pt]{0.350pt}{0.730pt}}
\put(450,263){\rule[-0.175pt]{0.350pt}{0.730pt}}
\put(451,266){\rule[-0.175pt]{0.350pt}{0.730pt}}
\put(452,269){\rule[-0.175pt]{0.350pt}{0.730pt}}
\put(453,272){\rule[-0.175pt]{0.350pt}{0.730pt}}
\put(454,275){\rule[-0.175pt]{0.350pt}{0.730pt}}
\put(455,278){\rule[-0.175pt]{0.350pt}{0.730pt}}
\put(456,281){\rule[-0.175pt]{0.350pt}{0.730pt}}
\put(457,284){\rule[-0.175pt]{0.350pt}{0.730pt}}
\put(458,287){\rule[-0.175pt]{0.350pt}{0.730pt}}
\put(415,202){\usebox{\plotpoint}}
\put(415,202){\rule[-0.175pt]{0.350pt}{0.723pt}}
\put(416,205){\rule[-0.175pt]{0.350pt}{0.723pt}}
\put(417,208){\rule[-0.175pt]{0.350pt}{0.723pt}}
\put(418,211){\rule[-0.175pt]{0.350pt}{0.723pt}}
\put(419,214){\rule[-0.175pt]{0.350pt}{0.723pt}}
\put(420,217){\rule[-0.175pt]{0.350pt}{0.723pt}}
\put(421,220){\rule[-0.175pt]{0.350pt}{0.723pt}}
\put(422,223){\rule[-0.175pt]{0.350pt}{0.723pt}}
\put(423,226){\rule[-0.175pt]{0.350pt}{0.723pt}}
\put(424,229){\rule[-0.175pt]{0.350pt}{0.723pt}}
\put(425,232){\rule[-0.175pt]{0.350pt}{0.723pt}}
\put(426,235){\rule[-0.175pt]{0.350pt}{0.723pt}}
\put(427,238){\rule[-0.175pt]{0.350pt}{0.723pt}}
\put(428,241){\rule[-0.175pt]{0.350pt}{0.723pt}}
\put(429,244){\rule[-0.175pt]{0.350pt}{0.723pt}}
\put(430,247){\rule[-0.175pt]{0.350pt}{0.723pt}}
\put(431,250){\rule[-0.175pt]{0.350pt}{0.723pt}}
\put(432,253){\rule[-0.175pt]{0.350pt}{0.723pt}}
\put(433,256){\rule[-0.175pt]{0.350pt}{0.723pt}}
\put(434,259){\rule[-0.175pt]{0.350pt}{0.723pt}}
\put(435,262){\rule[-0.175pt]{0.350pt}{0.723pt}}
\put(436,265){\rule[-0.175pt]{0.350pt}{0.723pt}}
\put(437,268){\rule[-0.175pt]{0.350pt}{0.723pt}}
\put(438,271){\rule[-0.175pt]{0.350pt}{0.723pt}}
\put(439,274){\rule[-0.175pt]{0.350pt}{0.723pt}}
\put(440,277){\rule[-0.175pt]{0.350pt}{0.723pt}}
\put(441,280){\rule[-0.175pt]{0.350pt}{0.723pt}}
\put(442,283){\rule[-0.175pt]{0.350pt}{0.723pt}}
\sbox{\plotpoint}{\rule[-0.250pt]{0.500pt}{0.500pt}}%
\end{picture}
	% GNUPLOT: LaTeX picture
\setlength{\unitlength}{0.240900pt}
\ifx\plotpoint\undefined\newsavebox{\plotpoint}\fi
\sbox{\plotpoint}{\rule[-0.175pt]{0.350pt}{0.350pt}}%
\begin{picture}(250,300)(0,100)
\tenrm
\sbox{\plotpoint}{\rule[-0.175pt]{0.350pt}{0.350pt}}%
\put(370,95){\makebox(0,0){$I_{5,12}$}}
\put(370,346){\circle*{40}}
\put(289,288){\circle*{40}}
\put(320,195){\circle*{40}}
\put(420,195){\circle*{40}}
\put(451,288){\circle*{40}}
\put(289,288){\usebox{\plotpoint}}
\put(289,288){\rule[-0.175pt]{39.026pt}{0.350pt}}
\put(420,195){\usebox{\plotpoint}}
\put(420,195){\rule[-0.175pt]{0.350pt}{0.723pt}}
\put(421,198){\rule[-0.175pt]{0.350pt}{0.723pt}}
\put(422,201){\rule[-0.175pt]{0.350pt}{0.723pt}}
\put(423,204){\rule[-0.175pt]{0.350pt}{0.723pt}}
\put(424,207){\rule[-0.175pt]{0.350pt}{0.723pt}}
\put(425,210){\rule[-0.175pt]{0.350pt}{0.723pt}}
\put(426,213){\rule[-0.175pt]{0.350pt}{0.723pt}}
\put(427,216){\rule[-0.175pt]{0.350pt}{0.723pt}}
\put(428,219){\rule[-0.175pt]{0.350pt}{0.723pt}}
\put(429,222){\rule[-0.175pt]{0.350pt}{0.723pt}}
\put(430,225){\rule[-0.175pt]{0.350pt}{0.723pt}}
\put(431,228){\rule[-0.175pt]{0.350pt}{0.723pt}}
\put(432,231){\rule[-0.175pt]{0.350pt}{0.723pt}}
\put(433,234){\rule[-0.175pt]{0.350pt}{0.723pt}}
\put(434,237){\rule[-0.175pt]{0.350pt}{0.723pt}}
\put(435,240){\rule[-0.175pt]{0.350pt}{0.723pt}}
\put(436,243){\rule[-0.175pt]{0.350pt}{0.723pt}}
\put(437,246){\rule[-0.175pt]{0.350pt}{0.723pt}}
\put(438,249){\rule[-0.175pt]{0.350pt}{0.723pt}}
\put(439,252){\rule[-0.175pt]{0.350pt}{0.723pt}}
\put(440,255){\rule[-0.175pt]{0.350pt}{0.723pt}}
\put(441,258){\rule[-0.175pt]{0.350pt}{0.723pt}}
\put(442,261){\rule[-0.175pt]{0.350pt}{0.723pt}}
\put(443,264){\rule[-0.175pt]{0.350pt}{0.723pt}}
\put(444,267){\rule[-0.175pt]{0.350pt}{0.723pt}}
\put(445,270){\rule[-0.175pt]{0.350pt}{0.723pt}}
\put(446,273){\rule[-0.175pt]{0.350pt}{0.723pt}}
\put(447,276){\rule[-0.175pt]{0.350pt}{0.723pt}}
\put(448,279){\rule[-0.175pt]{0.350pt}{0.723pt}}
\put(449,282){\rule[-0.175pt]{0.350pt}{0.723pt}}
\put(450,285){\rule[-0.175pt]{0.350pt}{0.723pt}}
\put(320,195){\usebox{\plotpoint}}
\put(320,195){\rule[-0.175pt]{24.090pt}{0.350pt}}
\put(289,288){\usebox{\plotpoint}}
\put(289,285){\rule[-0.175pt]{0.350pt}{0.723pt}}
\put(290,282){\rule[-0.175pt]{0.350pt}{0.723pt}}
\put(291,279){\rule[-0.175pt]{0.350pt}{0.723pt}}
\put(292,276){\rule[-0.175pt]{0.350pt}{0.723pt}}
\put(293,273){\rule[-0.175pt]{0.350pt}{0.723pt}}
\put(294,270){\rule[-0.175pt]{0.350pt}{0.723pt}}
\put(295,267){\rule[-0.175pt]{0.350pt}{0.723pt}}
\put(296,264){\rule[-0.175pt]{0.350pt}{0.723pt}}
\put(297,261){\rule[-0.175pt]{0.350pt}{0.723pt}}
\put(298,258){\rule[-0.175pt]{0.350pt}{0.723pt}}
\put(299,255){\rule[-0.175pt]{0.350pt}{0.723pt}}
\put(300,252){\rule[-0.175pt]{0.350pt}{0.723pt}}
\put(301,249){\rule[-0.175pt]{0.350pt}{0.723pt}}
\put(302,246){\rule[-0.175pt]{0.350pt}{0.723pt}}
\put(303,243){\rule[-0.175pt]{0.350pt}{0.723pt}}
\put(304,240){\rule[-0.175pt]{0.350pt}{0.723pt}}
\put(305,237){\rule[-0.175pt]{0.350pt}{0.723pt}}
\put(306,234){\rule[-0.175pt]{0.350pt}{0.723pt}}
\put(307,231){\rule[-0.175pt]{0.350pt}{0.723pt}}
\put(308,228){\rule[-0.175pt]{0.350pt}{0.723pt}}
\put(309,225){\rule[-0.175pt]{0.350pt}{0.723pt}}
\put(310,222){\rule[-0.175pt]{0.350pt}{0.723pt}}
\put(311,219){\rule[-0.175pt]{0.350pt}{0.723pt}}
\put(312,216){\rule[-0.175pt]{0.350pt}{0.723pt}}
\put(313,213){\rule[-0.175pt]{0.350pt}{0.723pt}}
\put(314,210){\rule[-0.175pt]{0.350pt}{0.723pt}}
\put(315,207){\rule[-0.175pt]{0.350pt}{0.723pt}}
\put(316,204){\rule[-0.175pt]{0.350pt}{0.723pt}}
\put(317,201){\rule[-0.175pt]{0.350pt}{0.723pt}}
\put(318,198){\rule[-0.175pt]{0.350pt}{0.723pt}}
\put(319,195){\rule[-0.175pt]{0.350pt}{0.723pt}}
\sbox{\plotpoint}{\rule[-0.250pt]{0.500pt}{0.500pt}}%
\end{picture}
	}
\end{figure}
\begin{figure}
	{\Large
	% GNUPLOT: LaTeX picture
\setlength{\unitlength}{0.240900pt}
\ifx\plotpoint\undefined\newsavebox{\plotpoint}\fi
\sbox{\plotpoint}{\rule[-0.175pt]{0.350pt}{0.350pt}}%
\begin{picture}(250,300)(0,100)
\tenrm
\sbox{\plotpoint}{\rule[-0.175pt]{0.350pt}{0.350pt}}%
\put(370,95){\makebox(0,0){$I_{5,13}$}}
\put(370,346){\circle*{40}}
\put(289,288){\circle*{40}}
\put(320,195){\circle*{40}}
\put(420,195){\circle*{40}}
\put(451,288){\circle*{40}}
\put(289,288){\usebox{\plotpoint}}
\put(289,288){\rule[-0.175pt]{39.026pt}{0.350pt}}
\put(370,346){\usebox{\plotpoint}}
\put(370,346){\usebox{\plotpoint}}
\put(371,345){\usebox{\plotpoint}}
\put(372,344){\usebox{\plotpoint}}
\put(374,343){\usebox{\plotpoint}}
\put(375,342){\usebox{\plotpoint}}
\put(376,341){\usebox{\plotpoint}}
\put(378,340){\usebox{\plotpoint}}
\put(379,339){\usebox{\plotpoint}}
\put(381,338){\usebox{\plotpoint}}
\put(382,337){\usebox{\plotpoint}}
\put(383,336){\usebox{\plotpoint}}
\put(385,335){\usebox{\plotpoint}}
\put(386,334){\usebox{\plotpoint}}
\put(388,333){\usebox{\plotpoint}}
\put(389,332){\usebox{\plotpoint}}
\put(390,331){\usebox{\plotpoint}}
\put(392,330){\usebox{\plotpoint}}
\put(393,329){\usebox{\plotpoint}}
\put(395,328){\usebox{\plotpoint}}
\put(396,327){\usebox{\plotpoint}}
\put(397,326){\usebox{\plotpoint}}
\put(399,325){\usebox{\plotpoint}}
\put(400,324){\usebox{\plotpoint}}
\put(402,323){\usebox{\plotpoint}}
\put(403,322){\usebox{\plotpoint}}
\put(404,321){\usebox{\plotpoint}}
\put(406,320){\usebox{\plotpoint}}
\put(407,319){\usebox{\plotpoint}}
\put(409,318){\usebox{\plotpoint}}
\put(410,317){\usebox{\plotpoint}}
\put(411,316){\usebox{\plotpoint}}
\put(413,315){\usebox{\plotpoint}}
\put(414,314){\usebox{\plotpoint}}
\put(416,313){\usebox{\plotpoint}}
\put(417,312){\usebox{\plotpoint}}
\put(418,311){\usebox{\plotpoint}}
\put(420,310){\usebox{\plotpoint}}
\put(421,309){\usebox{\plotpoint}}
\put(423,308){\usebox{\plotpoint}}
\put(424,307){\usebox{\plotpoint}}
\put(425,306){\usebox{\plotpoint}}
\put(427,305){\usebox{\plotpoint}}
\put(428,304){\usebox{\plotpoint}}
\put(430,303){\usebox{\plotpoint}}
\put(431,302){\usebox{\plotpoint}}
\put(432,301){\usebox{\plotpoint}}
\put(434,300){\usebox{\plotpoint}}
\put(435,299){\usebox{\plotpoint}}
\put(437,298){\usebox{\plotpoint}}
\put(438,297){\usebox{\plotpoint}}
\put(439,296){\usebox{\plotpoint}}
\put(441,295){\usebox{\plotpoint}}
\put(442,294){\usebox{\plotpoint}}
\put(444,293){\usebox{\plotpoint}}
\put(445,292){\usebox{\plotpoint}}
\put(446,291){\usebox{\plotpoint}}
\put(448,290){\usebox{\plotpoint}}
\put(449,289){\usebox{\plotpoint}}
\put(450,288){\usebox{\plotpoint}}
\put(370,354){\usebox{\plotpoint}}
\put(368,354){\usebox{\plotpoint}}
\put(367,353){\usebox{\plotpoint}}
\put(365,352){\usebox{\plotpoint}}
\put(364,351){\usebox{\plotpoint}}
\put(362,350){\usebox{\plotpoint}}
\put(361,349){\usebox{\plotpoint}}
\put(360,348){\usebox{\plotpoint}}
\put(358,347){\usebox{\plotpoint}}
\put(357,346){\usebox{\plotpoint}}
\put(355,345){\usebox{\plotpoint}}
\put(354,344){\usebox{\plotpoint}}
\put(353,343){\usebox{\plotpoint}}
\put(351,342){\usebox{\plotpoint}}
\put(350,341){\usebox{\plotpoint}}
\put(348,340){\usebox{\plotpoint}}
\put(347,339){\usebox{\plotpoint}}
\put(345,338){\usebox{\plotpoint}}
\put(344,337){\usebox{\plotpoint}}
\put(343,336){\usebox{\plotpoint}}
\put(341,335){\usebox{\plotpoint}}
\put(340,334){\usebox{\plotpoint}}
\put(338,333){\usebox{\plotpoint}}
\put(337,332){\usebox{\plotpoint}}
\put(336,331){\usebox{\plotpoint}}
\put(334,330){\usebox{\plotpoint}}
\put(333,329){\usebox{\plotpoint}}
\put(331,328){\usebox{\plotpoint}}
\put(330,327){\usebox{\plotpoint}}
\put(329,326){\usebox{\plotpoint}}
\put(327,325){\usebox{\plotpoint}}
\put(326,324){\usebox{\plotpoint}}
\put(324,323){\usebox{\plotpoint}}
\put(323,322){\usebox{\plotpoint}}
\put(321,321){\usebox{\plotpoint}}
\put(320,320){\usebox{\plotpoint}}
\put(319,319){\usebox{\plotpoint}}
\put(317,318){\usebox{\plotpoint}}
\put(316,317){\usebox{\plotpoint}}
\put(314,316){\usebox{\plotpoint}}
\put(313,315){\usebox{\plotpoint}}
\put(312,314){\usebox{\plotpoint}}
\put(310,313){\usebox{\plotpoint}}
\put(309,312){\usebox{\plotpoint}}
\put(307,311){\usebox{\plotpoint}}
\put(306,310){\usebox{\plotpoint}}
\put(305,309){\usebox{\plotpoint}}
\put(303,308){\usebox{\plotpoint}}
\put(302,307){\usebox{\plotpoint}}
\put(300,306){\usebox{\plotpoint}}
\put(299,305){\usebox{\plotpoint}}
\put(297,304){\usebox{\plotpoint}}
\put(296,303){\usebox{\plotpoint}}
\put(295,302){\usebox{\plotpoint}}
\put(293,301){\usebox{\plotpoint}}
\put(292,300){\usebox{\plotpoint}}
\put(290,299){\usebox{\plotpoint}}
\put(289,298){\usebox{\plotpoint}}
\put(288,297){\usebox{\plotpoint}}
\put(286,296){\usebox{\plotpoint}}
\put(285,295){\usebox{\plotpoint}}
\put(283,294){\usebox{\plotpoint}}
\put(282,293){\usebox{\plotpoint}}
\put(281,292){\usebox{\plotpoint}}
\put(281,291){\usebox{\plotpoint}}
\put(370,338){\usebox{\plotpoint}}
\put(368,338){\usebox{\plotpoint}}
\put(367,337){\usebox{\plotpoint}}
\put(365,336){\usebox{\plotpoint}}
\put(364,335){\usebox{\plotpoint}}
\put(362,334){\usebox{\plotpoint}}
\put(361,333){\usebox{\plotpoint}}
\put(360,332){\usebox{\plotpoint}}
\put(358,331){\usebox{\plotpoint}}
\put(357,330){\usebox{\plotpoint}}
\put(355,329){\usebox{\plotpoint}}
\put(354,328){\usebox{\plotpoint}}
\put(353,327){\usebox{\plotpoint}}
\put(351,326){\usebox{\plotpoint}}
\put(350,325){\usebox{\plotpoint}}
\put(348,324){\usebox{\plotpoint}}
\put(347,323){\usebox{\plotpoint}}
\put(346,322){\usebox{\plotpoint}}
\put(344,321){\usebox{\plotpoint}}
\put(343,320){\usebox{\plotpoint}}
\put(341,319){\usebox{\plotpoint}}
\put(340,318){\usebox{\plotpoint}}
\put(339,317){\usebox{\plotpoint}}
\put(337,316){\usebox{\plotpoint}}
\put(336,315){\usebox{\plotpoint}}
\put(334,314){\usebox{\plotpoint}}
\put(333,313){\usebox{\plotpoint}}
\put(332,312){\usebox{\plotpoint}}
\put(330,311){\usebox{\plotpoint}}
\put(329,310){\usebox{\plotpoint}}
\put(327,309){\usebox{\plotpoint}}
\put(326,308){\usebox{\plotpoint}}
\put(325,307){\usebox{\plotpoint}}
\put(323,306){\usebox{\plotpoint}}
\put(322,305){\usebox{\plotpoint}}
\put(320,304){\usebox{\plotpoint}}
\put(319,303){\usebox{\plotpoint}}
\put(318,302){\usebox{\plotpoint}}
\put(316,301){\usebox{\plotpoint}}
\put(315,300){\usebox{\plotpoint}}
\put(313,299){\usebox{\plotpoint}}
\put(312,298){\usebox{\plotpoint}}
\put(311,297){\usebox{\plotpoint}}
\put(309,296){\usebox{\plotpoint}}
\put(308,295){\usebox{\plotpoint}}
\put(306,294){\usebox{\plotpoint}}
\put(305,293){\usebox{\plotpoint}}
\put(304,292){\usebox{\plotpoint}}
\put(302,291){\usebox{\plotpoint}}
\put(301,290){\usebox{\plotpoint}}
\put(299,289){\usebox{\plotpoint}}
\put(298,288){\usebox{\plotpoint}}
\put(297,287){\usebox{\plotpoint}}
\put(297,286){\usebox{\plotpoint}}
\sbox{\plotpoint}{\rule[-0.250pt]{0.500pt}{0.500pt}}%
\end{picture}
	% GNUPLOT: LaTeX picture
\setlength{\unitlength}{0.240900pt}
\ifx\plotpoint\undefined\newsavebox{\plotpoint}\fi
\sbox{\plotpoint}{\rule[-0.175pt]{0.350pt}{0.350pt}}%
\begin{picture}(250,300)(0,100)
\tenrm
\sbox{\plotpoint}{\rule[-0.175pt]{0.350pt}{0.350pt}}%
\put(370,95){\makebox(0,0){$I_{5,14}$}}
\put(370,346){\circle*{40}}
\put(289,288){\circle*{40}}
\put(320,195){\circle*{40}}
\put(420,195){\circle*{40}}
\put(451,288){\circle*{40}}
\put(370,346){\usebox{\plotpoint}}
\put(368,346){\usebox{\plotpoint}}
\put(367,345){\usebox{\plotpoint}}
\put(365,344){\usebox{\plotpoint}}
\put(364,343){\usebox{\plotpoint}}
\put(363,342){\usebox{\plotpoint}}
\put(361,341){\usebox{\plotpoint}}
\put(360,340){\usebox{\plotpoint}}
\put(358,339){\usebox{\plotpoint}}
\put(357,338){\usebox{\plotpoint}}
\put(356,337){\usebox{\plotpoint}}
\put(354,336){\usebox{\plotpoint}}
\put(353,335){\usebox{\plotpoint}}
\put(351,334){\usebox{\plotpoint}}
\put(350,333){\usebox{\plotpoint}}
\put(349,332){\usebox{\plotpoint}}
\put(347,331){\usebox{\plotpoint}}
\put(346,330){\usebox{\plotpoint}}
\put(344,329){\usebox{\plotpoint}}
\put(343,328){\usebox{\plotpoint}}
\put(342,327){\usebox{\plotpoint}}
\put(340,326){\usebox{\plotpoint}}
\put(339,325){\usebox{\plotpoint}}
\put(337,324){\usebox{\plotpoint}}
\put(336,323){\usebox{\plotpoint}}
\put(335,322){\usebox{\plotpoint}}
\put(333,321){\usebox{\plotpoint}}
\put(332,320){\usebox{\plotpoint}}
\put(330,319){\usebox{\plotpoint}}
\put(329,318){\usebox{\plotpoint}}
\put(328,317){\usebox{\plotpoint}}
\put(326,316){\usebox{\plotpoint}}
\put(325,315){\usebox{\plotpoint}}
\put(323,314){\usebox{\plotpoint}}
\put(322,313){\usebox{\plotpoint}}
\put(321,312){\usebox{\plotpoint}}
\put(319,311){\usebox{\plotpoint}}
\put(318,310){\usebox{\plotpoint}}
\put(316,309){\usebox{\plotpoint}}
\put(315,308){\usebox{\plotpoint}}
\put(314,307){\usebox{\plotpoint}}
\put(312,306){\usebox{\plotpoint}}
\put(311,305){\usebox{\plotpoint}}
\put(309,304){\usebox{\plotpoint}}
\put(308,303){\usebox{\plotpoint}}
\put(307,302){\usebox{\plotpoint}}
\put(305,301){\usebox{\plotpoint}}
\put(304,300){\usebox{\plotpoint}}
\put(302,299){\usebox{\plotpoint}}
\put(301,298){\usebox{\plotpoint}}
\put(300,297){\usebox{\plotpoint}}
\put(298,296){\usebox{\plotpoint}}
\put(297,295){\usebox{\plotpoint}}
\put(295,294){\usebox{\plotpoint}}
\put(294,293){\usebox{\plotpoint}}
\put(293,292){\usebox{\plotpoint}}
\put(291,291){\usebox{\plotpoint}}
\put(290,290){\usebox{\plotpoint}}
\put(289,289){\usebox{\plotpoint}}
\put(289,288){\usebox{\plotpoint}}
\put(315,188){\usebox{\plotpoint}}
\put(315,188){\rule[-0.175pt]{26.499pt}{0.350pt}}
\put(320,195){\usebox{\plotpoint}}
\put(320,195){\rule[-0.175pt]{24.090pt}{0.350pt}}
\put(325,202){\usebox{\plotpoint}}
\put(325,202){\rule[-0.175pt]{21.681pt}{0.350pt}}
\sbox{\plotpoint}{\rule[-0.250pt]{0.500pt}{0.500pt}}%
\end{picture}
	% GNUPLOT: LaTeX picture
\setlength{\unitlength}{0.240900pt}
\ifx\plotpoint\undefined\newsavebox{\plotpoint}\fi
\sbox{\plotpoint}{\rule[-0.175pt]{0.350pt}{0.350pt}}%
\begin{picture}(250,300)(0,100)
\tenrm
\sbox{\plotpoint}{\rule[-0.175pt]{0.350pt}{0.350pt}}%
\put(370,95){\makebox(0,0){$I_{5,15}$}}
\put(370,346){\circle*{40}}
\put(289,288){\circle*{40}}
\put(320,195){\circle*{40}}
\put(420,195){\circle*{40}}
\put(451,288){\circle*{40}}
\put(310,181){\usebox{\plotpoint}}
\put(310,181){\rule[-0.175pt]{28.908pt}{0.350pt}}
\put(320,195){\usebox{\plotpoint}}
\put(320,195){\rule[-0.175pt]{24.090pt}{0.350pt}}
\put(330,208){\usebox{\plotpoint}}
\put(330,208){\rule[-0.175pt]{19.272pt}{0.350pt}}
\put(315,188){\usebox{\plotpoint}}
\put(315,188){\rule[-0.175pt]{26.499pt}{0.350pt}}
\put(325,202){\usebox{\plotpoint}}
\put(325,202){\rule[-0.175pt]{21.681pt}{0.350pt}}
\end{picture}
	% GNUPLOT: LaTeX picture
\setlength{\unitlength}{0.240900pt}
\ifx\plotpoint\undefined\newsavebox{\plotpoint}\fi
\sbox{\plotpoint}{\rule[-0.175pt]{0.350pt}{0.350pt}}%
\begin{picture}(250,300)(0,100)
\tenrm
\sbox{\plotpoint}{\rule[-0.175pt]{0.350pt}{0.350pt}}%
\put(370,95){\makebox(0,0){$I_{5,16}$}}
\put(370,346){\circle*{40}}
\put(289,288){\circle*{40}}
\put(320,195){\circle*{40}}
\put(420,195){\circle*{40}}
\put(451,288){\circle*{40}}
\put(370,346){\usebox{\plotpoint}}
\put(368,346){\usebox{\plotpoint}}
\put(367,345){\usebox{\plotpoint}}
\put(365,344){\usebox{\plotpoint}}
\put(364,343){\usebox{\plotpoint}}
\put(363,342){\usebox{\plotpoint}}
\put(361,341){\usebox{\plotpoint}}
\put(360,340){\usebox{\plotpoint}}
\put(358,339){\usebox{\plotpoint}}
\put(357,338){\usebox{\plotpoint}}
\put(356,337){\usebox{\plotpoint}}
\put(354,336){\usebox{\plotpoint}}
\put(353,335){\usebox{\plotpoint}}
\put(351,334){\usebox{\plotpoint}}
\put(350,333){\usebox{\plotpoint}}
\put(349,332){\usebox{\plotpoint}}
\put(347,331){\usebox{\plotpoint}}
\put(346,330){\usebox{\plotpoint}}
\put(344,329){\usebox{\plotpoint}}
\put(343,328){\usebox{\plotpoint}}
\put(342,327){\usebox{\plotpoint}}
\put(340,326){\usebox{\plotpoint}}
\put(339,325){\usebox{\plotpoint}}
\put(337,324){\usebox{\plotpoint}}
\put(336,323){\usebox{\plotpoint}}
\put(335,322){\usebox{\plotpoint}}
\put(333,321){\usebox{\plotpoint}}
\put(332,320){\usebox{\plotpoint}}
\put(330,319){\usebox{\plotpoint}}
\put(329,318){\usebox{\plotpoint}}
\put(328,317){\usebox{\plotpoint}}
\put(326,316){\usebox{\plotpoint}}
\put(325,315){\usebox{\plotpoint}}
\put(323,314){\usebox{\plotpoint}}
\put(322,313){\usebox{\plotpoint}}
\put(321,312){\usebox{\plotpoint}}
\put(319,311){\usebox{\plotpoint}}
\put(318,310){\usebox{\plotpoint}}
\put(316,309){\usebox{\plotpoint}}
\put(315,308){\usebox{\plotpoint}}
\put(314,307){\usebox{\plotpoint}}
\put(312,306){\usebox{\plotpoint}}
\put(311,305){\usebox{\plotpoint}}
\put(309,304){\usebox{\plotpoint}}
\put(308,303){\usebox{\plotpoint}}
\put(307,302){\usebox{\plotpoint}}
\put(305,301){\usebox{\plotpoint}}
\put(304,300){\usebox{\plotpoint}}
\put(302,299){\usebox{\plotpoint}}
\put(301,298){\usebox{\plotpoint}}
\put(300,297){\usebox{\plotpoint}}
\put(298,296){\usebox{\plotpoint}}
\put(297,295){\usebox{\plotpoint}}
\put(295,294){\usebox{\plotpoint}}
\put(294,293){\usebox{\plotpoint}}
\put(293,292){\usebox{\plotpoint}}
\put(291,291){\usebox{\plotpoint}}
\put(290,290){\usebox{\plotpoint}}
\put(289,289){\usebox{\plotpoint}}
\put(289,288){\usebox{\plotpoint}}
\put(370,346){\usebox{\plotpoint}}
\put(370,346){\usebox{\plotpoint}}
\put(371,345){\usebox{\plotpoint}}
\put(372,344){\usebox{\plotpoint}}
\put(374,343){\usebox{\plotpoint}}
\put(375,342){\usebox{\plotpoint}}
\put(376,341){\usebox{\plotpoint}}
\put(378,340){\usebox{\plotpoint}}
\put(379,339){\usebox{\plotpoint}}
\put(381,338){\usebox{\plotpoint}}
\put(382,337){\usebox{\plotpoint}}
\put(383,336){\usebox{\plotpoint}}
\put(385,335){\usebox{\plotpoint}}
\put(386,334){\usebox{\plotpoint}}
\put(388,333){\usebox{\plotpoint}}
\put(389,332){\usebox{\plotpoint}}
\put(390,331){\usebox{\plotpoint}}
\put(392,330){\usebox{\plotpoint}}
\put(393,329){\usebox{\plotpoint}}
\put(395,328){\usebox{\plotpoint}}
\put(396,327){\usebox{\plotpoint}}
\put(397,326){\usebox{\plotpoint}}
\put(399,325){\usebox{\plotpoint}}
\put(400,324){\usebox{\plotpoint}}
\put(402,323){\usebox{\plotpoint}}
\put(403,322){\usebox{\plotpoint}}
\put(404,321){\usebox{\plotpoint}}
\put(406,320){\usebox{\plotpoint}}
\put(407,319){\usebox{\plotpoint}}
\put(409,318){\usebox{\plotpoint}}
\put(410,317){\usebox{\plotpoint}}
\put(411,316){\usebox{\plotpoint}}
\put(413,315){\usebox{\plotpoint}}
\put(414,314){\usebox{\plotpoint}}
\put(416,313){\usebox{\plotpoint}}
\put(417,312){\usebox{\plotpoint}}
\put(418,311){\usebox{\plotpoint}}
\put(420,310){\usebox{\plotpoint}}
\put(421,309){\usebox{\plotpoint}}
\put(423,308){\usebox{\plotpoint}}
\put(424,307){\usebox{\plotpoint}}
\put(425,306){\usebox{\plotpoint}}
\put(427,305){\usebox{\plotpoint}}
\put(428,304){\usebox{\plotpoint}}
\put(430,303){\usebox{\plotpoint}}
\put(431,302){\usebox{\plotpoint}}
\put(432,301){\usebox{\plotpoint}}
\put(434,300){\usebox{\plotpoint}}
\put(435,299){\usebox{\plotpoint}}
\put(437,298){\usebox{\plotpoint}}
\put(438,297){\usebox{\plotpoint}}
\put(439,296){\usebox{\plotpoint}}
\put(441,295){\usebox{\plotpoint}}
\put(442,294){\usebox{\plotpoint}}
\put(444,293){\usebox{\plotpoint}}
\put(445,292){\usebox{\plotpoint}}
\put(446,291){\usebox{\plotpoint}}
\put(448,290){\usebox{\plotpoint}}
\put(449,289){\usebox{\plotpoint}}
\put(450,288){\usebox{\plotpoint}}
\put(289,288){\usebox{\plotpoint}}
\put(289,288){\rule[-0.175pt]{39.026pt}{0.350pt}}
\put(315,188){\usebox{\plotpoint}}
\put(315,188){\rule[-0.175pt]{26.499pt}{0.350pt}}
\put(325,202){\usebox{\plotpoint}}
\put(325,202){\rule[-0.175pt]{21.681pt}{0.350pt}}
\end{picture}
	% GNUPLOT: LaTeX picture
\setlength{\unitlength}{0.240900pt}
\ifx\plotpoint\undefined\newsavebox{\plotpoint}\fi
\sbox{\plotpoint}{\rule[-0.175pt]{0.350pt}{0.350pt}}%
\begin{picture}(250,300)(0,100)
\tenrm
\sbox{\plotpoint}{\rule[-0.175pt]{0.350pt}{0.350pt}}%
\put(370,95){\makebox(0,0){$I_{5,17}$}}
\put(370,346){\circle*{40}}
\put(289,288){\circle*{40}}
\put(320,195){\circle*{40}}
\put(420,195){\circle*{40}}
\put(451,288){\circle*{40}}
\put(370,346){\usebox{\plotpoint}}
\put(368,346){\usebox{\plotpoint}}
\put(367,345){\usebox{\plotpoint}}
\put(365,344){\usebox{\plotpoint}}
\put(364,343){\usebox{\plotpoint}}
\put(363,342){\usebox{\plotpoint}}
\put(361,341){\usebox{\plotpoint}}
\put(360,340){\usebox{\plotpoint}}
\put(358,339){\usebox{\plotpoint}}
\put(357,338){\usebox{\plotpoint}}
\put(356,337){\usebox{\plotpoint}}
\put(354,336){\usebox{\plotpoint}}
\put(353,335){\usebox{\plotpoint}}
\put(351,334){\usebox{\plotpoint}}
\put(350,333){\usebox{\plotpoint}}
\put(349,332){\usebox{\plotpoint}}
\put(347,331){\usebox{\plotpoint}}
\put(346,330){\usebox{\plotpoint}}
\put(344,329){\usebox{\plotpoint}}
\put(343,328){\usebox{\plotpoint}}
\put(342,327){\usebox{\plotpoint}}
\put(340,326){\usebox{\plotpoint}}
\put(339,325){\usebox{\plotpoint}}
\put(337,324){\usebox{\plotpoint}}
\put(336,323){\usebox{\plotpoint}}
\put(335,322){\usebox{\plotpoint}}
\put(333,321){\usebox{\plotpoint}}
\put(332,320){\usebox{\plotpoint}}
\put(330,319){\usebox{\plotpoint}}
\put(329,318){\usebox{\plotpoint}}
\put(328,317){\usebox{\plotpoint}}
\put(326,316){\usebox{\plotpoint}}
\put(325,315){\usebox{\plotpoint}}
\put(323,314){\usebox{\plotpoint}}
\put(322,313){\usebox{\plotpoint}}
\put(321,312){\usebox{\plotpoint}}
\put(319,311){\usebox{\plotpoint}}
\put(318,310){\usebox{\plotpoint}}
\put(316,309){\usebox{\plotpoint}}
\put(315,308){\usebox{\plotpoint}}
\put(314,307){\usebox{\plotpoint}}
\put(312,306){\usebox{\plotpoint}}
\put(311,305){\usebox{\plotpoint}}
\put(309,304){\usebox{\plotpoint}}
\put(308,303){\usebox{\plotpoint}}
\put(307,302){\usebox{\plotpoint}}
\put(305,301){\usebox{\plotpoint}}
\put(304,300){\usebox{\plotpoint}}
\put(302,299){\usebox{\plotpoint}}
\put(301,298){\usebox{\plotpoint}}
\put(300,297){\usebox{\plotpoint}}
\put(298,296){\usebox{\plotpoint}}
\put(297,295){\usebox{\plotpoint}}
\put(295,294){\usebox{\plotpoint}}
\put(294,293){\usebox{\plotpoint}}
\put(293,292){\usebox{\plotpoint}}
\put(291,291){\usebox{\plotpoint}}
\put(290,290){\usebox{\plotpoint}}
\put(289,289){\usebox{\plotpoint}}
\put(289,288){\usebox{\plotpoint}}
\put(289,288){\usebox{\plotpoint}}
\put(289,285){\rule[-0.175pt]{0.350pt}{0.723pt}}
\put(290,282){\rule[-0.175pt]{0.350pt}{0.723pt}}
\put(291,279){\rule[-0.175pt]{0.350pt}{0.723pt}}
\put(292,276){\rule[-0.175pt]{0.350pt}{0.723pt}}
\put(293,273){\rule[-0.175pt]{0.350pt}{0.723pt}}
\put(294,270){\rule[-0.175pt]{0.350pt}{0.723pt}}
\put(295,267){\rule[-0.175pt]{0.350pt}{0.723pt}}
\put(296,264){\rule[-0.175pt]{0.350pt}{0.723pt}}
\put(297,261){\rule[-0.175pt]{0.350pt}{0.723pt}}
\put(298,258){\rule[-0.175pt]{0.350pt}{0.723pt}}
\put(299,255){\rule[-0.175pt]{0.350pt}{0.723pt}}
\put(300,252){\rule[-0.175pt]{0.350pt}{0.723pt}}
\put(301,249){\rule[-0.175pt]{0.350pt}{0.723pt}}
\put(302,246){\rule[-0.175pt]{0.350pt}{0.723pt}}
\put(303,243){\rule[-0.175pt]{0.350pt}{0.723pt}}
\put(304,240){\rule[-0.175pt]{0.350pt}{0.723pt}}
\put(305,237){\rule[-0.175pt]{0.350pt}{0.723pt}}
\put(306,234){\rule[-0.175pt]{0.350pt}{0.723pt}}
\put(307,231){\rule[-0.175pt]{0.350pt}{0.723pt}}
\put(308,228){\rule[-0.175pt]{0.350pt}{0.723pt}}
\put(309,225){\rule[-0.175pt]{0.350pt}{0.723pt}}
\put(310,222){\rule[-0.175pt]{0.350pt}{0.723pt}}
\put(311,219){\rule[-0.175pt]{0.350pt}{0.723pt}}
\put(312,216){\rule[-0.175pt]{0.350pt}{0.723pt}}
\put(313,213){\rule[-0.175pt]{0.350pt}{0.723pt}}
\put(314,210){\rule[-0.175pt]{0.350pt}{0.723pt}}
\put(315,207){\rule[-0.175pt]{0.350pt}{0.723pt}}
\put(316,204){\rule[-0.175pt]{0.350pt}{0.723pt}}
\put(317,201){\rule[-0.175pt]{0.350pt}{0.723pt}}
\put(318,198){\rule[-0.175pt]{0.350pt}{0.723pt}}
\put(319,195){\rule[-0.175pt]{0.350pt}{0.723pt}}
\put(320,195){\usebox{\plotpoint}}
\put(320,195){\rule[-0.175pt]{24.090pt}{0.350pt}}
\put(420,195){\usebox{\plotpoint}}
\put(420,195){\rule[-0.175pt]{0.350pt}{0.723pt}}
\put(421,198){\rule[-0.175pt]{0.350pt}{0.723pt}}
\put(422,201){\rule[-0.175pt]{0.350pt}{0.723pt}}
\put(423,204){\rule[-0.175pt]{0.350pt}{0.723pt}}
\put(424,207){\rule[-0.175pt]{0.350pt}{0.723pt}}
\put(425,210){\rule[-0.175pt]{0.350pt}{0.723pt}}
\put(426,213){\rule[-0.175pt]{0.350pt}{0.723pt}}
\put(427,216){\rule[-0.175pt]{0.350pt}{0.723pt}}
\put(428,219){\rule[-0.175pt]{0.350pt}{0.723pt}}
\put(429,222){\rule[-0.175pt]{0.350pt}{0.723pt}}
\put(430,225){\rule[-0.175pt]{0.350pt}{0.723pt}}
\put(431,228){\rule[-0.175pt]{0.350pt}{0.723pt}}
\put(432,231){\rule[-0.175pt]{0.350pt}{0.723pt}}
\put(433,234){\rule[-0.175pt]{0.350pt}{0.723pt}}
\put(434,237){\rule[-0.175pt]{0.350pt}{0.723pt}}
\put(435,240){\rule[-0.175pt]{0.350pt}{0.723pt}}
\put(436,243){\rule[-0.175pt]{0.350pt}{0.723pt}}
\put(437,246){\rule[-0.175pt]{0.350pt}{0.723pt}}
\put(438,249){\rule[-0.175pt]{0.350pt}{0.723pt}}
\put(439,252){\rule[-0.175pt]{0.350pt}{0.723pt}}
\put(440,255){\rule[-0.175pt]{0.350pt}{0.723pt}}
\put(441,258){\rule[-0.175pt]{0.350pt}{0.723pt}}
\put(442,261){\rule[-0.175pt]{0.350pt}{0.723pt}}
\put(443,264){\rule[-0.175pt]{0.350pt}{0.723pt}}
\put(444,267){\rule[-0.175pt]{0.350pt}{0.723pt}}
\put(445,270){\rule[-0.175pt]{0.350pt}{0.723pt}}
\put(446,273){\rule[-0.175pt]{0.350pt}{0.723pt}}
\put(447,276){\rule[-0.175pt]{0.350pt}{0.723pt}}
\put(448,279){\rule[-0.175pt]{0.350pt}{0.723pt}}
\put(449,282){\rule[-0.175pt]{0.350pt}{0.723pt}}
\put(450,285){\rule[-0.175pt]{0.350pt}{0.723pt}}
\put(370,346){\usebox{\plotpoint}}
\put(370,346){\usebox{\plotpoint}}
\put(371,345){\usebox{\plotpoint}}
\put(372,344){\usebox{\plotpoint}}
\put(374,343){\usebox{\plotpoint}}
\put(375,342){\usebox{\plotpoint}}
\put(376,341){\usebox{\plotpoint}}
\put(378,340){\usebox{\plotpoint}}
\put(379,339){\usebox{\plotpoint}}
\put(381,338){\usebox{\plotpoint}}
\put(382,337){\usebox{\plotpoint}}
\put(383,336){\usebox{\plotpoint}}
\put(385,335){\usebox{\plotpoint}}
\put(386,334){\usebox{\plotpoint}}
\put(388,333){\usebox{\plotpoint}}
\put(389,332){\usebox{\plotpoint}}
\put(390,331){\usebox{\plotpoint}}
\put(392,330){\usebox{\plotpoint}}
\put(393,329){\usebox{\plotpoint}}
\put(395,328){\usebox{\plotpoint}}
\put(396,327){\usebox{\plotpoint}}
\put(397,326){\usebox{\plotpoint}}
\put(399,325){\usebox{\plotpoint}}
\put(400,324){\usebox{\plotpoint}}
\put(402,323){\usebox{\plotpoint}}
\put(403,322){\usebox{\plotpoint}}
\put(404,321){\usebox{\plotpoint}}
\put(406,320){\usebox{\plotpoint}}
\put(407,319){\usebox{\plotpoint}}
\put(409,318){\usebox{\plotpoint}}
\put(410,317){\usebox{\plotpoint}}
\put(411,316){\usebox{\plotpoint}}
\put(413,315){\usebox{\plotpoint}}
\put(414,314){\usebox{\plotpoint}}
\put(416,313){\usebox{\plotpoint}}
\put(417,312){\usebox{\plotpoint}}
\put(418,311){\usebox{\plotpoint}}
\put(420,310){\usebox{\plotpoint}}
\put(421,309){\usebox{\plotpoint}}
\put(423,308){\usebox{\plotpoint}}
\put(424,307){\usebox{\plotpoint}}
\put(425,306){\usebox{\plotpoint}}
\put(427,305){\usebox{\plotpoint}}
\put(428,304){\usebox{\plotpoint}}
\put(430,303){\usebox{\plotpoint}}
\put(431,302){\usebox{\plotpoint}}
\put(432,301){\usebox{\plotpoint}}
\put(434,300){\usebox{\plotpoint}}
\put(435,299){\usebox{\plotpoint}}
\put(437,298){\usebox{\plotpoint}}
\put(438,297){\usebox{\plotpoint}}
\put(439,296){\usebox{\plotpoint}}
\put(441,295){\usebox{\plotpoint}}
\put(442,294){\usebox{\plotpoint}}
\put(444,293){\usebox{\plotpoint}}
\put(445,292){\usebox{\plotpoint}}
\put(446,291){\usebox{\plotpoint}}
\put(448,290){\usebox{\plotpoint}}
\put(449,289){\usebox{\plotpoint}}
\put(450,288){\usebox{\plotpoint}}
\end{picture}
	\input{5.18.tex}
	}
\end{figure}
\begin{figure}
	{\Large
	% GNUPLOT: LaTeX picture
\setlength{\unitlength}{0.240900pt}
\ifx\plotpoint\undefined\newsavebox{\plotpoint}\fi
\sbox{\plotpoint}{\rule[-0.175pt]{0.350pt}{0.350pt}}%
\begin{picture}(250,300)(0,100)
\tenrm
\sbox{\plotpoint}{\rule[-0.175pt]{0.350pt}{0.350pt}}%
\put(370,95){\makebox(0,0){$I_{5,19}$}}
\put(370,346){\circle*{40}}
\put(289,288){\circle*{40}}
\put(320,195){\circle*{40}}
\put(420,195){\circle*{40}}
\put(451,288){\circle*{40}}
\put(370,346){\usebox{\plotpoint}}
\put(368,346){\usebox{\plotpoint}}
\put(367,345){\usebox{\plotpoint}}
\put(365,344){\usebox{\plotpoint}}
\put(364,343){\usebox{\plotpoint}}
\put(363,342){\usebox{\plotpoint}}
\put(361,341){\usebox{\plotpoint}}
\put(360,340){\usebox{\plotpoint}}
\put(358,339){\usebox{\plotpoint}}
\put(357,338){\usebox{\plotpoint}}
\put(356,337){\usebox{\plotpoint}}
\put(354,336){\usebox{\plotpoint}}
\put(353,335){\usebox{\plotpoint}}
\put(351,334){\usebox{\plotpoint}}
\put(350,333){\usebox{\plotpoint}}
\put(349,332){\usebox{\plotpoint}}
\put(347,331){\usebox{\plotpoint}}
\put(346,330){\usebox{\plotpoint}}
\put(344,329){\usebox{\plotpoint}}
\put(343,328){\usebox{\plotpoint}}
\put(342,327){\usebox{\plotpoint}}
\put(340,326){\usebox{\plotpoint}}
\put(339,325){\usebox{\plotpoint}}
\put(337,324){\usebox{\plotpoint}}
\put(336,323){\usebox{\plotpoint}}
\put(335,322){\usebox{\plotpoint}}
\put(333,321){\usebox{\plotpoint}}
\put(332,320){\usebox{\plotpoint}}
\put(330,319){\usebox{\plotpoint}}
\put(329,318){\usebox{\plotpoint}}
\put(328,317){\usebox{\plotpoint}}
\put(326,316){\usebox{\plotpoint}}
\put(325,315){\usebox{\plotpoint}}
\put(323,314){\usebox{\plotpoint}}
\put(322,313){\usebox{\plotpoint}}
\put(321,312){\usebox{\plotpoint}}
\put(319,311){\usebox{\plotpoint}}
\put(318,310){\usebox{\plotpoint}}
\put(316,309){\usebox{\plotpoint}}
\put(315,308){\usebox{\plotpoint}}
\put(314,307){\usebox{\plotpoint}}
\put(312,306){\usebox{\plotpoint}}
\put(311,305){\usebox{\plotpoint}}
\put(309,304){\usebox{\plotpoint}}
\put(308,303){\usebox{\plotpoint}}
\put(307,302){\usebox{\plotpoint}}
\put(305,301){\usebox{\plotpoint}}
\put(304,300){\usebox{\plotpoint}}
\put(302,299){\usebox{\plotpoint}}
\put(301,298){\usebox{\plotpoint}}
\put(300,297){\usebox{\plotpoint}}
\put(298,296){\usebox{\plotpoint}}
\put(297,295){\usebox{\plotpoint}}
\put(295,294){\usebox{\plotpoint}}
\put(294,293){\usebox{\plotpoint}}
\put(293,292){\usebox{\plotpoint}}
\put(291,291){\usebox{\plotpoint}}
\put(290,290){\usebox{\plotpoint}}
\put(289,289){\usebox{\plotpoint}}
\put(289,288){\usebox{\plotpoint}}
\put(289,288){\usebox{\plotpoint}}
\put(289,288){\rule[-0.175pt]{39.026pt}{0.350pt}}
\put(370,346){\usebox{\plotpoint}}
\put(370,346){\usebox{\plotpoint}}
\put(371,345){\usebox{\plotpoint}}
\put(372,344){\usebox{\plotpoint}}
\put(374,343){\usebox{\plotpoint}}
\put(375,342){\usebox{\plotpoint}}
\put(376,341){\usebox{\plotpoint}}
\put(378,340){\usebox{\plotpoint}}
\put(379,339){\usebox{\plotpoint}}
\put(381,338){\usebox{\plotpoint}}
\put(382,337){\usebox{\plotpoint}}
\put(383,336){\usebox{\plotpoint}}
\put(385,335){\usebox{\plotpoint}}
\put(386,334){\usebox{\plotpoint}}
\put(388,333){\usebox{\plotpoint}}
\put(389,332){\usebox{\plotpoint}}
\put(390,331){\usebox{\plotpoint}}
\put(392,330){\usebox{\plotpoint}}
\put(393,329){\usebox{\plotpoint}}
\put(395,328){\usebox{\plotpoint}}
\put(396,327){\usebox{\plotpoint}}
\put(397,326){\usebox{\plotpoint}}
\put(399,325){\usebox{\plotpoint}}
\put(400,324){\usebox{\plotpoint}}
\put(402,323){\usebox{\plotpoint}}
\put(403,322){\usebox{\plotpoint}}
\put(404,321){\usebox{\plotpoint}}
\put(406,320){\usebox{\plotpoint}}
\put(407,319){\usebox{\plotpoint}}
\put(409,318){\usebox{\plotpoint}}
\put(410,317){\usebox{\plotpoint}}
\put(411,316){\usebox{\plotpoint}}
\put(413,315){\usebox{\plotpoint}}
\put(414,314){\usebox{\plotpoint}}
\put(416,313){\usebox{\plotpoint}}
\put(417,312){\usebox{\plotpoint}}
\put(418,311){\usebox{\plotpoint}}
\put(420,310){\usebox{\plotpoint}}
\put(421,309){\usebox{\plotpoint}}
\put(423,308){\usebox{\plotpoint}}
\put(424,307){\usebox{\plotpoint}}
\put(425,306){\usebox{\plotpoint}}
\put(427,305){\usebox{\plotpoint}}
\put(428,304){\usebox{\plotpoint}}
\put(430,303){\usebox{\plotpoint}}
\put(431,302){\usebox{\plotpoint}}
\put(432,301){\usebox{\plotpoint}}
\put(434,300){\usebox{\plotpoint}}
\put(435,299){\usebox{\plotpoint}}
\put(437,298){\usebox{\plotpoint}}
\put(438,297){\usebox{\plotpoint}}
\put(439,296){\usebox{\plotpoint}}
\put(441,295){\usebox{\plotpoint}}
\put(442,294){\usebox{\plotpoint}}
\put(444,293){\usebox{\plotpoint}}
\put(445,292){\usebox{\plotpoint}}
\put(446,291){\usebox{\plotpoint}}
\put(448,290){\usebox{\plotpoint}}
\put(449,289){\usebox{\plotpoint}}
\put(450,288){\usebox{\plotpoint}}
\put(370,354){\usebox{\plotpoint}}
\put(368,354){\usebox{\plotpoint}}
\put(367,353){\usebox{\plotpoint}}
\put(365,352){\usebox{\plotpoint}}
\put(364,351){\usebox{\plotpoint}}
\put(362,350){\usebox{\plotpoint}}
\put(361,349){\usebox{\plotpoint}}
\put(360,348){\usebox{\plotpoint}}
\put(358,347){\usebox{\plotpoint}}
\put(357,346){\usebox{\plotpoint}}
\put(355,345){\usebox{\plotpoint}}
\put(354,344){\usebox{\plotpoint}}
\put(353,343){\usebox{\plotpoint}}
\put(351,342){\usebox{\plotpoint}}
\put(350,341){\usebox{\plotpoint}}
\put(348,340){\usebox{\plotpoint}}
\put(347,339){\usebox{\plotpoint}}
\put(345,338){\usebox{\plotpoint}}
\put(344,337){\usebox{\plotpoint}}
\put(343,336){\usebox{\plotpoint}}
\put(341,335){\usebox{\plotpoint}}
\put(340,334){\usebox{\plotpoint}}
\put(338,333){\usebox{\plotpoint}}
\put(337,332){\usebox{\plotpoint}}
\put(336,331){\usebox{\plotpoint}}
\put(334,330){\usebox{\plotpoint}}
\put(333,329){\usebox{\plotpoint}}
\put(331,328){\usebox{\plotpoint}}
\put(330,327){\usebox{\plotpoint}}
\put(329,326){\usebox{\plotpoint}}
\put(327,325){\usebox{\plotpoint}}
\put(326,324){\usebox{\plotpoint}}
\put(324,323){\usebox{\plotpoint}}
\put(323,322){\usebox{\plotpoint}}
\put(321,321){\usebox{\plotpoint}}
\put(320,320){\usebox{\plotpoint}}
\put(319,319){\usebox{\plotpoint}}
\put(317,318){\usebox{\plotpoint}}
\put(316,317){\usebox{\plotpoint}}
\put(314,316){\usebox{\plotpoint}}
\put(313,315){\usebox{\plotpoint}}
\put(312,314){\usebox{\plotpoint}}
\put(310,313){\usebox{\plotpoint}}
\put(309,312){\usebox{\plotpoint}}
\put(307,311){\usebox{\plotpoint}}
\put(306,310){\usebox{\plotpoint}}
\put(305,309){\usebox{\plotpoint}}
\put(303,308){\usebox{\plotpoint}}
\put(302,307){\usebox{\plotpoint}}
\put(300,306){\usebox{\plotpoint}}
\put(299,305){\usebox{\plotpoint}}
\put(297,304){\usebox{\plotpoint}}
\put(296,303){\usebox{\plotpoint}}
\put(295,302){\usebox{\plotpoint}}
\put(293,301){\usebox{\plotpoint}}
\put(292,300){\usebox{\plotpoint}}
\put(290,299){\usebox{\plotpoint}}
\put(289,298){\usebox{\plotpoint}}
\put(288,297){\usebox{\plotpoint}}
\put(286,296){\usebox{\plotpoint}}
\put(285,295){\usebox{\plotpoint}}
\put(283,294){\usebox{\plotpoint}}
\put(282,293){\usebox{\plotpoint}}
\put(281,292){\usebox{\plotpoint}}
\put(281,291){\usebox{\plotpoint}}
\put(370,338){\usebox{\plotpoint}}
\put(368,338){\usebox{\plotpoint}}
\put(367,337){\usebox{\plotpoint}}
\put(365,336){\usebox{\plotpoint}}
\put(364,335){\usebox{\plotpoint}}
\put(362,334){\usebox{\plotpoint}}
\put(361,333){\usebox{\plotpoint}}
\put(360,332){\usebox{\plotpoint}}
\put(358,331){\usebox{\plotpoint}}
\put(357,330){\usebox{\plotpoint}}
\put(355,329){\usebox{\plotpoint}}
\put(354,328){\usebox{\plotpoint}}
\put(353,327){\usebox{\plotpoint}}
\put(351,326){\usebox{\plotpoint}}
\put(350,325){\usebox{\plotpoint}}
\put(348,324){\usebox{\plotpoint}}
\put(347,323){\usebox{\plotpoint}}
\put(346,322){\usebox{\plotpoint}}
\put(344,321){\usebox{\plotpoint}}
\put(343,320){\usebox{\plotpoint}}
\put(341,319){\usebox{\plotpoint}}
\put(340,318){\usebox{\plotpoint}}
\put(339,317){\usebox{\plotpoint}}
\put(337,316){\usebox{\plotpoint}}
\put(336,315){\usebox{\plotpoint}}
\put(334,314){\usebox{\plotpoint}}
\put(333,313){\usebox{\plotpoint}}
\put(332,312){\usebox{\plotpoint}}
\put(330,311){\usebox{\plotpoint}}
\put(329,310){\usebox{\plotpoint}}
\put(327,309){\usebox{\plotpoint}}
\put(326,308){\usebox{\plotpoint}}
\put(325,307){\usebox{\plotpoint}}
\put(323,306){\usebox{\plotpoint}}
\put(322,305){\usebox{\plotpoint}}
\put(320,304){\usebox{\plotpoint}}
\put(319,303){\usebox{\plotpoint}}
\put(318,302){\usebox{\plotpoint}}
\put(316,301){\usebox{\plotpoint}}
\put(315,300){\usebox{\plotpoint}}
\put(313,299){\usebox{\plotpoint}}
\put(312,298){\usebox{\plotpoint}}
\put(311,297){\usebox{\plotpoint}}
\put(309,296){\usebox{\plotpoint}}
\put(308,295){\usebox{\plotpoint}}
\put(306,294){\usebox{\plotpoint}}
\put(305,293){\usebox{\plotpoint}}
\put(304,292){\usebox{\plotpoint}}
\put(302,291){\usebox{\plotpoint}}
\put(301,290){\usebox{\plotpoint}}
\put(299,289){\usebox{\plotpoint}}
\put(298,288){\usebox{\plotpoint}}
\put(297,287){\usebox{\plotpoint}}
\put(297,286){\usebox{\plotpoint}}
\end{picture}
	% GNUPLOT: LaTeX picture
\setlength{\unitlength}{0.240900pt}
\ifx\plotpoint\undefined\newsavebox{\plotpoint}\fi
\sbox{\plotpoint}{\rule[-0.175pt]{0.350pt}{0.350pt}}%
\begin{picture}(250,300)(0,100)
\tenrm
\sbox{\plotpoint}{\rule[-0.175pt]{0.350pt}{0.350pt}}%
\put(370,95){\makebox(0,0){$I_{5,20}$}}
\put(370,346){\circle*{40}}
\put(289,288){\circle*{40}}
\put(320,195){\circle*{40}}
\put(420,195){\circle*{40}}
\put(451,288){\circle*{40}}
\put(370,354){\usebox{\plotpoint}}
\put(368,354){\usebox{\plotpoint}}
\put(367,353){\usebox{\plotpoint}}
\put(365,352){\usebox{\plotpoint}}
\put(364,351){\usebox{\plotpoint}}
\put(362,350){\usebox{\plotpoint}}
\put(361,349){\usebox{\plotpoint}}
\put(360,348){\usebox{\plotpoint}}
\put(358,347){\usebox{\plotpoint}}
\put(357,346){\usebox{\plotpoint}}
\put(355,345){\usebox{\plotpoint}}
\put(354,344){\usebox{\plotpoint}}
\put(353,343){\usebox{\plotpoint}}
\put(351,342){\usebox{\plotpoint}}
\put(350,341){\usebox{\plotpoint}}
\put(348,340){\usebox{\plotpoint}}
\put(347,339){\usebox{\plotpoint}}
\put(345,338){\usebox{\plotpoint}}
\put(344,337){\usebox{\plotpoint}}
\put(343,336){\usebox{\plotpoint}}
\put(341,335){\usebox{\plotpoint}}
\put(340,334){\usebox{\plotpoint}}
\put(338,333){\usebox{\plotpoint}}
\put(337,332){\usebox{\plotpoint}}
\put(336,331){\usebox{\plotpoint}}
\put(334,330){\usebox{\plotpoint}}
\put(333,329){\usebox{\plotpoint}}
\put(331,328){\usebox{\plotpoint}}
\put(330,327){\usebox{\plotpoint}}
\put(329,326){\usebox{\plotpoint}}
\put(327,325){\usebox{\plotpoint}}
\put(326,324){\usebox{\plotpoint}}
\put(324,323){\usebox{\plotpoint}}
\put(323,322){\usebox{\plotpoint}}
\put(321,321){\usebox{\plotpoint}}
\put(320,320){\usebox{\plotpoint}}
\put(319,319){\usebox{\plotpoint}}
\put(317,318){\usebox{\plotpoint}}
\put(316,317){\usebox{\plotpoint}}
\put(314,316){\usebox{\plotpoint}}
\put(313,315){\usebox{\plotpoint}}
\put(312,314){\usebox{\plotpoint}}
\put(310,313){\usebox{\plotpoint}}
\put(309,312){\usebox{\plotpoint}}
\put(307,311){\usebox{\plotpoint}}
\put(306,310){\usebox{\plotpoint}}
\put(305,309){\usebox{\plotpoint}}
\put(303,308){\usebox{\plotpoint}}
\put(302,307){\usebox{\plotpoint}}
\put(300,306){\usebox{\plotpoint}}
\put(299,305){\usebox{\plotpoint}}
\put(297,304){\usebox{\plotpoint}}
\put(296,303){\usebox{\plotpoint}}
\put(295,302){\usebox{\plotpoint}}
\put(293,301){\usebox{\plotpoint}}
\put(292,300){\usebox{\plotpoint}}
\put(290,299){\usebox{\plotpoint}}
\put(289,298){\usebox{\plotpoint}}
\put(288,297){\usebox{\plotpoint}}
\put(286,296){\usebox{\plotpoint}}
\put(285,295){\usebox{\plotpoint}}
\put(283,294){\usebox{\plotpoint}}
\put(282,293){\usebox{\plotpoint}}
\put(281,292){\usebox{\plotpoint}}
\put(281,291){\usebox{\plotpoint}}
\put(370,338){\usebox{\plotpoint}}
\put(368,338){\usebox{\plotpoint}}
\put(367,337){\usebox{\plotpoint}}
\put(365,336){\usebox{\plotpoint}}
\put(364,335){\usebox{\plotpoint}}
\put(362,334){\usebox{\plotpoint}}
\put(361,333){\usebox{\plotpoint}}
\put(360,332){\usebox{\plotpoint}}
\put(358,331){\usebox{\plotpoint}}
\put(357,330){\usebox{\plotpoint}}
\put(355,329){\usebox{\plotpoint}}
\put(354,328){\usebox{\plotpoint}}
\put(353,327){\usebox{\plotpoint}}
\put(351,326){\usebox{\plotpoint}}
\put(350,325){\usebox{\plotpoint}}
\put(348,324){\usebox{\plotpoint}}
\put(347,323){\usebox{\plotpoint}}
\put(346,322){\usebox{\plotpoint}}
\put(344,321){\usebox{\plotpoint}}
\put(343,320){\usebox{\plotpoint}}
\put(341,319){\usebox{\plotpoint}}
\put(340,318){\usebox{\plotpoint}}
\put(339,317){\usebox{\plotpoint}}
\put(337,316){\usebox{\plotpoint}}
\put(336,315){\usebox{\plotpoint}}
\put(334,314){\usebox{\plotpoint}}
\put(333,313){\usebox{\plotpoint}}
\put(332,312){\usebox{\plotpoint}}
\put(330,311){\usebox{\plotpoint}}
\put(329,310){\usebox{\plotpoint}}
\put(327,309){\usebox{\plotpoint}}
\put(326,308){\usebox{\plotpoint}}
\put(325,307){\usebox{\plotpoint}}
\put(323,306){\usebox{\plotpoint}}
\put(322,305){\usebox{\plotpoint}}
\put(320,304){\usebox{\plotpoint}}
\put(319,303){\usebox{\plotpoint}}
\put(318,302){\usebox{\plotpoint}}
\put(316,301){\usebox{\plotpoint}}
\put(315,300){\usebox{\plotpoint}}
\put(313,299){\usebox{\plotpoint}}
\put(312,298){\usebox{\plotpoint}}
\put(311,297){\usebox{\plotpoint}}
\put(309,296){\usebox{\plotpoint}}
\put(308,295){\usebox{\plotpoint}}
\put(306,294){\usebox{\plotpoint}}
\put(305,293){\usebox{\plotpoint}}
\put(304,292){\usebox{\plotpoint}}
\put(302,291){\usebox{\plotpoint}}
\put(301,290){\usebox{\plotpoint}}
\put(299,289){\usebox{\plotpoint}}
\put(298,288){\usebox{\plotpoint}}
\put(297,287){\usebox{\plotpoint}}
\put(297,286){\usebox{\plotpoint}}
\put(370,354){\usebox{\plotpoint}}
\put(370,354){\usebox{\plotpoint}}
\put(371,353){\usebox{\plotpoint}}
\put(372,352){\usebox{\plotpoint}}
\put(374,351){\usebox{\plotpoint}}
\put(375,350){\usebox{\plotpoint}}
\put(377,349){\usebox{\plotpoint}}
\put(378,348){\usebox{\plotpoint}}
\put(379,347){\usebox{\plotpoint}}
\put(381,346){\usebox{\plotpoint}}
\put(382,345){\usebox{\plotpoint}}
\put(384,344){\usebox{\plotpoint}}
\put(385,343){\usebox{\plotpoint}}
\put(386,342){\usebox{\plotpoint}}
\put(388,341){\usebox{\plotpoint}}
\put(389,340){\usebox{\plotpoint}}
\put(391,339){\usebox{\plotpoint}}
\put(392,338){\usebox{\plotpoint}}
\put(394,337){\usebox{\plotpoint}}
\put(395,336){\usebox{\plotpoint}}
\put(396,335){\usebox{\plotpoint}}
\put(398,334){\usebox{\plotpoint}}
\put(399,333){\usebox{\plotpoint}}
\put(401,332){\usebox{\plotpoint}}
\put(402,331){\usebox{\plotpoint}}
\put(403,330){\usebox{\plotpoint}}
\put(405,329){\usebox{\plotpoint}}
\put(406,328){\usebox{\plotpoint}}
\put(408,327){\usebox{\plotpoint}}
\put(409,326){\usebox{\plotpoint}}
\put(410,325){\usebox{\plotpoint}}
\put(412,324){\usebox{\plotpoint}}
\put(413,323){\usebox{\plotpoint}}
\put(415,322){\usebox{\plotpoint}}
\put(416,321){\usebox{\plotpoint}}
\put(418,320){\usebox{\plotpoint}}
\put(419,319){\usebox{\plotpoint}}
\put(420,318){\usebox{\plotpoint}}
\put(422,317){\usebox{\plotpoint}}
\put(423,316){\usebox{\plotpoint}}
\put(425,315){\usebox{\plotpoint}}
\put(426,314){\usebox{\plotpoint}}
\put(427,313){\usebox{\plotpoint}}
\put(429,312){\usebox{\plotpoint}}
\put(430,311){\usebox{\plotpoint}}
\put(432,310){\usebox{\plotpoint}}
\put(433,309){\usebox{\plotpoint}}
\put(434,308){\usebox{\plotpoint}}
\put(436,307){\usebox{\plotpoint}}
\put(437,306){\usebox{\plotpoint}}
\put(439,305){\usebox{\plotpoint}}
\put(440,304){\usebox{\plotpoint}}
\put(442,303){\usebox{\plotpoint}}
\put(443,302){\usebox{\plotpoint}}
\put(444,301){\usebox{\plotpoint}}
\put(446,300){\usebox{\plotpoint}}
\put(447,299){\usebox{\plotpoint}}
\put(449,298){\usebox{\plotpoint}}
\put(450,297){\usebox{\plotpoint}}
\put(451,296){\usebox{\plotpoint}}
\put(453,295){\usebox{\plotpoint}}
\put(454,294){\usebox{\plotpoint}}
\put(456,293){\usebox{\plotpoint}}
\put(457,292){\usebox{\plotpoint}}
\put(458,291){\usebox{\plotpoint}}
\put(370,338){\usebox{\plotpoint}}
\put(370,338){\usebox{\plotpoint}}
\put(371,337){\usebox{\plotpoint}}
\put(372,336){\usebox{\plotpoint}}
\put(374,335){\usebox{\plotpoint}}
\put(375,334){\usebox{\plotpoint}}
\put(377,333){\usebox{\plotpoint}}
\put(378,332){\usebox{\plotpoint}}
\put(379,331){\usebox{\plotpoint}}
\put(381,330){\usebox{\plotpoint}}
\put(382,329){\usebox{\plotpoint}}
\put(384,328){\usebox{\plotpoint}}
\put(385,327){\usebox{\plotpoint}}
\put(386,326){\usebox{\plotpoint}}
\put(388,325){\usebox{\plotpoint}}
\put(389,324){\usebox{\plotpoint}}
\put(391,323){\usebox{\plotpoint}}
\put(392,322){\usebox{\plotpoint}}
\put(393,321){\usebox{\plotpoint}}
\put(395,320){\usebox{\plotpoint}}
\put(396,319){\usebox{\plotpoint}}
\put(398,318){\usebox{\plotpoint}}
\put(399,317){\usebox{\plotpoint}}
\put(400,316){\usebox{\plotpoint}}
\put(402,315){\usebox{\plotpoint}}
\put(403,314){\usebox{\plotpoint}}
\put(405,313){\usebox{\plotpoint}}
\put(406,312){\usebox{\plotpoint}}
\put(407,311){\usebox{\plotpoint}}
\put(409,310){\usebox{\plotpoint}}
\put(410,309){\usebox{\plotpoint}}
\put(412,308){\usebox{\plotpoint}}
\put(413,307){\usebox{\plotpoint}}
\put(414,306){\usebox{\plotpoint}}
\put(416,305){\usebox{\plotpoint}}
\put(417,304){\usebox{\plotpoint}}
\put(419,303){\usebox{\plotpoint}}
\put(420,302){\usebox{\plotpoint}}
\put(421,301){\usebox{\plotpoint}}
\put(423,300){\usebox{\plotpoint}}
\put(424,299){\usebox{\plotpoint}}
\put(426,298){\usebox{\plotpoint}}
\put(427,297){\usebox{\plotpoint}}
\put(428,296){\usebox{\plotpoint}}
\put(430,295){\usebox{\plotpoint}}
\put(431,294){\usebox{\plotpoint}}
\put(433,293){\usebox{\plotpoint}}
\put(434,292){\usebox{\plotpoint}}
\put(435,291){\usebox{\plotpoint}}
\put(437,290){\usebox{\plotpoint}}
\put(438,289){\usebox{\plotpoint}}
\put(440,288){\usebox{\plotpoint}}
\put(441,287){\usebox{\plotpoint}}
\put(442,286){\usebox{\plotpoint}}
\put(289,288){\usebox{\plotpoint}}
\put(289,288){\rule[-0.175pt]{39.026pt}{0.350pt}}
\end{picture}
	% GNUPLOT: LaTeX picture
\setlength{\unitlength}{0.240900pt}
\ifx\plotpoint\undefined\newsavebox{\plotpoint}\fi
\sbox{\plotpoint}{\rule[-0.175pt]{0.350pt}{0.350pt}}%
\begin{picture}(250,300)(0,100)
\tenrm
\sbox{\plotpoint}{\rule[-0.175pt]{0.350pt}{0.350pt}}%
\put(370,95){\makebox(0,0){$I_{5,21}$}}
\put(370,346){\circle*{40}}
\put(289,288){\circle*{40}}
\put(320,195){\circle*{40}}
\put(420,195){\circle*{40}}
\put(451,288){\circle*{40}}
\put(370,346){\usebox{\plotpoint}}
\put(368,346){\usebox{\plotpoint}}
\put(367,345){\usebox{\plotpoint}}
\put(365,344){\usebox{\plotpoint}}
\put(364,343){\usebox{\plotpoint}}
\put(363,342){\usebox{\plotpoint}}
\put(361,341){\usebox{\plotpoint}}
\put(360,340){\usebox{\plotpoint}}
\put(358,339){\usebox{\plotpoint}}
\put(357,338){\usebox{\plotpoint}}
\put(356,337){\usebox{\plotpoint}}
\put(354,336){\usebox{\plotpoint}}
\put(353,335){\usebox{\plotpoint}}
\put(351,334){\usebox{\plotpoint}}
\put(350,333){\usebox{\plotpoint}}
\put(349,332){\usebox{\plotpoint}}
\put(347,331){\usebox{\plotpoint}}
\put(346,330){\usebox{\plotpoint}}
\put(344,329){\usebox{\plotpoint}}
\put(343,328){\usebox{\plotpoint}}
\put(342,327){\usebox{\plotpoint}}
\put(340,326){\usebox{\plotpoint}}
\put(339,325){\usebox{\plotpoint}}
\put(337,324){\usebox{\plotpoint}}
\put(336,323){\usebox{\plotpoint}}
\put(335,322){\usebox{\plotpoint}}
\put(333,321){\usebox{\plotpoint}}
\put(332,320){\usebox{\plotpoint}}
\put(330,319){\usebox{\plotpoint}}
\put(329,318){\usebox{\plotpoint}}
\put(328,317){\usebox{\plotpoint}}
\put(326,316){\usebox{\plotpoint}}
\put(325,315){\usebox{\plotpoint}}
\put(323,314){\usebox{\plotpoint}}
\put(322,313){\usebox{\plotpoint}}
\put(321,312){\usebox{\plotpoint}}
\put(319,311){\usebox{\plotpoint}}
\put(318,310){\usebox{\plotpoint}}
\put(316,309){\usebox{\plotpoint}}
\put(315,308){\usebox{\plotpoint}}
\put(314,307){\usebox{\plotpoint}}
\put(312,306){\usebox{\plotpoint}}
\put(311,305){\usebox{\plotpoint}}
\put(309,304){\usebox{\plotpoint}}
\put(308,303){\usebox{\plotpoint}}
\put(307,302){\usebox{\plotpoint}}
\put(305,301){\usebox{\plotpoint}}
\put(304,300){\usebox{\plotpoint}}
\put(302,299){\usebox{\plotpoint}}
\put(301,298){\usebox{\plotpoint}}
\put(300,297){\usebox{\plotpoint}}
\put(298,296){\usebox{\plotpoint}}
\put(297,295){\usebox{\plotpoint}}
\put(295,294){\usebox{\plotpoint}}
\put(294,293){\usebox{\plotpoint}}
\put(293,292){\usebox{\plotpoint}}
\put(291,291){\usebox{\plotpoint}}
\put(290,290){\usebox{\plotpoint}}
\put(289,289){\usebox{\plotpoint}}
\put(289,288){\usebox{\plotpoint}}
\put(310,181){\usebox{\plotpoint}}
\put(310,181){\rule[-0.175pt]{28.908pt}{0.350pt}}
\put(330,208){\usebox{\plotpoint}}
\put(330,208){\rule[-0.175pt]{19.272pt}{0.350pt}}
\put(315,188){\usebox{\plotpoint}}
\put(315,188){\rule[-0.175pt]{26.499pt}{0.350pt}}
\put(325,202){\usebox{\plotpoint}}
\put(325,202){\rule[-0.175pt]{21.681pt}{0.350pt}}
\end{picture}
	% GNUPLOT: LaTeX picture
\setlength{\unitlength}{0.240900pt}
\ifx\plotpoint\undefined\newsavebox{\plotpoint}\fi
\sbox{\plotpoint}{\rule[-0.175pt]{0.350pt}{0.350pt}}%
\begin{picture}(250,300)(0,100)
\tenrm
\sbox{\plotpoint}{\rule[-0.175pt]{0.350pt}{0.350pt}}%
\put(370,95){\makebox(0,0){$I_{5,22}$}}
\put(370,346){\circle*{40}}
\put(289,288){\circle*{40}}
\put(320,195){\circle*{40}}
\put(420,195){\circle*{40}}
\put(451,288){\circle*{40}}
\put(370,354){\usebox{\plotpoint}}
\put(370,354){\usebox{\plotpoint}}
\put(371,353){\usebox{\plotpoint}}
\put(372,352){\usebox{\plotpoint}}
\put(374,351){\usebox{\plotpoint}}
\put(375,350){\usebox{\plotpoint}}
\put(377,349){\usebox{\plotpoint}}
\put(378,348){\usebox{\plotpoint}}
\put(379,347){\usebox{\plotpoint}}
\put(381,346){\usebox{\plotpoint}}
\put(382,345){\usebox{\plotpoint}}
\put(384,344){\usebox{\plotpoint}}
\put(385,343){\usebox{\plotpoint}}
\put(386,342){\usebox{\plotpoint}}
\put(388,341){\usebox{\plotpoint}}
\put(389,340){\usebox{\plotpoint}}
\put(391,339){\usebox{\plotpoint}}
\put(392,338){\usebox{\plotpoint}}
\put(394,337){\usebox{\plotpoint}}
\put(395,336){\usebox{\plotpoint}}
\put(396,335){\usebox{\plotpoint}}
\put(398,334){\usebox{\plotpoint}}
\put(399,333){\usebox{\plotpoint}}
\put(401,332){\usebox{\plotpoint}}
\put(402,331){\usebox{\plotpoint}}
\put(403,330){\usebox{\plotpoint}}
\put(405,329){\usebox{\plotpoint}}
\put(406,328){\usebox{\plotpoint}}
\put(408,327){\usebox{\plotpoint}}
\put(409,326){\usebox{\plotpoint}}
\put(410,325){\usebox{\plotpoint}}
\put(412,324){\usebox{\plotpoint}}
\put(413,323){\usebox{\plotpoint}}
\put(415,322){\usebox{\plotpoint}}
\put(416,321){\usebox{\plotpoint}}
\put(418,320){\usebox{\plotpoint}}
\put(419,319){\usebox{\plotpoint}}
\put(420,318){\usebox{\plotpoint}}
\put(422,317){\usebox{\plotpoint}}
\put(423,316){\usebox{\plotpoint}}
\put(425,315){\usebox{\plotpoint}}
\put(426,314){\usebox{\plotpoint}}
\put(427,313){\usebox{\plotpoint}}
\put(429,312){\usebox{\plotpoint}}
\put(430,311){\usebox{\plotpoint}}
\put(432,310){\usebox{\plotpoint}}
\put(433,309){\usebox{\plotpoint}}
\put(434,308){\usebox{\plotpoint}}
\put(436,307){\usebox{\plotpoint}}
\put(437,306){\usebox{\plotpoint}}
\put(439,305){\usebox{\plotpoint}}
\put(440,304){\usebox{\plotpoint}}
\put(442,303){\usebox{\plotpoint}}
\put(443,302){\usebox{\plotpoint}}
\put(444,301){\usebox{\plotpoint}}
\put(446,300){\usebox{\plotpoint}}
\put(447,299){\usebox{\plotpoint}}
\put(449,298){\usebox{\plotpoint}}
\put(450,297){\usebox{\plotpoint}}
\put(451,296){\usebox{\plotpoint}}
\put(453,295){\usebox{\plotpoint}}
\put(454,294){\usebox{\plotpoint}}
\put(456,293){\usebox{\plotpoint}}
\put(457,292){\usebox{\plotpoint}}
\put(458,291){\usebox{\plotpoint}}
\put(370,338){\usebox{\plotpoint}}
\put(370,338){\usebox{\plotpoint}}
\put(371,337){\usebox{\plotpoint}}
\put(372,336){\usebox{\plotpoint}}
\put(374,335){\usebox{\plotpoint}}
\put(375,334){\usebox{\plotpoint}}
\put(377,333){\usebox{\plotpoint}}
\put(378,332){\usebox{\plotpoint}}
\put(379,331){\usebox{\plotpoint}}
\put(381,330){\usebox{\plotpoint}}
\put(382,329){\usebox{\plotpoint}}
\put(384,328){\usebox{\plotpoint}}
\put(385,327){\usebox{\plotpoint}}
\put(386,326){\usebox{\plotpoint}}
\put(388,325){\usebox{\plotpoint}}
\put(389,324){\usebox{\plotpoint}}
\put(391,323){\usebox{\plotpoint}}
\put(392,322){\usebox{\plotpoint}}
\put(393,321){\usebox{\plotpoint}}
\put(395,320){\usebox{\plotpoint}}
\put(396,319){\usebox{\plotpoint}}
\put(398,318){\usebox{\plotpoint}}
\put(399,317){\usebox{\plotpoint}}
\put(400,316){\usebox{\plotpoint}}
\put(402,315){\usebox{\plotpoint}}
\put(403,314){\usebox{\plotpoint}}
\put(405,313){\usebox{\plotpoint}}
\put(406,312){\usebox{\plotpoint}}
\put(407,311){\usebox{\plotpoint}}
\put(409,310){\usebox{\plotpoint}}
\put(410,309){\usebox{\plotpoint}}
\put(412,308){\usebox{\plotpoint}}
\put(413,307){\usebox{\plotpoint}}
\put(414,306){\usebox{\plotpoint}}
\put(416,305){\usebox{\plotpoint}}
\put(417,304){\usebox{\plotpoint}}
\put(419,303){\usebox{\plotpoint}}
\put(420,302){\usebox{\plotpoint}}
\put(421,301){\usebox{\plotpoint}}
\put(423,300){\usebox{\plotpoint}}
\put(424,299){\usebox{\plotpoint}}
\put(426,298){\usebox{\plotpoint}}
\put(427,297){\usebox{\plotpoint}}
\put(428,296){\usebox{\plotpoint}}
\put(430,295){\usebox{\plotpoint}}
\put(431,294){\usebox{\plotpoint}}
\put(433,293){\usebox{\plotpoint}}
\put(434,292){\usebox{\plotpoint}}
\put(435,291){\usebox{\plotpoint}}
\put(437,290){\usebox{\plotpoint}}
\put(438,289){\usebox{\plotpoint}}
\put(440,288){\usebox{\plotpoint}}
\put(441,287){\usebox{\plotpoint}}
\put(442,286){\usebox{\plotpoint}}
\put(370,354){\usebox{\plotpoint}}
\put(368,354){\usebox{\plotpoint}}
\put(367,353){\usebox{\plotpoint}}
\put(365,352){\usebox{\plotpoint}}
\put(364,351){\usebox{\plotpoint}}
\put(362,350){\usebox{\plotpoint}}
\put(361,349){\usebox{\plotpoint}}
\put(360,348){\usebox{\plotpoint}}
\put(358,347){\usebox{\plotpoint}}
\put(357,346){\usebox{\plotpoint}}
\put(355,345){\usebox{\plotpoint}}
\put(354,344){\usebox{\plotpoint}}
\put(353,343){\usebox{\plotpoint}}
\put(351,342){\usebox{\plotpoint}}
\put(350,341){\usebox{\plotpoint}}
\put(348,340){\usebox{\plotpoint}}
\put(347,339){\usebox{\plotpoint}}
\put(345,338){\usebox{\plotpoint}}
\put(344,337){\usebox{\plotpoint}}
\put(343,336){\usebox{\plotpoint}}
\put(341,335){\usebox{\plotpoint}}
\put(340,334){\usebox{\plotpoint}}
\put(338,333){\usebox{\plotpoint}}
\put(337,332){\usebox{\plotpoint}}
\put(336,331){\usebox{\plotpoint}}
\put(334,330){\usebox{\plotpoint}}
\put(333,329){\usebox{\plotpoint}}
\put(331,328){\usebox{\plotpoint}}
\put(330,327){\usebox{\plotpoint}}
\put(329,326){\usebox{\plotpoint}}
\put(327,325){\usebox{\plotpoint}}
\put(326,324){\usebox{\plotpoint}}
\put(324,323){\usebox{\plotpoint}}
\put(323,322){\usebox{\plotpoint}}
\put(321,321){\usebox{\plotpoint}}
\put(320,320){\usebox{\plotpoint}}
\put(319,319){\usebox{\plotpoint}}
\put(317,318){\usebox{\plotpoint}}
\put(316,317){\usebox{\plotpoint}}
\put(314,316){\usebox{\plotpoint}}
\put(313,315){\usebox{\plotpoint}}
\put(312,314){\usebox{\plotpoint}}
\put(310,313){\usebox{\plotpoint}}
\put(309,312){\usebox{\plotpoint}}
\put(307,311){\usebox{\plotpoint}}
\put(306,310){\usebox{\plotpoint}}
\put(305,309){\usebox{\plotpoint}}
\put(303,308){\usebox{\plotpoint}}
\put(302,307){\usebox{\plotpoint}}
\put(300,306){\usebox{\plotpoint}}
\put(299,305){\usebox{\plotpoint}}
\put(297,304){\usebox{\plotpoint}}
\put(296,303){\usebox{\plotpoint}}
\put(295,302){\usebox{\plotpoint}}
\put(293,301){\usebox{\plotpoint}}
\put(292,300){\usebox{\plotpoint}}
\put(290,299){\usebox{\plotpoint}}
\put(289,298){\usebox{\plotpoint}}
\put(288,297){\usebox{\plotpoint}}
\put(286,296){\usebox{\plotpoint}}
\put(285,295){\usebox{\plotpoint}}
\put(283,294){\usebox{\plotpoint}}
\put(282,293){\usebox{\plotpoint}}
\put(281,292){\usebox{\plotpoint}}
\put(281,291){\usebox{\plotpoint}}
\put(370,338){\usebox{\plotpoint}}
\put(368,338){\usebox{\plotpoint}}
\put(367,337){\usebox{\plotpoint}}
\put(365,336){\usebox{\plotpoint}}
\put(364,335){\usebox{\plotpoint}}
\put(362,334){\usebox{\plotpoint}}
\put(361,333){\usebox{\plotpoint}}
\put(360,332){\usebox{\plotpoint}}
\put(358,331){\usebox{\plotpoint}}
\put(357,330){\usebox{\plotpoint}}
\put(355,329){\usebox{\plotpoint}}
\put(354,328){\usebox{\plotpoint}}
\put(353,327){\usebox{\plotpoint}}
\put(351,326){\usebox{\plotpoint}}
\put(350,325){\usebox{\plotpoint}}
\put(348,324){\usebox{\plotpoint}}
\put(347,323){\usebox{\plotpoint}}
\put(346,322){\usebox{\plotpoint}}
\put(344,321){\usebox{\plotpoint}}
\put(343,320){\usebox{\plotpoint}}
\put(341,319){\usebox{\plotpoint}}
\put(340,318){\usebox{\plotpoint}}
\put(339,317){\usebox{\plotpoint}}
\put(337,316){\usebox{\plotpoint}}
\put(336,315){\usebox{\plotpoint}}
\put(334,314){\usebox{\plotpoint}}
\put(333,313){\usebox{\plotpoint}}
\put(332,312){\usebox{\plotpoint}}
\put(330,311){\usebox{\plotpoint}}
\put(329,310){\usebox{\plotpoint}}
\put(327,309){\usebox{\plotpoint}}
\put(326,308){\usebox{\plotpoint}}
\put(325,307){\usebox{\plotpoint}}
\put(323,306){\usebox{\plotpoint}}
\put(322,305){\usebox{\plotpoint}}
\put(320,304){\usebox{\plotpoint}}
\put(319,303){\usebox{\plotpoint}}
\put(318,302){\usebox{\plotpoint}}
\put(316,301){\usebox{\plotpoint}}
\put(315,300){\usebox{\plotpoint}}
\put(313,299){\usebox{\plotpoint}}
\put(312,298){\usebox{\plotpoint}}
\put(311,297){\usebox{\plotpoint}}
\put(309,296){\usebox{\plotpoint}}
\put(308,295){\usebox{\plotpoint}}
\put(306,294){\usebox{\plotpoint}}
\put(305,293){\usebox{\plotpoint}}
\put(304,292){\usebox{\plotpoint}}
\put(302,291){\usebox{\plotpoint}}
\put(301,290){\usebox{\plotpoint}}
\put(299,289){\usebox{\plotpoint}}
\put(298,288){\usebox{\plotpoint}}
\put(297,287){\usebox{\plotpoint}}
\put(297,286){\usebox{\plotpoint}}
\put(320,195){\usebox{\plotpoint}}
\put(320,195){\rule[-0.175pt]{24.090pt}{0.350pt}}
\end{picture}
	\input{5.23.tex}
	% GNUPLOT: LaTeX picture
\setlength{\unitlength}{0.240900pt}
\ifx\plotpoint\undefined\newsavebox{\plotpoint}\fi
\sbox{\plotpoint}{\rule[-0.175pt]{0.350pt}{0.350pt}}%
\begin{picture}(250,300)(0,100)
\tenrm
\sbox{\plotpoint}{\rule[-0.175pt]{0.350pt}{0.350pt}}%
\put(370,95){\makebox(0,0){$I_{5,24}$}}
\put(370,346){\circle*{40}}
\put(289,288){\circle*{40}}
\put(320,195){\circle*{40}}
\put(420,195){\circle*{40}}
\put(451,288){\circle*{40}}
\put(370,346){\usebox{\plotpoint}}
\put(368,346){\usebox{\plotpoint}}
\put(367,345){\usebox{\plotpoint}}
\put(365,344){\usebox{\plotpoint}}
\put(364,343){\usebox{\plotpoint}}
\put(363,342){\usebox{\plotpoint}}
\put(361,341){\usebox{\plotpoint}}
\put(360,340){\usebox{\plotpoint}}
\put(358,339){\usebox{\plotpoint}}
\put(357,338){\usebox{\plotpoint}}
\put(356,337){\usebox{\plotpoint}}
\put(354,336){\usebox{\plotpoint}}
\put(353,335){\usebox{\plotpoint}}
\put(351,334){\usebox{\plotpoint}}
\put(350,333){\usebox{\plotpoint}}
\put(349,332){\usebox{\plotpoint}}
\put(347,331){\usebox{\plotpoint}}
\put(346,330){\usebox{\plotpoint}}
\put(344,329){\usebox{\plotpoint}}
\put(343,328){\usebox{\plotpoint}}
\put(342,327){\usebox{\plotpoint}}
\put(340,326){\usebox{\plotpoint}}
\put(339,325){\usebox{\plotpoint}}
\put(337,324){\usebox{\plotpoint}}
\put(336,323){\usebox{\plotpoint}}
\put(335,322){\usebox{\plotpoint}}
\put(333,321){\usebox{\plotpoint}}
\put(332,320){\usebox{\plotpoint}}
\put(330,319){\usebox{\plotpoint}}
\put(329,318){\usebox{\plotpoint}}
\put(328,317){\usebox{\plotpoint}}
\put(326,316){\usebox{\plotpoint}}
\put(325,315){\usebox{\plotpoint}}
\put(323,314){\usebox{\plotpoint}}
\put(322,313){\usebox{\plotpoint}}
\put(321,312){\usebox{\plotpoint}}
\put(319,311){\usebox{\plotpoint}}
\put(318,310){\usebox{\plotpoint}}
\put(316,309){\usebox{\plotpoint}}
\put(315,308){\usebox{\plotpoint}}
\put(314,307){\usebox{\plotpoint}}
\put(312,306){\usebox{\plotpoint}}
\put(311,305){\usebox{\plotpoint}}
\put(309,304){\usebox{\plotpoint}}
\put(308,303){\usebox{\plotpoint}}
\put(307,302){\usebox{\plotpoint}}
\put(305,301){\usebox{\plotpoint}}
\put(304,300){\usebox{\plotpoint}}
\put(302,299){\usebox{\plotpoint}}
\put(301,298){\usebox{\plotpoint}}
\put(300,297){\usebox{\plotpoint}}
\put(298,296){\usebox{\plotpoint}}
\put(297,295){\usebox{\plotpoint}}
\put(295,294){\usebox{\plotpoint}}
\put(294,293){\usebox{\plotpoint}}
\put(293,292){\usebox{\plotpoint}}
\put(291,291){\usebox{\plotpoint}}
\put(290,290){\usebox{\plotpoint}}
\put(289,289){\usebox{\plotpoint}}
\put(289,288){\usebox{\plotpoint}}
\put(370,346){\usebox{\plotpoint}}
\put(370,346){\usebox{\plotpoint}}
\put(371,345){\usebox{\plotpoint}}
\put(372,344){\usebox{\plotpoint}}
\put(374,343){\usebox{\plotpoint}}
\put(375,342){\usebox{\plotpoint}}
\put(376,341){\usebox{\plotpoint}}
\put(378,340){\usebox{\plotpoint}}
\put(379,339){\usebox{\plotpoint}}
\put(381,338){\usebox{\plotpoint}}
\put(382,337){\usebox{\plotpoint}}
\put(383,336){\usebox{\plotpoint}}
\put(385,335){\usebox{\plotpoint}}
\put(386,334){\usebox{\plotpoint}}
\put(388,333){\usebox{\plotpoint}}
\put(389,332){\usebox{\plotpoint}}
\put(390,331){\usebox{\plotpoint}}
\put(392,330){\usebox{\plotpoint}}
\put(393,329){\usebox{\plotpoint}}
\put(395,328){\usebox{\plotpoint}}
\put(396,327){\usebox{\plotpoint}}
\put(397,326){\usebox{\plotpoint}}
\put(399,325){\usebox{\plotpoint}}
\put(400,324){\usebox{\plotpoint}}
\put(402,323){\usebox{\plotpoint}}
\put(403,322){\usebox{\plotpoint}}
\put(404,321){\usebox{\plotpoint}}
\put(406,320){\usebox{\plotpoint}}
\put(407,319){\usebox{\plotpoint}}
\put(409,318){\usebox{\plotpoint}}
\put(410,317){\usebox{\plotpoint}}
\put(411,316){\usebox{\plotpoint}}
\put(413,315){\usebox{\plotpoint}}
\put(414,314){\usebox{\plotpoint}}
\put(416,313){\usebox{\plotpoint}}
\put(417,312){\usebox{\plotpoint}}
\put(418,311){\usebox{\plotpoint}}
\put(420,310){\usebox{\plotpoint}}
\put(421,309){\usebox{\plotpoint}}
\put(423,308){\usebox{\plotpoint}}
\put(424,307){\usebox{\plotpoint}}
\put(425,306){\usebox{\plotpoint}}
\put(427,305){\usebox{\plotpoint}}
\put(428,304){\usebox{\plotpoint}}
\put(430,303){\usebox{\plotpoint}}
\put(431,302){\usebox{\plotpoint}}
\put(432,301){\usebox{\plotpoint}}
\put(434,300){\usebox{\plotpoint}}
\put(435,299){\usebox{\plotpoint}}
\put(437,298){\usebox{\plotpoint}}
\put(438,297){\usebox{\plotpoint}}
\put(439,296){\usebox{\plotpoint}}
\put(441,295){\usebox{\plotpoint}}
\put(442,294){\usebox{\plotpoint}}
\put(444,293){\usebox{\plotpoint}}
\put(445,292){\usebox{\plotpoint}}
\put(446,291){\usebox{\plotpoint}}
\put(448,290){\usebox{\plotpoint}}
\put(449,289){\usebox{\plotpoint}}
\put(450,288){\usebox{\plotpoint}}
\put(310,181){\usebox{\plotpoint}}
\put(310,181){\rule[-0.175pt]{28.908pt}{0.350pt}}
\put(320,195){\usebox{\plotpoint}}
\put(320,195){\rule[-0.175pt]{24.090pt}{0.350pt}}
\put(330,208){\usebox{\plotpoint}}
\put(330,208){\rule[-0.175pt]{19.272pt}{0.350pt}}
\end{picture}
	}
\end{figure}
%\end{document}


\begin{thebibliography}{99}
\bibitem{dbl_float} In C this constant $d$ is conventionally called
DBL\_DIG and can be found in a standard include header file called
float.h.
\end{thebibliography}
 \end{document}